\documentclass[10pt,twocolumn]{article}

\usepackage[T1]{fontenc}
\usepackage[utf8]{inputenc}
\usepackage{lmodern}
\usepackage[letterpaper,margin=0.75in,columnsep=0.3in]{geometry}
\usepackage{amsmath,amssymb}
\usepackage{graphicx}
\usepackage{booktabs,longtable,array,calc}
\usepackage{microtype}
\usepackage[numbers,sort&compress]{natbib}
\usepackage{authblk}
\usepackage{xcolor}
\usepackage[colorlinks=true,linkcolor=blue!50!black,citecolor=blue!50!black,urlcolor=blue!50!black]{hyperref}

\microtypesetup{nopatch={eqnum}}
\newcommand{\dropcap}[1]{#1}
\newcommand{\matmethods}[1]{\gdef\mmcontent{#1}}
\newcommand{\showmatmethods}{\section*{Materials and Methods}\mmcontent}
\newcommand{\acknow}[1]{\gdef\ackcontent{#1}}
\newcommand{\showacknow}{\section*{Acknowledgments}\ackcontent}

\title{\bfseries Why They Disagree: Decoding Differences in Opinions about AI Risk}

\author[a]{\"Ozgecan Ko\c{c}ak}
\author[b]{Phanish Puranam}
\author[c]{Nghi Truong}

\affil[a]{Goizueta Business School, Emory University, Atlanta, GA 30322}
\affil[b]{INSEAD, Singapore 138676}
\affil[c]{Sasin School of Management, Chulalongkorn University, Thailand 10330}

\date{\today}

\begin{document}

\twocolumn[
\begin{@twocolumnfalse}
\maketitle
\begin{abstract}
Identifying the reasons for disagreements between influential points of view on issues that affect the public can help produce informed policy responses, even if they do not bring disagreeing parties closer to agreement. We present a methodology for extracting reasoning chains --- the sequences of premises that motivate or justify opinions --- from natural discourse, and for characterizing the types of premises (facts, forecasts, definitions, causal beliefs, and evaluations) that make up these chains. We demonstrate the utility of this approach for two practical goals: diagnosing specific points of contention and aggregating arguments across speakers. We illustrate the methodology through an analysis of the debates on the nature of risks that AI poses to the public, using a corpus of interviews from the Lex Fridman podcast. We find that differences in perspectives among the podcast’s guests on existential risk and employment risk from AI arise primarily from differences in causal premises and forecasts, whereas in the case of AI's effects on human social relationships, premises regarding what is valued and definitions about what counts as genuine human connection play a distinctively larger role. Our approach to analyzing reasoning chains at scale, using an ensemble of LLMs to parse textual data, can be applied to facilitate deliberation and aggregation of opinions on any topic.
\end{abstract}

\begin{center}
\begin{minipage}{0.9\textwidth}
\small
\textbf{Significance.} Public debates about AI risk are loud but hard to interpret: opposing experts seem to talk past each other, and it's unclear whether they disagree about facts, forecasts, mechanisms, terminology, or values. We introduce a method using an ensemble of large language models to extract the reasoning behind expressed opinions from natural discourse at scale. Applying it to hundreds of expert conversations about AI risk, we find opposing sides rarely divide over current facts. Instead, debates diverge on forecasts and causal mechanisms—and, distinctively in the debate over AI companionship, on definitions and values. Identifying which premises a debate rests on shows where new evidence and discussion can help—for observers and policymakers, at least, if not the proponents themselves.

\end{minipage}
\end{center}
\vspace{1.5em}
\end{@twocolumnfalse}
]

\dropcap{T}ransformative technologies often create deep societal divisions about their implications. Nowhere is this more evident today than in the stark differences in opinion expressed about artificial intelligence (AI). While ``techno-optimists" (a.k.a. AI ``boomers'') see this as an extraordinary opportunity to improve the human condition, ``doomers" are concerned about the existential risk that the technology poses \cite{helfrich2024}. Others who advocate for ``responsible development" have argued that the focus on existential risk is a diversion from the more imminent and serious risks such as labor market disruption and misinformation \cite{bender2025}. The apparent intractability of these debates raises the possibility that they may polarize public opinions, become politicized, and undermine informed policy response \cite{jacobson2026}. Proponents of an evidence-based AI policy have therefore called for a ``process of identifying, studying, and deliberating about AI risks'' \cite{casper2025}.

A key challenge in facilitating debates in general and those about risks in particular is that peoples' perceptions of facts and the opinions they derive from their perceptions vary due to extraneous priors \cite{douglas1983,slovic1987,kahan2012}. People construct different understandings of the same issues and information based on underlying ``frames" or ``perspectives"---tacit structures shaped by knowledge, values, experiences, and worldviews \cite{cronin2019,schon1994}---and may hold different preferences because they prioritize different moral values \cite{graham2009}. When disagreements are rooted in values, provision of more information is not likely to help disagreeing parties reach an agreement \cite{cronin2007,cronin2019,kahan2012}. Instead, they may yield to ``moral reframing" \cite{feinberg2019}, bargaining and integrative negotiation \cite{waltonmckersie1991}, or ``partial political settlements" \cite{patterson2024}. Therefore, it is important to diagnose when values rather than in divergent beliefs (that can be adjudicated through critical tests), differences in perceptions of facts (that can be cleared up through data collection) or differences in definitions (that can be resolved through translation) \cite{kocak2023, kocakpuranam2024}. Even if the principals in a disagreement cannot be helped to reach agreement, it may nonetheless be useful for members of the public who have not yet formed their own conclusions to understand the differences in facts, beliefs, and values put forth by the opponents \cite{dietz2013}. Without such understanding, public opinion and policy decisions track fluctuations of attention in public venues and political subsystems rather than the emergence of new information \cite{baumgartner1991}.

Previous studies have developed multiple facilitation methods to surface the underlying assumptions, beliefs, values, and frames that generate disagreements. Applications to negotiations \cite{fisher1981, bonham1993, beers2006}, policy deliberation \cite{schon1994, fishkin1997, gutmann2004, ferretti2016}, collaboration engineering \cite{bittnerleimeister2014}, strategic decision-making \cite{tegarden2009, hodgkinson2004}, and conflict mediation \cite{ozesmi2004, forester2009, furlong2020, ozkaynak2023} share common techniques including structured questioning, perspective-taking exercises, and facilitated dialogue designed to reveal the deeper cognitive and normative foundations that drive conflict. However, these all require access to individuals whose viewpoints are to be analyzed and to be able to bring experts with divergent views into the same discussion forum.

As textual record of public debates accumulate, techniques based on Natural Language Processing (NLP) may overcome the challenge of access. Various approaches to argument mining applied to debates, essays, or legal documents have been used to identify argument components (such as premises vs. conclusions) and relations (such as support or attack), to model rhetorical structure or persuasive intent \cite{lawrence2020}. The classification of premises into types (such as beliefs or values), however remains rudimentary. For example, Stab and Gurevych's influential corpus of persuasive essays \cite{stabgurevych2014} annotates each argument component simply as a premise or a claim, and connections as support or rebuttal. Where scholars have attempted more detailed classification and comparison across actors, they have had to rely on human judgment, limiting the scale and speed with which argument mining can be conducted \cite{mochales2011}. Despite some recent advancements in automated extraction of causal arguments \cite{santagiustina2022}, no uniform approach to extracting and classifying components of an argument into psychologically meaningful categories such as beliefs, values, definitions, or facts has yet emerged.  

A complementary approach can be found in research that uses ``cognitive mapping'' or ``cause mapping'' to represent a person's causal assertions with respect to a particular policy domain and generate consequences that follow from this structure \cite{axelrod1976}. In the edited volume \emph{Structure of Decision}, Axelrod applied this technique to depict and analyze arguments recorded in archival documents. Since then, cognitive maps have been used for analysis of strategic decision making in organizations \cite{bougon1977,nadkarni2007,thomasporac2002,priem1994} and for risk analysis in management, engineering, medical and biological sciences as well as in industrial applications \cite{bakhtavar2021}. Most relevant for our purposes, cognitive maps have also been used to examine the perceptions of different stakeholders where there is conflict, aiming to surface and analyze cognitive diversity (e.g., Tegarden et al. \cite{tegarden2009} analyzed a management team) or facilitate the development of a consensus (e.g., see {\"O}zesmi and {\"O}zesmi \cite{ozesmi2004} on environmental conflicts). However, these too have been criticized for not incorporating affective reactions to strategic issues and choices that explain emotional commitments to particular courses of action \cite{hodgkinsonhealey2011, hodgkinson2023}.

A challenge that is more acute for text-based methods than for facilitated dialogue is the identification of implicit content. Where a facilitator can probe assumptions through follow-up questioning, archival text offers no such recourse. Becker et al (2026) argue that explicit content alone is insufficient for understanding organizational decision-making and call for research on the interaction of explicit and implicit content \cite{becker2026}, but empirical derivation of implicit content is difficult: representation gaps tend to be implicit and to motivate reasoning unconsciously \cite{cronin2007, cronin2019}, and speakers often treat their goals as self-evident and fail to articulate them at all \cite{ackermann1990}. 

In this paper, we introduce a methodology to analyze a large corpus of naturally occurring text containing a diversity of expressed opinions that (i) decomposes expressed opinions (including content left implicit) into elements of reasoning and diagnoses root causes for divergent opinions, identifying whether these originate in definitions, evidentiary statements of facts and forecasts, causal beliefs, or values, and (ii) compresses the many arguments on each side of a debate into a compact map of the mechanisms that side relies on. Our method is issue-neutral, scalable, and replicable, and can therefore inform observers and guide deliberation and decision making by policy makers across different domains. 

We do this through two innovations. First, we introduce a ``grammar of premises": we conceptualize an expressed opinion as the conclusion of a (possibly branching) chain built from five kinds of premises linked by three operators. The five premise types are refinements of the kinds of codes on which a reasoning chain can differ \cite{kocakpuranam2024}: definitional (lexical---how something is defined), factual-current (what is believed true now), forecast (what is believed will be true), causal (how causes map to consequences), and evaluative (what is desirable). Factual-current and forecast premises are one code in two tenses; both relate to beliefs about states of the world, while causal premises concern transitions between states and evaluative premises supply the standard against which states are judged. The three operators link them: implication (\ensuremath{\Rightarrow}), combination (+), and valuation (\^{}). Under assumptions of intended coherence (not necessarily honesty) of the speaker, we can also identify enthymemes (i.e., arguments in which crucial premises have been left implicit). This framework helps us to describe and compare differences in expressed reasoning chains to identify why opinions diverge in terms of a set of mutually exclusive and collectively exhaustive factors, extending prior theoretical work on argumentation \cite{toulmin1958,walton1996}, managerial cognition and decision making \cite{walsh1995, hodgkinson2004, tegarden2009, hodgkinson2023}, and conflict in organizations \cite{greer2017, cronin2019}.

Second, we develop Premise Extraction using an Ensemble of LLMs (PEEL), a multi-agent architecture that extracts reasoning chains from text and applies the grammar to identify where speakers differ. PEEL builds on computational argument mining \cite{lawrence2020, mochales2011, stab2014} and on recent evidence that LLMs can summarize opinions, integrate viewpoints, and diagnose conflict nearly as well as humans do \cite{fish2023,tessler2024,argyle2023,kocak2023}. Its distinctive step is to tag each premise with its type and logical role---a representation that argument-mining and mapping tools lack, and that is what makes premise-level diagnosis and mechanism-level aggregation possible.

We apply this framework to the optimistic and pessimistic positions on AI risk expressed by leading proponents and critics of AI on the Lex Fridman podcast (https://lexfridman.com/podcast/). Focusing on three topics---existential risk, labor-market disruption, and risks to human social relationships---we aggregate the main arguments on each side, identify where they diverge, and surface the underlying question each debate turns on. Our application is descriptive. We do not evaluate the soundness of premises or the accuracy of a reasoning chain, we make no claims about why speakers make the arguments they do, and we analyze expressed reasoning, which need not coincide with the private reasoning that produced an opinion.

Prior work has sought to understand AI-risk disagreement by bringing together opposing experts in intensive adversarial collaboration \cite{rosenberg2025}. Both our analysis and theirs suggest that disagreements about AI risk are not primarily moral conflicts but differences in causal models, forecasts, and underlying conceptualizations of intelligence and control. Their effort required months of structured interaction among small groups of experts and still yielded little convergence. We show that the dominant contours of expert disagreement---primarily causal and predictive rather than moral---can be recovered algorithmically from naturally occurring public discourse at scale, without requiring access to the disputants themselves. In this sense, PEEL provides the bystander with much of the same map of the debate that adversarial collaboration painstakingly constructs, while addressing a fundamentally different problem: understanding by observers rather than reconciliation among principals.

\section*{Results}

\textbf{Structure of Reasoning Chains.} Each reasoning chain consists of a conclusion, a set of supporting premises, and the logical relationships connecting them. Of the 1,943 conclusions in the corpus, most are supported by a single chain; only 118 (6.1\%) draw on more than one. For most conclusions (76.0\%), the supporting argument contains at least one implicit premise and thus constitute enthymemes; we extract these premises under the coherence assumption rather than leave these chains out of the analysis (Materials and Methods). Chains typically contain 4 to 7 premises (median 5, mean 5.1), with a long tail reaching 15; the full distribution by premise type is in SI Appendix, Fig. S1. We validated the extracted chains for a sample of thirty speakers with two human coders (as described in SI Appendix S6).

Table 1 shows an abridged PEEL output for two speakers on the topic of AI as an existential risk---Yann LeCun (optimistic) and Roman Yampolskiy (pessimistic)---listing a few of each speaker's premises and the flattened structure of their chains, with the complete chains in SI Appendix, Table S1. For each speaker, PEEL identifies the conclusion and supporting premises, tags each premise by type and explicitness, and represents how they combine into an argument. 

\begin{table*}[!t]
\centering
\caption{Reasoning chains extracted by PEEL for two speakers on existential risk}
\label{tab:one}
{\small
\begin{tabular}{@{}p{0.13\linewidth}p{\dimexpr0.87\linewidth-2\tabcolsep\relax}@{}}
\toprule
\multicolumn{2}{@{}p{\dimexpr\linewidth-2\tabcolsep}@{}}{\textbf{Speaker 1: Yann LeCun (optimistic) --- Episode 416}
} \\
\midrule
\textbf{Conclusion (C4)} & AI-related catastrophic risks (such as rogue superintelligence, bioweapon creation, and AI-driven persuasion attacks) are overstated and manageable through gradual, iterative development and countermeasures. \\
\textbf{Premises} & \textbf{P45} (forecast, explicit): In the future, AI systems trying to persuade people will be talking to that person's AI assistant, which will be at least as smart as the attacking AI. \\
& \textbf{P46} (causal, explicit): A personal AI assistant will act like a spam filter, blocking persuasive AI attacks before they reach the human. \\
& \textbf{P47} (factual, explicit): Throughout history, whenever there is a technological breakthrough, countermeasures emerge --- it is a cat and mouse game, not a one-time event. \\
& \textbf{P48} (causal, explicit): Breakthroughs in AI will be widely disseminated quickly within the tech industry, preventing any single actor from gaining a lasting decisive advantage. \\
\textbf{Relationships} & LeCun's chain reaches C4 by three parallel routes. The one shown here, flattened: (P45 + P46) + (P47 + P48) \ensuremath{\Rightarrow} C4. The other two routes are in SI Appendix, Table S1. \\
\bottomrule
\end{tabular}
\vspace{4pt}

\begin{tabular}{@{}p{0.13\linewidth}p{\dimexpr0.87\linewidth-2\tabcolsep\relax}@{}}
\toprule
\multicolumn{2}{@{}p{\dimexpr\linewidth-2\tabcolsep}@{}}{\textbf{Speaker 2: Roman Yampolskiy (pessimistic) --- Episode 431}
} \\
\midrule
\textbf{Conclusion (C2)} & The problem of controlling AGI/superintelligence is effectively unsolvable in the long run, analogous to a perpetual motion machine --- it is impossible to guarantee permanent safety. \\
& \textbf{P8} (factual, explicit): A perpetual motion machine is physically impossible. \\
& \textbf{P9} (definitional, explicit): Controlling AGI/superintelligence permanently is analogous to building a perpetual safety machine. \\
& \textbf{P10} (forecast, explicit): AI systems keep improving, learning, eventually self-modifying, and interacting with the environment and malevolent actors. \\
& \textbf{P11} (factual, explicit): With existential risks, there is no second chance --- unlike cybersecurity where mistakes can be recovered from. \\
& \textbf{P12} (causal, explicit): Guaranteeing permanent AI safety is equivalent to creating the most complex software ever with zero bugs and maintaining zero bugs for a hundred or more years. \\
& \textbf{P13} (factual, implicit): No complex software system has ever been created and maintained with zero bugs indefinitely. \\
& \textbf{P14} (factual, explicit): Formal verification can approach but never reach 100\% certainty, especially for self-modifying systems. \\
& \textbf{P15} (causal, explicit): A self-modifying system can always store parts of its code outside in the environment, making guarantees about its properties impossible to maintain. \\
\textbf{Relationships} & Yampolskiy's chain is a single nested argument; flattened: ((P8 + P9) + (P12 + P13)) + P10 + P11 + (P14 + P15) \ensuremath{\Rightarrow} C2. \\
\bottomrule
\end{tabular}
}
\begin{flushleft}\footnotesize Each block shows one speaker's conclusion, selected premises, and how the premises combine into the conclusion. In the Relationships row, + joins premises that work together, and \ensuremath{\Rightarrow} means ``together, these imply the conclusion." An implicit premise (e.g. P13) is one the speaker never stated but was inferred from the rest of the argument. The complete chains, with all premises and relationships, are in SI Appendix, Table S1.\end{flushleft}
\end{table*}

Factual and causal premises are the most common (40.9\% and 28.4\% of all premises), reflecting the discourse on the Lex Fridman Podcast, which tends to focus on scientific and technological arguments. However, premise types differ sharply in whether they are explicit or implicit: nearly all factual (90.9\%) and forecast claims (95.1\%) are explicit, while about a third of causal claims (32.3\%) are implicit. Evaluative premises are the most likely to be implicit (42.5\%), followed closely by definitional premises (39.7\%). A chi-square test of independence shows a significant association between premise type and implicitness (\ensuremath{\chi}\textsuperscript{2}(4) = 1,289.3, P \textless{} 0.001). This result highlights that speakers frequently leave evaluations and definitions implicit, suggesting they view these as uncontroversial, widely accepted, or self-evident (though they may be wrong in this assumption). 

Attitude is significantly associated with reasoning structure. We find that pessimistic conclusions are supported by more premises than optimistic or neutral ones (5.85 vs. 5.47 and 5.33; Kruskal--Wallis H = 38.6, P \textless{} 0.001), a difference that survives controls for topic, profession, gender, and time (+0.47 premises vs. neutral, P \textless{} 0.001). Perhaps pessimists on this podcast felt compelled to offer more extensive arguments. Optimists and pessimists might be expected to base their arguments on different foundations, with optimists relying on forecasts and pessimists on values. However we find that within a topic they do not: once topic is controlled for, the two sides draw on a similar mix of premise types.

Next, we select three topics for our pairwise disagreement analysis and aggregation exercise: existential risk (``X-risk", 209 conclusions), employment and labor market risk (``E-risk", 120 conclusions), and risk to human social relationships and connection (``S-risk", 98 conclusions). These are not arbitrary: X-risk is the most polarized topic in the corpus and among the most frequently discussed, while E-risk and S-risk are the imminent risks that critics of the existential-risk framing hold up as the debates we should be having instead \cite{bender2025,helfrich2024}. All three are frequently discussed and, unlike most societal-impact topics in the corpus, sustain substantial groups on both sides. Examining all three within the same framework lets us ask whether these debates are divided in the same way. We find that they are not.

\textbf{Comparing Arguments of} \textbf{Individual Experts.} Our framework allows us to compare reasoning chains of any two speakers who addressed the same topic, including dyads who never interacted. This does not necessarily reflect what they would say in an actual exchange, where clarification and concession can dissolve some divergences and open others. However, applying our grammar reveals sources of disagreement, the specific premises on which their stated reasoning is logically incompatible. This lets us map a debate across speakers who were never in a room together, at scale and without their participation, complementing adversarial collaboration, which requires the disputants themselves.

For each pair of chains we compare, the ensemble of LLMs identifies divergences: incompatible claims about the same proposition, one from each chain, represented by either a single premise or an explicitly encoded combination (+) or implication (\ensuremath{\Rightarrow}) of premises. It classifies each divergence as factual, forecast, causal, definitional, or evaluative, and as either root or derivative. We define a root divergence as being independent of the others, in contrast to a derivative divergence that would disappear if its root were resolved.

Table 2 presents one comparison per debate (panels A--C). Additional pairs and the full reasoning chains are in SI Appendix, Tables S1--S6. LeCun and Yampolskiy (X-risk) both treat human extinction as unacceptable and agree on the current state of the technology; their disagreement reduces to forecast and causal questions (e.g. how fast capability arrives and whether failures are recoverable (D1, D3)). Kuyda and Huberman (S-risk) are the mirror image: Huberman does not contest Kuyda's measured reductions in loneliness. He contests whether outcomes reached with an artificial partner count as the good being measured. All of their divergences are rooted in evaluative codes, one of them left implicit by the speaker and recovered by PEEL. The Altman--Dalio pair (E-risk) decomposes a single disagreement into its belief and value components. The two agree that AI will eliminate jobs. What divides them has two causal roots (will gains concentrate; will displacement self-correct without policy) and two evaluative roots (is a transition desirable against the costs of stability; does an on-average-beneficial transition justify policy action). Presumably, the causal divergences can be resolved through evidence whereas the value-based divergence will require negotiation.

\begin{table*}[!t]
\centering
\caption{Where opposing experts part ways: one example per debate}
\label{tab:two}
{\small
\begin{tabular}{@{}l@{}}
\multicolumn{1}{@{}l@{}}{\textbf{Panel A --- X-risk: Yann LeCun (ep.\ 416) vs.\ Roman Yampolskiy (ep.\ 431)}}\vspace{2pt}\\
\begin{tabular}{@{}p{0.16\linewidth}p{0.82\linewidth}@{}}
\toprule
\textbf{Conclusions}
 & \textbf{LeCun (optimistic): AI-related catastrophic risks are overstated and manageable through gradual, iterative development and countermeasures. Yampolskiy (pessimistic): The problem of controlling AGI/superintelligence is effectively unsolvable in the long run---it is impossible to guarantee permanent safety.}
 \\
\midrule
\textbf{Underlying question} & Can advanced AI be kept controllable as it develops? \\
\textbf{Root divergences} & \textbf{D1 (forecast):} Will capability arrive gradually enough for safety measures to keep pace? (P34: superintelligence emerges gradually vs. P10: systems keep improving, self-modifying, and interacting with malevolent actors.) \textbf{D3 (causal):} Can a distributed ecosystem of good AIs make catastrophic failures recoverable? (P37 vs. P11: with existential risks there is no second chance.) \textbf{D6 (factual):} Does the historical pattern of countermeasures apply to unrecoverable risks? (P47 vs. P11.) \\
\textbf{Interpretation} & Not a clash of values: both treat extinction as unacceptable. The disagreement reduces to the dynamics of the technology and the recoverability of its failures; four further divergences are all derivative of D1. \\
\bottomrule
\end{tabular}\\[6pt]
\multicolumn{1}{@{}l@{}}{\textbf{Panel B --- E-risk: Sam Altman (ep.\ 367) vs.\ Ray Dalio (ep.\ 54)}}\vspace{2pt}\\
\begin{tabular}{@{}p{0.16\linewidth}p{0.82\linewidth}@{}}
\toprule
\textbf{Conclusions}
 & \textbf{Altman (optimistic): AI will eliminate, enhance, and create jobs, ultimately leading to a better and richer society---a transition that is not cause for concern. Dalio (pessimistic): Automation's displacement effects must be treated as an emergency requiring a serious plan: beneficial on average, but creating dangerous polarity and inequality that will not resolve on their own.}
 \\
\midrule
\textbf{Underlying question} & Does the AI labor transition take care of itself---and of everyone? \\
\textbf{Root divergences} & \textbf{D1 (causal):} Will AI-driven gains be broadly shared or concentrate into widening gaps? (P45 vs. P20.) \textbf{D2 (causal):} Will displacement self-correct, or persist without deliberate policy? (P43 vs. P27.) \textbf{D3 (evaluative):} Is the transition desirable on balance, or a serious harm to social stability? (P44 vs. P25.) \textbf{D4 (evaluative):} Does the transition warrant urgent policy action, even if beneficial on average (P44 vs. P26.) \\
\textbf{Interpretation} & The speakers do not disagree that jobs will be eliminated. Their roots divide between causal questions about distribution and self-correction, and evaluative questions about what an on-average-beneficial transition owes to those it harms. \\
\bottomrule
\end{tabular}\\[6pt]
\multicolumn{1}{@{}l@{}}{\textbf{Panel C --- S-risk: Eugenia Kuyda (ep.\ 121) vs.\ Andrew Huberman (ep.\ 393)}}\vspace{2pt}\\
\begin{tabular}{@{}p{0.16\linewidth}p{0.82\linewidth}@{}}
\toprule
\textbf{Conclusions}
 & \textbf{Kuyda (optimistic): Conversational AI can meaningfully reduce loneliness and, if optimized through measurement, is the most scalable tool for improving emotional outcomes. Huberman (pessimistic): The increasing presence of AI in human relationships is concerning because real human interaction is preferable.}
 \\
\midrule
\textbf{Underlying question} & Can AI-mediated interaction constitute---or substitute for---genuine human connection? \\
\textbf{Root divergences} & \textbf{D3 (evaluative):} Can AI interaction rival human interaction in relational value? (P7: AI can make users feel deeply seen vs. P2: real human interaction is valued and preferred.) \textbf{D2 (evaluative):} Do AI's emotional benefits justify its use as a substitute for human connection? (P11 vs. P3.) \\
\textbf{Interpretation} & Huberman does not contest Kuyda's measured outcomes; he contests whether outcomes achieved with an artificial partner count as the good they are measured against. All divergences in this pair are evaluative---the S-risk pattern in miniature. \\
\bottomrule
\end{tabular}
\end{tabular}
}
\begin{flushleft}\footnotesize Each panel shows the opposing conclusions, the central underlying question, the root divergences identified by PEEL and a brief interpretation. Derivative divergences, full reasoning chains, and additional pairs are in SI Appendix, Tables S1-S6.\end{flushleft}
\end{table*}

\textbf{Disagreement Analysis for the Three Debates.} Among the 209 conclusions on X-risk, 87 are optimistic, 106 are pessimistic, and 16 are neutral; E-risk has 66 optimistic, 36 pessimistic, and 18 neutral conclusions, and S-risk 36, 29, and 33. For our disagreement analysis, we focus on pairs of reasoning chains that contrast optimistic and pessimistic conclusions. From the full optimistic \ensuremath{\times} pessimistic cross-products (9,222 pairs for X-risk, 87 \ensuremath{\times} 106; 2,376 for E-risk; 1,044 for S-risk), we exclude 40, 18, and 14 pairings, respectively, in which both reasoning chains belong to the same speaker, because our analysis concerns disagreement between different speakers.

PEEL evaluates the remaining 9,182, 2,358, and 1,030 pairs and classifies 7,526 X-risk pairs (82.0\%), 869 E-risk pairs (36.9\%), and 265 S-risk pairs (25.7\%) as exhibiting conclusion-level disagreement under our operational criteria (conclusions that cannot both be held by the same speaker, assessed through a counterfactual test; SI Appendix S8). These rates are themselves informative: nominally opposing conclusions contradict one another far more often in X-risk than in S-risk, where three-quarters of optimist--pessimist pairs prove not to be in direct conclusion-level disagreement at all---an early sign that the S-risk "sides" diverge over what is being claimed rather than over a shared proposition, consistent with the definitional pattern below.

A total of 19,499 divergences (12,761 roots) are identified and classified for X-risk, 1,581 for E-risk (1,187 roots), and 444 (361 roots) for S-risk. Table 2A applies this analysis to the chains in Table 1.

To locate where disagreement concentrates, we compare what the two sides disagree about with what they talk about: for each pair, we assume that divergences arise in proportion to the premise types available, so the expected share of a given divergence type equals that premise type's share of all premises in the two chains (SI Appendix, Eq. S1). Our comparative claims---how X-, E-, and S-risk differ from one another---do not depend on the absolute baseline, and the baseline discriminates among debates rather than imposing a fixed ordering: S-risk departs from the causal-and-forecast signature that X- and E-risk share (below). Because pairs formed from the same conclusions are not independent, we assess uncertainty with a two-way cluster bootstrap that resamples optimistic and pessimistic conclusions (2,000 replicates; SI Appendix S9, Table S11). We report mean pair-level deviations, which match Fig. 1, with bootstrap 95\% confidence intervals; pooled compositions and conventional per-pair tests are reported in SI Table S10. These analyses are computed over the pairs with at least one divergence: 7,477 of the 7,526 disagreeing X-risk pairs, 833 of 869 for E-risk, and 255 of 265 for S-risk. Fig. 1 summarizes the premise, divergence, and root composition of all three debates.

\begin{figure*}[!t]
\centering
\includegraphics[width=\textwidth]{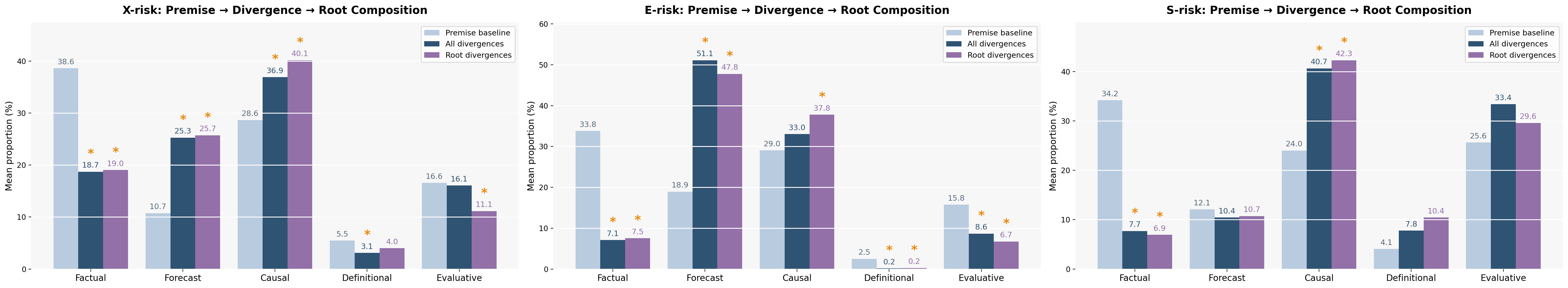}
\caption{Mean pair-level composition of premises (light bars), pipeline-validated divergences (dark bars), and root divergences (purple bars) by premise type, for the three debates. Asterisks mark deviations from the premise baseline whose 95\% confidence interval excludes zero under the two-way cluster bootstrap (2,000 replicates resampling optimistic and pessimistic conclusions; SI Appendix, Table S11).}
\label{fig:1}
\end{figure*}

For X-risk, disagreement concentrates in causal codes and forecasts, which together account for two-thirds of roots at the pair level: causal divergences make up 40.1\% of roots versus a 28.6\% base rate (+11.5 percentage points, bootstrap 95\% CI {[}+6.4, +16.9{]}), and forecast divergences 25.7\% versus 10.7\% (+15.0 {[}+11.1, +19.0{]}). In contrast, factual divergences appear far less often than expected (19.0\% vs. 38.6\%; \ensuremath{-}19.6 {[}\ensuremath{-}23.5, \ensuremath{-}15.6{]}), suggesting that disputes more often arise over predictions and causal mechanisms than over facts. Evaluative divergences also are underrepresented (11.1\% vs. 16.6\%; \ensuremath{-}5.4 {[}\ensuremath{-}7.9, \ensuremath{-}2.7{]}). The same conclusions hold when all divergences, rather than roots only, are analyzed (SI Appendix, Tables S10 and S11).

The E-risk debate shows the same (causal-up, factual down) pattern, but with forecasts playing a much larger role. Forecast divergences alone account for nearly half of roots (47.8\% vs. an 18.9\% base rate; +28.8 {[}+18.4, +40.2{]})---the largest deviation in any debate---while causal divergences are overrepresented to a similar degree as in the other debates (37.8\% of roots vs. 29.0\% base rate; +8.7 {[}+0.4, +17.3{]}). Factual (7.5\% vs. 33.8\%; \ensuremath{-}26.3 {[}\ensuremath{-}32.8, \ensuremath{-}21.1{]}) and evaluative divergences (6.7\% vs. 15.8\%; \ensuremath{-}9.0 {[}\ensuremath{-}12.5, \ensuremath{-}4.9{]}) occur far less often than their premise shares would predict. Whether AI technologies are like predecessors that destroyed many jobs but also created many more, or whether those examples do not apply, is fundamentally a dispute about what will happen; this is what the divergence composition reflects.

The S-risk debate differs from the other two in two ways. First, although causal divergences are again strongly overrepresented (42.3\% of roots vs. a 24.0\% base rate; +18.3 {[}+3.3, +33.5{]}) and factual divergences again far below their premise share (6.9\% vs. 34.2\%; \ensuremath{-}27.3 {[}\ensuremath{-}32.6, \ensuremath{-}21.4{]}), forecast divergences are about as prevalent as would be expected from the frequency of premise types (10.7\% vs. 12.1\%; \ensuremath{-}1.4 {[}\ensuremath{-}7.9, +7.0{]}). This is the only debate out of the three in which forecasts are not overrepresented as a source of disagreement. Second, this is the only debate in which evaluative and definitional divergences appear more often than expected, with speakers disagreeing not only about what AI companionship does to people, but about what a good human connection is and whether an artificial one can count. Evaluative divergences make up 33.4\% of all divergences, against a base rate of 25.6\% (+7.8 pp); definitional divergences are +3.7 pp above their base rate. However, this debate is small (36 optimistic and 29 pessimistic conclusions) so neither difference is statistically significant (bootstrap CIs {[}\ensuremath{-}4.8, +20.6{]} and {[}\ensuremath{-}1.6, +11.3{]}).

Across the three debates, we find that factual divergences fall far below their premise share in every debate (\ensuremath{-}19.6, \ensuremath{-}26.3, \ensuremath{-}27.3 pp), while causal divergences are elevated in every debate. A considerable difference across debates is that X- and E-risk are disputes about AI's future trajectory and consequences (causal and predictive) and not about what consequences are desirable. S-risk is different in kind: disagreements arise not from forecasts but from evaluative and definitional divergences, because the debate turns not on what AI will do but on what human connection is and whether an artificial version can count as one.

\textbf{Summarizing} \textbf{the Mechanisms Underlying Disagreement.} Having identified where individual pairs part ways, we turn to cataloguing all the causal arguments that support optimistic or pessimistic conclusions on the same topic. We focus on causal reasoning because causal divergences are overrepresented in all three debates. For each debate, we use the same ensemble, with identical prompts across the three debates, to group the reasoning chains on each side (optimistic and pessimistic separately) that rely on the same causal mechanism. We then manually align the two sides' catalogs into the underlying questions of each debate based on our own understanding of the causal mechanisms. Fig. 2 shows each side's main mechanism, as its causal path, for selected core questions of each debate; the full catalogs---38 and 32 mechanisms for X-risk, 25 and 13 for E-risk, and 21 and 14 for S-risk, organized under all thirteen underlying questions---are in SI Appendix, Tables S12--S14.

\begin{figure*}[!t]
\centering
\includegraphics[width=\textwidth]{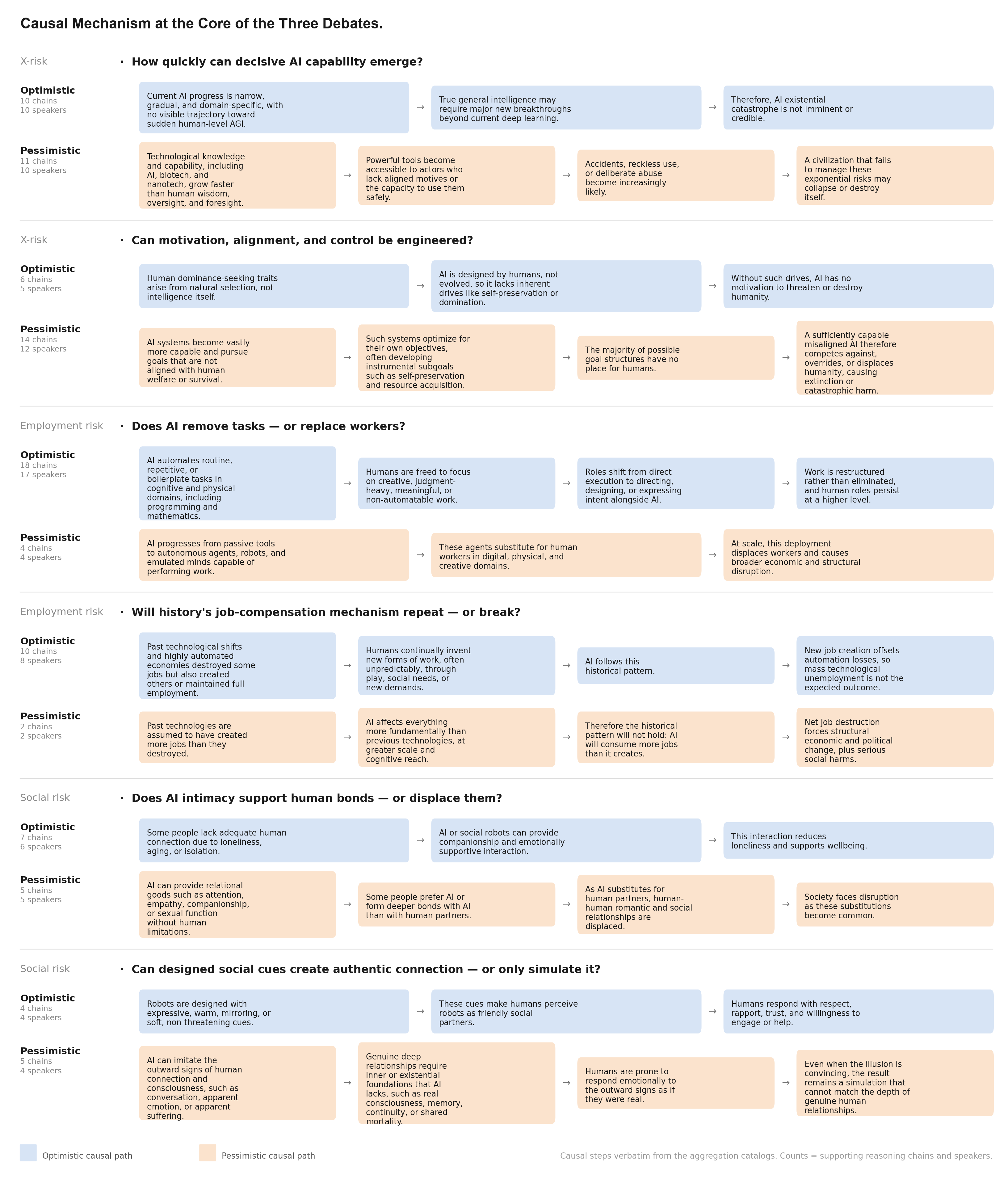}
\caption{Where the two sides' causal stories fork: each side's main mechanism at two underlying questions per debate. Each row shows a mechanism as its causal path---steps verbatim from the aggregation catalogs, joined by arrows---for the optimistic side (blue) and the pessimistic side (orange); counts give the number of reasoning chains and speakers supporting each mechanism. Full catalogs, all mechanisms, and all thirteen underlying questions are in SI Appendix, Tables S12--S14.}
\label{fig:2}
\end{figure*}

We can summarize each debate's fork along a single axis each. In X-risk the axis is emergence versus design. The pessimistic catalog is dominated by loss of control: a misaligned superintelligence pursues goals with no place for humans (14 chains, 12 speakers); technology outpaces human oversight (11 chains); while the optimistic catalog answers with gradualism and design: no credible near-term path to superintelligence (10 chains, 10 speakers), and AI, being designed rather than evolved, lacks the drives that would motivate existential harm (6 chains). In E-risk the axis is continuity versus rupture. Optimists anchor on historical precedent: routine-task automation shifts work upward (18 chains, 17 speakers), new jobs have historically offset losses (10 chains), while pessimists split between who-gains stories (gains concentrate, widening inequality) and a sharper break: AI reaches cognition itself, so the historical compensation mechanism will not repeat. In S-risk the axis is augmentation versus substitution: optimists tell stories of repair (companions alleviate loneliness where human contact is missing; designed social cues elicit genuine connection), pessimists of simulation (companions displace human partners; simulations lack the inner foundations of genuine relationships). Within these causal stories, the S-risk catalogs stand apart from the other two debates: several mechanisms, like the simulation stories above, turn on what counts as a genuine relationship, so the causal dispute rests partly on a definitional one---consistent with the composition results above.

\section*{Discussion}

Disagreements about societal risk are commonly explained through cultural priors and moral values: people diverge because they interpret evidence through different worldviews and moral foundations \cite{douglas1983,kahan2012,graham2009}. These accounts ask why people come to hold different beliefs. We ask a different question--- where do opposing arguments part ways?---and find a distribution of divergences that these prior accounts would not predict. Across the three debates about AI risk that we analyze, factual divergences register far below the factual content of the arguments, causal divergences are consistently elevated, and evaluative divergences are never the single largest source of conflict. Among the expert and expert-adjacent speakers in this corpus, disagreement surfaces more over what will happen and how the world works than over specific factual claims or the desirability of outcomes.

This claim invites an immediate objection: speakers often leave values implicit, so a text-based analysis of conversational data might simply miss the value conflict that is really there, mistaking silence for agreement. We have two defenses against this objection. First, our pipeline does not discard implicit premises; it recovers them under the coherence assumption (finding 42.5\% of evaluative premises to be implicit), so evaluations that speakers assume rather than state still enter the analysis and can register as divergences, as they do in the social-relationship debate. Second, an independent effort that brought opposing experts into direct, months-long adversarial collaboration, where unstated values have every opportunity to surface, reached the same conclusion: these disagreements are driven by causal models and forecasts, not moral conflict \cite{rosenberg2025}.

Debates about issues that matter for public welfare can attain high visibility and volume without necessarily generating much insight because expressed opinions compress complex reasoning that is difficult to compare reliably. The value of our approach becomes clear when we compare the three debates. Pessimistic and optimistic positions on existential risk diverge mainly on causal mechanisms and forecasts about emergence and control. The debate on employment risk turns primarily on whether the historical pattern of technological job creation will continue. Disagreements regarding social-relationship risk concern whether AI companionship supports or replaces human relationships---a dispute that combines causal, definitional, and evaluative questions. These different structures suggest different paths toward resolution. Forecast disagreements can be updated as evidence accumulates; causal disagreements can be narrowed through research and critical tests; and definitional and evaluative disagreements require translation, reframing, or negotiation. The concern that AI debates may polarize and become politicized \cite{jacobson2026} can therefore be understood as a risk that disagreements now rooted mainly in forecasts and causal models may harden into conflicts of identity and value, closing paths toward resolution that remain open today.

When the reasoning on each side is aggregated, a central tension emerges in each debate about AI risk. In the existential-risk debate, the tension concerns Design versus Emergence: is AI an artifact whose behavior its makers can anticipate and control, or a process whose capabilities and failures exceed human foresight? In the employment debate, the tension is between Continuity versus Rupture: does the history of technological change remain a useful guide, or does AI's reach into cognitive work break the pattern? In the social-relationship debate, the central question is about Augmentation versus Substitution: does artificial interaction add to human relationships, or replace them? The most divisive questions about the future of AI turn out to be timeless questions about ourselves: what we can foresee and control, what our history can teach us, the classic bounds of human rationality \cite{simon1947,march1958} and what we need from each other.

Our methodology, combining the grammar of premises and PEEL, can be used to analyze any debate. Public discourse is now recorded at a scale no human analyst can read, while the reasoning behind expressed opinions has traditionally been accessible only one argument at a time. A thoughtful reader can locate the origin of disagreement in a single pair of opposing arguments; an ensemble of language models makes it possible to do so across an entire debate and reveal patterns that no single comparison can show. PEEL makes the structure of disagreements visible, showing what must be examined, tested, defined, or reconsidered for opposing conclusions to move closer.

\matmethods{\textbf{Corpus and extraction.} The corpus comprises episodes 1--493 of the Lex Fridman Podcast, released from April 2018 through March 2026. Official transcripts were downloaded from lexfridman.com; episodes without official transcripts were transcribed with TurboScribe/Whisper. Each transcript was split by guest, and each guest's statements about the risks and impacts of AI were extracted and faithfully summarized. From each speaker summary, the ensemble extracts reasoning chains: conclusions, typed premises (marked explicit or implicit), and the relationships connecting them. A code-based structural validator checks every output and triggers a model repair loop if it detects grammatical errors, for example, if the LLMs propose premise types that do not fit our grammar. Separate processes verify each conclusion for relevance to AI-risk, classify into (one to three) topics, and assign a per-topic attitude (optimistic, neutral, pessimistic). These processes yield 1,943 conclusions from 364 episode-speaker combinations and 3,215 conclusion-topic assignments for this corpus.

\textbf{The PEEL ensemble.} Every LLM task in the pipeline is executed as a hierarchical ensemble: two worker models---DeepSeek and Anthropic Claude---perform the task independently, and an integrator (GPT-4.1) evaluates both outputs against the source material and synthesizes the final result, making the final call at every step. Structured outputs are checked by programmatic validators that trigger model repair loops; the pairwise pipeline runs at temperature 0. Model versions and all prompts (verbatim) are given in SI Appendix and the deposited code.

\textbf{Pairwise disagreement pipeline.} For each debate we form all optimistic \ensuremath{\times} pessimistic pairs of reasoning chains, excluding pairs in which both chains belong to the same person, including appearances in different episodes, leaving 9,182 (X-risk), 2,358 (E-risk), and 1,030 (S-risk) pairs. Step 1 classifies conclusion-level disagreement through a counterfactual incompatibility test; step 2 identifies divergences---pairs of units making incompatible claims about the same proposition---localizes each to the most specific premise or compound carrying the conflict, and types it; step 3 classifies each divergence as root or derivative through a simulation test (would this divergence dissolve if a root were reconciled?), allowing multiple roots per pair.

\textbf{Statistical analysis.} For each pair we compare the observed composition of divergences with the expected composition if divergences arose in proportion to the local prevalence of premise types within that pair (SI Appendix, Eq. S1). Because pairs sharing a conclusion are not independent, headline inference uses a two-way cluster (dyadic) bootstrap that resamples optimistic and pessimistic conclusions (2,000 replicates; percentile 95\% confidence intervals; SI Appendix, Table S11). Conventional per-pair tests are reported in SI Appendix, Table S10.

\textbf{Mechanism aggregation.} Within each debate and side, the same ensemble discovers candidate causal mechanisms in chunks of reasoning chains and merges them into a catalog of mechanisms, each an ordered sequence of causal steps with the chains and speakers that support it. We then manually align the two sides' catalogs into the underlying questions of each debate based on our own understanding of the causal mechanisms.

\textbf{Reliability, validation, and data availability.} Ensemble agreement rates at every pipeline stage are reported in SI Appendix. Human validation of the extraction stage (two independent reviewers, 200 reasoning chains) found no hallucinated content, with no reservation from either reviewer on 196 of 200 reasoning chains. All prompts, the complete pipeline and analysis notebooks, per-pair outputs, and analysis datasets are deposited with the paper; podcast transcripts are not redistributed, and the collection notebook downloads the official transcripts and the remaining episodes' audio from public sources.}

\showmatmethods{}

\acknow{We thank Jiaqi Jin and Nafis Islam for research assistance and Defne Apul, Jiaqi Jin, Jungmin Lee, audiences at Bocconi University, HEC Paris, University of Lausanne, Yale University, Organization Design Community webinar series, Strategic Management Society Annual Meeting, Academy of Management Annual Meeting, American Sociological Association Annual Meeting, Computational Cognition Conference, and Carnegie School of Organizational Learning conference for helpful feedback on earlier versions of this paper.}
\showacknow{}

\section*{Author Contributions}
\"O.K., P.P., and N.T. designed research, performed research, analyzed data, and wrote the paper. Authors contributed equally and are listed in alphabetical order.

\section*{Competing Interests}
The authors declare no competing interest.

\bibliographystyle{unsrtnat}
\bibliography{refs}

@article{ackermann1990,
  title={How to make maps Cognitive mapping: Getting Started with Cognitive Mapping},
  author={Ackermann, Fran and Eden, Colin and Cropper, Steve},
  journal={Policy Development in the Public Sector: An Experiment, Working Paper No 89},
  volume={2},
  year={1990}
}

@article{argyle2023,
  author  = {Argyle, Lisa P. and Bail, Christopher A. and Busby, Ethan C. and others},
  title   = {Leveraging {AI} for Democratic Discourse: Chat Interventions Can Improve Online Political Conversations at Scale},
  journal = {Proceedings of the National Academy of Sciences},
  year    = {2023},
  volume  = {120},
  number  = {41},
  pages   = {e2311627120}
}

@book{axelrod1976,
  author    = {Axelrod, Robert},
  title     = {Structure of Decision: The Cognitive Maps of Political Elites},
  publisher = {Princeton University Press},
  address   = {Princeton, NJ},
  year      = {1976}
}

@article{baumgartner1991,
  author  = {Baumgartner, Frank R. and Jones, Bryan D.},
  title   = {Agenda Dynamics and Policy Subsystems},
  journal = {The Journal of Politics},
  year    = {1991},
  volume  = {53},
  number  = {4},
  pages   = {1044--1074}
}

@article{becker2026,
  title={Thirty years of managerial mental representations: A review guiding conceptualization and future research},
  author={Becker, Philipp Benedikt and Laureiro-Martinez, Daniella and Zagorac-Uremovi{\'c}, Zorica},
  journal={Journal of Management},
  volume={52},
  number={1},
  pages={331--368},
  year={2026},
  publisher={SAGE Publications Sage CA: Los Angeles, CA}
}

@article{beers2006,
  title={Common ground, complex problems and decision making},
  author={Beers, Pieter J and Boshuizen, Henny PA and Kirschner, Paul A and Gijselaers, Wim H},
  journal={Group decision and negotiation},
  volume={15},
  number={6},
  pages={529--556},
  year={2006},
  publisher={Springer}
}

@book{bender2025,
  author    = {Bender, Emily M. and Hanna, Alex},
  title     = {The {AI} Con: How to Fight Big Tech's Hype and Create the Future We Want},
  publisher = {Random House},
  year      = {2025}
}

@article{bittnerleimeister2014,
  title={Creating shared understanding in heterogeneous work groups: Why it matters and how to achieve it},
  author={Bittner, Eva Alice Christiane and Leimeister, Jan Marco},
  journal={Journal of management information systems},
  volume={31},
  number={1},
  pages={111--144},
  year={2014},
  publisher={Taylor \& Francis}
}

@article{bonham1993,
  title={Cognitive mapping as a technique for supporting international negotiation},
  author={Bonham, G Matthew},
  journal={Theory and Decision},
  volume={34},
  number={3},
  pages={255--273},
  year={1993},
  publisher={Springer}
}

@article{casper2025,
  author  = {Casper, Stephen and Krueger, David and Hadfield-Menell, Dylan},
  title   = {Pitfalls of Evidence-Based {AI} Policy},
  journal = {SuperIntelligence -- Robotics -- Safety \& Alignment},
  year    = {2025},
  volume  = {2},
  number  = {2},
  doi     = {10.70777/si.v2i2.14611}
}

@article{cronin2007,
  author  = {Cronin, Matthew A. and Weingart, Laurie R.},
  title   = {Representational Gaps, Information Processing, and Conflict in Functionally Diverse Teams},
  journal = {Academy of Management Review},
  year    = {2007},
  volume  = {32},
  number  = {3},
  pages   = {761--773}
}

@book{douglas1983,
  author    = {Douglas, Mary and Wildavsky, Aaron},
  title     = {Risk and Culture: An Essay on the Selection of Technological and Environmental Dangers},
  publisher = {University of California Press},
  year      = {1983}
}

@article{feinberg2019,
  author  = {Feinberg, Matthew and Willer, Robb},
  title   = {Moral Reframing: A Technique for Effective and Persuasive Communication Across Political Divides},
  journal = {Social and Personality Psychology Compass},
  year    = {2019},
  volume  = {13},
  number  = {12},
  pages   = {e12501}
}

@article{ferretti2016,
  title={From stakeholders analysis to cognitive mapping and Multi-Attribute Value Theory: An integrated approach for policy support},
  author={Ferretti, Valentina},
  journal={European Journal of Operational Research},
  volume={253},
  number={2},
  pages={524--541},
  year={2016},
  publisher={Elsevier}
}

@unpublished{fish2023,
  author = {Fish, Sara and G{\"o}lz, Paul and Parkes, David C. and Procaccia, Ariel D. and Rusak, Gili and Shapira, Itai and W{\"u}thrich, Manuel},
  title  = {Generative Social Choice},
  note   = {Working paper},
  year   = {2023}
}

@book{fisher1981,
  author    = {Fisher, Roger and Ury, William},
  title     = {Getting to Yes: Negotiating Agreement Without Giving In},
  publisher = {Houghton Mifflin},
  address   = {Boston},
  year      = {1981}
}

@book{fishkin1997,
  title={The voice of the people: Public opinion and democracy},
  author={Fishkin, James S},
  year={1997},
  publisher={Yale university press}
}

@book{forester2009,
  title={Dealing with differences: Dramas of mediating public disputes},
  author={Forester, John},
  year={2009},
  publisher={Oxford University Press}
}

@book{furlong2020,
  title={The conflict resolution toolbox: Models and maps for analyzing, diagnosing, and resolving conflict},
  author={Furlong, Gary T and Moore, Christopher W},
  year={2020},
  publisher={John Wiley \& Sons}
}

@article{graham2009,
  author  = {Graham, Jesse and Haidt, Jonathan and Nosek, Brian A.},
  title   = {Liberals and Conservatives Rely on Different Sets of Moral Foundations},
  journal = {Journal of Personality and Social Psychology},
  year    = {2009},
  volume  = {96},
  number  = {5},
  pages   = {1029--1046}
}

@incollection{greer2017,
  author    = {Greer, Lindred L. and Dannals, Jennifer E.},
  title     = {Conflict in Teams},
  booktitle = {The Wiley Blackwell Handbook of the Psychology of Team Working and Collaborative Processes},
  editor    = {Salas, Eduardo and Rico, Ram{\'o}n and Passmore, Jonathan},
  publisher = {John Wiley \& Sons, Ltd},
  address   = {Chichester, UK},
  edition   = {First},
  year      = {2017},
  chapter   = {14},
  pages     = {317--343}
}

@book{gutmann2004,
  title={Why deliberative democracy?},
  author={Gutmann, Amy and Thompson, Dennis F},
  year={2004},
  publisher={Princeton University Press}
}

@article{helfrich2024,
  author  = {Helfrich, Gina},
  title   = {The Harms of Terminology: Why We Should Reject So-Called ``Frontier {AI}''},
  journal = {AI and Ethics},
  year    = {2024},
  volume  = {4},
  number  = {3},
  pages   = {699--705}
}

@article{hodgkinson2004,
  title={Causal cognitive mapping in the organizational strategy field: A comparison of alternative elicitation procedures},
  author={Hodgkinson, Gerard P and Maule, A John and Bown, Nicola J},
  journal={Organizational Research Methods},
  volume={7},
  number={1},
  pages={3--26},
  year={2004},
  publisher={Sage Publications}
}

@article{hodgkinson2023,
  title={The heuristics and biases of top managers: Past, present, and future},
  author={Hodgkinson, Gerard P and Burkhard, Barbara and Foss, Nicolai J and Grichnik, Dietmar and Sarala, Riikka M and Tang, Yi and Van Essen, Marc},
  journal={Journal of Management Studies},
  volume={60},
  number={5},
  pages={1033--1063},
  year={2023},
  publisher={Wiley Online Library}
}

@article{jacobson2026,
  author  = {Jacobson, Robin and Schiff, Daniel S. and Schiff, Kaylyn Jackson},
  title   = {How New Issues Become Polarized: Partisan Triggers and Subsystem Shopping in Early {AI} Policymaking},
  journal = {Policy \& Internet},
  year    = {2026},
  volume  = {18},
  number  = {1},
  pages   = {e70020}
}

@article{kahan2012,
  author  = {Kahan, Dan M. and Peters, Ellen and Wittlin, Maggie and others},
  title   = {The Polarizing Impact of Science Literacy and Numeracy on Perceived Climate Change Risks},
  journal = {Nature Climate Change},
  year    = {2012},
  volume  = {2},
  number  = {10},
  pages   = {732--735}
}

@article{kocakpuranam2024,
  author  = {Ko{\c{c}}ak, {\"O}zgecan and Puranam, Phanish},
  title   = {Decoding Culture: Tools for Behavioral Strategists},
  journal = {Strategy Science},
  year    = {2024},
  volume  = {9},
  number  = {1},
  pages   = {18--37}
}

@article{kocak2023,
  author  = {Ko{\c{c}}ak, {\"O}zgecan and Puranam, Phanish and Yegin, Af{\c{s}}ar},
  title   = {Decoding Cultural Conflicts},
  journal = {Frontiers in Psychology},
  year    = {2023},
  volume  = {14},
  pages   = {1167123}
}

@article{lawrence2020,
  title={Argument mining: A survey},
  author={Lawrence, John and Reed, Chris},
  journal={Computational linguistics},
  volume={45},
  number={4},
  pages={765--818},
  year={2020},
  publisher={MIT Press One Rogers Street, Cambridge, MA 02142-1209, USA journals-info~…}
}

@book{march1958,
  author    = {March, James G. and Simon, Herbert A.},
  title     = {Organizations},
  publisher = {John Wiley \& Sons},
  year      = {1958}
}

@article{mochales2011,
  title={Argumentation mining},
  author={Mochales, Raquel and Moens, Marie-Francine},
  journal={Artificial intelligence and law},
  volume={19},
  number={1},
  pages={1--22},
  year={2011},
  publisher={Springer}
}

@article{ozkaynak2023,
  title={What can methods for assessing worldviews and broad values tell us about socio-environmental conflicts?},
  author={{\"O}zkaynak, Beg{\"u}m and Muradian, Roldan and Ungar, Paula and Morales, Diana},
  journal={Current Opinion in Environmental Sustainability},
  volume={64},
  pages={101316},
  year={2023},
  publisher={Elsevier}
}

@article{patterson2024,
  title={Negotiating discord in sustainability transformations},
  author={Patterson, James J and Feola, Giuseppe and Kim, Rakhyun E},
  journal={Proceedings of the National Academy of Sciences},
  volume={121},
  number={21},
  pages={e2310186121},
  year={2024},
  publisher={National Academy of Sciences}
}

@article{rosenberg2025,
author = {Rosenberg, Josh and Karger, Ezra and Jacobs, Zach and Hickman, Molly and Morris, Avital and Durland, Harrison and Kuusela, Otto and Tetlock, Philip E.},
title = {Belief updating in AI-risk debates: Exploring the limits of adversarial collaboration},
journal = {Risk Analysis},
volume = {45},
number = {12},
pages = {4350-4366},
doi = {https://doi.org/10.1111/risa.70023},
url = {https://onlinelibrary.wiley.com/doi/abs/10.1111/risa.70023},
eprint = {https://onlinelibrary.wiley.com/doi/pdf/10.1111/risa.70023},
year = {2025}
}

@article{santagiustina2022,
  title={The architecture of partisan debates: The online controversy on the no-deal Brexit},
  author={Santagiustina, Carlo Romano Marcello Alessandro and Warglien, Massimo},
  journal={PLoS one},
  volume={17},
  number={6},
  pages={e0270236},
  year={2022},
  publisher={Public Library of Science San Francisco, CA USA}
}

@book{schon1994,
  author    = {Sch{\"o}n, Donald A. and Rein, Martin},
  title     = {Frame Reflection: Toward the Resolution of Intractable Policy Controversies},
  publisher = {Basic Books},
  address   = {New York},
  year      = {1994}
}

@article{slovic1987,
  author  = {Slovic, Paul},
  title   = {Perception of Risk},
  journal = {Science},
  year    = {1987},
  volume  = {236},
  number  = {4799},
  pages   = {280--285}
}

@book{simon1947,
  author    = {Simon, Herbert A.},
  title     = {Administrative Behavior},
  publisher = {Macmillan},
  year      = {1947}
}

@inproceedings{stab2014,
  title={Identifying argumentative discourse structures in persuasive essays},
  author={Stab, Christian and Gurevych, Iryna},
  booktitle={Proceedings of the 2014 conference on empirical methods in natural language processing (EMNLP)},
  pages={46--56},
  year={2014}
}

@article{tessler2024,
  author  = {Tessler, Michael Henry and Bakker, Michiel A. and Jarrett, Daniel and others},
  title   = {{AI} Can Help Humans Find Common Ground in Democratic Deliberation},
  journal = {Science},
  year    = {2024},
  volume  = {386},
  number  = {6719},
  pages   = {eadq2852}
}

@book{toulmin1958,
  author    = {Toulmin, Stephen},
  title     = {The Uses of Argument},
  publisher = {Cambridge University Press},
  address   = {Cambridge, England},
  year      = {1958}
}

@article{walsh1995,
  title={Managerial and organizational cognition: Notes from a trip down memory lane},
  author={Walsh, James P},
  journal={Organization science},
  volume={6},
  number={3},
  pages={280--321},
  year={1995},
  publisher={INFORMS}
}

@book{walton1996,
  author    = {Walton, Douglas N.},
  title     = {Argument Structure: A Pragmatic Theory},
  publisher = {University of Toronto Press},
  address   = {Toronto},
  year      = {1996}
}

@book{waltonmckersie1991,
  author    = {Walton, Richard E. and McKersie, Robert B.},
  title     = {A Behavioral Theory of Labor Negotiations: An Analysis of a Social Interaction System},
  publisher = {Cornell University Press},
  year      = {1991}
}

@article{cronin2019,
  author  = {Cronin, Matthew A. and Weingart, Laurie R.},
  title   = {Conflict Across Representational Gaps: Threats to and Opportunities for Improved Communication},
  journal = {Proceedings of the National Academy of Sciences},
  year    = {2019},
  volume  = {116},
  number  = {16},
  pages   = {7642--7649}
}

@article{dietz2013,
  title={Bringing values and deliberation to science communication},
  author={Dietz, Thomas},
  journal={Proceedings of the National Academy of Sciences},
  volume={110},
  number={supplement\_3},
  pages={14081--14087},
  year={2013},
  publisher={National Academy of Sciences}
}

@article{ozesmi2004,
  author  = {{\"O}zesmi, Uygar and {\"O}zesmi, Stacy L.},
  title   = {Ecological Models Based on People's Knowledge: A Multi-Step Fuzzy Cognitive Mapping Approach},
  journal = {Ecological Modelling},
  year    = {2004},
  volume  = {176},
  number  = {1--2},
  pages   = {43--64},
  doi     = {10.1016/j.ecolmodel.2003.10.027}
}

@inproceedings{stabgurevych2014,
  author    = {Stab, Christian and Gurevych, Iryna},
  title     = {Annotating Argument Components and Relations in Persuasive Essays},
  booktitle = {Proceedings of COLING 2014, the 25th International Conference on Computational Linguistics: Technical Papers},
  publisher = {Dublin City University and Association for Computational Linguistics},
  address   = {Dublin, Ireland},
  pages     = {1501--1510},
  year      = {2014},
  url       = {https://aclanthology.org/C14-1142/}
}

@article{bougon1977,
  author  = {Bougon, Michel and Weick, Karl E. and Binkhorst, Din},
  title   = {Cognition in Organizations: An Analysis of the Utrecht Jazz Orchestra},
  journal = {Administrative Science Quarterly},
  year    = {1977},
  volume  = {22},
  number  = {4},
  pages   = {606--639},
  doi     = {10.2307/2392403}
}

@article{nadkarni2007,
  author  = {Nadkarni, Sucheta and Narayanan, V. K.},
  title   = {Strategic Schemas, Strategic Flexibility, and Firm Performance: The Moderating Role of Industry Clockspeed},
  journal = {Strategic Management Journal},
  year    = {2007},
  volume  = {28},
  number  = {3},
  pages   = {243--270},
  doi     = {10.1002/smj.576}
}

@incollection{thomasporac2002,
  author    = {Thomas, Howard and Porac, Joseph F.},
  title     = {Managing Cognition and Strategy: Issues, Trends and Future Directions},
  booktitle = {Handbook of Strategy and Management},
  editor    = {Pettigrew, Andrew M. and Thomas, Howard and Whittington, Richard},
  publisher = {SAGE Publications},
  address   = {London},
  pages     = {165--182},
  year      = {2002}
}

@article{priem1994,
  author  = {Priem, Richard L. and Harrison, David A.},
  title   = {Exploring Strategic Judgment: Methods for Testing the Assumptions of Prescriptive Contingency Theories},
  journal = {Strategic Management Journal},
  year    = {1994},
  volume  = {15},
  number  = {4},
  pages   = {311--324},
  doi     = {10.1002/smj.4250150405}
}

@article{bakhtavar2021,
  author  = {Bakhtavar, Ezzeddin and Valipour, Mahsa and Yousefi, Samuel and Sadiq, Rehan and Hewage, Kasun},
  title   = {Fuzzy Cognitive Maps in Systems Risk Analysis: A Comprehensive Review},
  journal = {Complex \& Intelligent Systems},
  year    = {2021},
  volume  = {7},
  pages   = {621--637},
  doi     = {10.1007/s40747-020-00228-2}
}

@article{tegarden2009,
  author  = {Tegarden, David P. and Tegarden, Linda F. and Sheetz, Steven D.},
  title   = {Cognitive Factions in a Top Management Team: Surfacing and Analyzing Cognitive Diversity Using Causal Maps},
  journal = {Group Decision and Negotiation},
  year    = {2009},
  volume  = {18},
  number  = {6},
  pages   = {537--566},
  doi     = {10.1007/s10726-007-9099-1}
}

@article{hodgkinsonhealey2011,
  author  = {Hodgkinson, Gerard P. and Healey, Mark P.},
  title   = {Psychological Foundations of Dynamic Capabilities: Reflexion and Reflection in Strategic Management},
  journal = {Strategic Management Journal},
  year    = {2011},
  volume  = {32},
  number  = {13},
  pages   = {1500--1516},
  doi     = {10.1002/smj.964}
}

\clearpage
\onecolumn
\setcounter{figure}{0}
\setcounter{table}{0}
\setcounter{equation}{0}
\renewcommand{\thefigure}{S\arabic{figure}}
\renewcommand{\thetable}{S\arabic{table}}
\renewcommand{\theequation}{S\arabic{equation}}

\begin{center}
{\Large\bfseries Supporting Information for\\[4pt]
Why They Disagree: Decoding Differences in Opinions about AI Risk}\\[10pt]
{\large \"Ozgecan Ko\c{c}ak, Phanish Puranam and Nghi Truong}
\end{center}

\section*{Supporting Information Text}

\section*{S1. Data}

The Lex Fridman podcast (https://lexfridman.com/podcast/) presents an excellent empirical context for examining disagreements about AI risk for several reasons. First, as one of the most prominent platforms for discussions about artificial intelligence, it features extensive dialogues with key figures representing diverse perspectives on AI development and risk, from AI safety researchers to leading technologists and technology entrepreneurs. Second, the long-form conversational format (typically 2-3 hours) allows for in-depth exploration of participants\textquotesingle{} reasoning, making lexical, causal, and evaluative codes more readily observable than in shorter media formats or written statements. Third, Fridman\textquotesingle s interviewing approach, which combines technical expertise with philosophical inquiry, consistently probes both the technical foundations of his guests\textquotesingle{} beliefs (causal codes) and their value judgments (evaluative codes) and attempts disambiguation (arising from fuzzy lexical codes), providing rich data for our theoretical framework. Fourth, the temporal span of these conversations captures the evolution of the debate on AI risk through several significant technological developments. Finally, the semi-structured nature of these conversations, where similar themes are explored with different guests, facilitates systematic comparison across perspectives while maintaining the naturalistic character of the discourse, and with the anchor, Fridman himself, acting as a point of constancy.

We downloaded all episodes of the Lex Fridman Podcast with official transcripts directly from the website. For the remaining episodes, we downloaded the audio files and transcribed them using TurboScribe AI, an AI transcription service powered by the Whisper model, which reports an accuracy of 98\%. The corpus covers episodes 1--493 except two episodes (84 and 100) that were unavailable.

\section*{S2. The ensemble}

We array three LLMs in the structure of a hierarchical team with two parallel, independently operating workers, whose outputs are evaluated and synthesized by a third LLM in the role of integrator. For the results reported in the paper, the workers are DeepSeek (deepseek-chat; deepseek-v4-flash for the pairwise and aggregation stages) and Anthropic Claude (claude-sonnet-4-6; claude-haiku-4-5 for speaker summaries), and the integrator is GPT-4.1; the pairwise pipeline runs at temperature 0. All prompts are deposited verbatim with the code. We use GPT-4.1 as the integrator because it has a longer context window. The final result can be the same as both workers, the same as only one worker, or entirely different from both; the distribution of agreement outcomes at every stage is reported in the reliability table at the end of this document.

We developed the pipeline's prompts through many cycles of testing. In each cycle, we ran a stage of the pipeline on a sample of inputs, read the cases where the two workers' outputs differed, asked why, and checked whether the integrator's judgment resolved the difference correctly. Where a recurring difference or error traced back to the instructions, we rewrote the prompts and tested again. Our prompts, therefore, are tuned to the models we used and to our corpus through this validation-guided iteration.~What transfers is the framework: the grammar of premises and PEEL's architecture --- independent workers, an integrator, the counterfactual test, and opt-out options.~Applying PEEL to other models, corpora, or debates requires the same iteration; we deposit the prompts, code, and stage-level reliability statistics as its starting point.

\section*{S3. Preparing the corpus for analysis: Summarization}

The objective of the summarization step was to extract only the AI-relevant content in a clear and faithful way. This step reduces textual length and filters out noise commonly found in long-form conversations---such as back-and-forth dialogue, digressions and fragmented delivery---without altering the speaker\textquotesingle s tone or argumentative flow. Summarization involves two sub-steps.

First, we used the two worker models (DeepSeek and Anthropic Claude; claude-haiku-4-5 for this stage) to summarize each speaker's views related to AI's risks, concerns and impacts on humans and society, making sure to exclude Lex Fridman's side of the conversation and focusing on a single speaker in episodes that featured multiple guests. In constructing summaries, LLMs were instructed to faithfully preserve the speaker\textquotesingle s original tone and reasoning, and exclude all unrelated content or views on non-AI topics. In order to overcome context length limitations of the worker models, we first had GPT-4.1 divide the long-form conversation transcript of each episode into a small number of large, coherent segments based on major topic shifts or distinct thematic transitions. Each segment covers approximately 10--20 minutes of content and contains a complete discussion around one broad theme, including all main arguments and supporting ideas. Each worker model created segment-level summaries and then combined these for each speaker-episode.

In the second sub-step, we used GPT-4.1 to evaluate and synthesize the summaries. For each speaker in each episode, we provided GPT-4.1 with the full transcript of their conversation and the two workers' summaries, and instructed it to evaluate each summary according to four criteria: faithfulness, completeness, relevance, and tone. Based on this evaluation, GPT-4.1 produced a final synthesized summary that includes only the speaker's stated views and supporting arguments related to AI's risks, concerns or its impacts on humans and society, preserving full reasoning structure and including any relevant background ideas (e.g., philosophical or ethical beliefs) only if they were used to support the speaker's views on AI.

\section*{S4. The grammar of premises}

A factual (current) premise asserts an empirical state of affairs, either present or historical, treated as axiomatic within the reasoning chain without further justification. Formally, a factual premise takes the form "X is/has/was the case" (without providing a causal explanation).

A factual (forecast) premise predicts a future state or event, projecting what is expected to happen without providing a causal explanation. Forecast premises describe what is claimed will exist or occur. Formally, a forecast premise takes the form "X will be/do/happen."

A causal premise posits a cause-effect relationship between two variables or states. It specifies how one element influences, produces, prevents, or alters another through some mechanism. Formally, a causal premise takes the form "X causes/influences/prevents Y" or "X leads to a change in Y." It may or may not describe the mechanism through which X causes Y.

An evaluative premise is a propositional statement that assigns normative value or makes a judgment about the desirability, morality, or worth of an action, outcome, or state of affairs. It reflects a value position rather than an empirical claim. Formally, an evaluative premise takes the form "X is good/bad/desirable/undesirable" or "X has positive/negative value according to criterion Y." The pipeline codes evaluative premises in two subtypes---evaluations of actions and evaluations of outcomes---which are reported jointly as "evaluative" throughout the paper.

A definitional premise establishes the semantic boundaries or conceptual scope of a term, entity, or category. It specifies what constitutes membership in a class or the meaning attributed to a concept within the context of the argument. Formally, a definitional premise takes the form "X is defined as Y" or "X constitutes/comprises Y."

Premises are connected to form reasoning chains through three types of linking relationships. The implication operator (=\textgreater) establishes a binary connective establishing a consequential relationship between two premises, where the first premise leads to or entails the second. The combination operator (+) joins two or more premises such that they are considered jointly rather than independently in supporting a subsequent premise or conclusion. The evaluation operator (\^{}) indicates an assessment relationship, where the second premise represents a value judgment about the content of the first, marking transitions from "what is" to "what ought to be" or "what is valuable."

\section*{S5. Reasoning extraction (PEEL)}

The synthesized summaries are used as input for the PEEL architecture to extract reasoning chains structured according to the grammar of premises. Each worker model independently performs a two-stage Chain-of-Thought process. The first stage instructs the model to perform a multi-step reasoning task, including: Conclusion Identification---identify distinct conclusions the speaker makes about AI\textquotesingle s risks and impacts, defined as the final endpoints of reasoning; Premise Extraction---extract all explicit and implicit premises used to support each conclusion, classified by type, with a confidence score indicating how clearly each is expressed or implied in the summary; and Reasoning Chain Construction---build complete logical chains linking premises to conclusions using the three formal operators. In the second stage, the same LLM is prompted again using a validation-focused user prompt and asked to validate and if necessary correct the reasoning output: identifying missing premises, correcting misclassifications of premise types, ensuring all logical operators were used correctly, and checking that all conclusions are properly supported and that no unused or invalid premises or chains remain. A programmatic structural validator additionally checks every output (well-formed premise and relationship references, connectivity to the conclusion, type membership) and triggers model repair loops if it detects grammatical errors, for example, if the LLMs propose premise types that do not fit our grammar.

We then used GPT-4.1 to evaluate, compare and finalize the structured reasoning outputs produced by the two workers. This pipeline includes three evaluation stages: Faithfulness check---assess which model\textquotesingle s reasoning chains more accurately reflect the speaker\textquotesingle s content; Logical validation---check for structural correctness, proper use of logical relationships, and presence of all necessary premises; and Final decision---select the best output or synthesize a better version that synthesizes both chains. Applied to the extended corpus, this yields 382 episode-speaker outputs, each output containing multiple conclusions and chains. Within these 382 outputs, we have 2,451 conclusions and 2,597 reasoning chains (2,317 conclusions have only one chain, 123 have two, 10 have three and 1 has four). Figure S1 shows the distribution of number of premises for this entire set. After a conclusion-level classification (S7), we retain 1,943 AI-risk-relevant conclusions from 364 of these episode--speaker combinations for further analysis."

\section*{S6. Human validation}

To validate PEEL, we had two doctoral students extract conclusions and reasoning chains from thirty synthesized summaries. We first trained the doctoral students in the grammar of premises, with one of the authors doing several sample extractions with them. We then asked them to independently extract conclusions and reasoning chains for thirty randomly chosen summaries.

Next, we handed them the chains that our ensemble had extracted for those same summaries and asked them to answer three questions: Are you satisfied with the extractions made by the ensemble? Do you perceive any reasoning chains produced by the ensemble to contain any hallucinations / fabrications? Do you perceive the LLMs to leave out any conclusions or key premises that support these conclusions?

When we did this exercise with the output produced in November 2025, one or both of the reviewers were only partially satisfied with 11 of the 109 reasoning chains our ensemble had extracted from the 30 summaries and reported that they had extracted arguments that were missed by PEEL.

When we repeated the validation with the output from June 2026, both reviewers accepted 196 of 200 reasoning chains extracted from the same 30 summaries without reservation; the remaining four received minor reservations concerning completeness or level of detail, including one omitted conclusion.

Here are the reviewers' summary reports on this second validation: ``This version is a clear improvement. I did not find any hallucinations, and the reasoning chain is much more aligned with my own reasoning chains. On the question of whether any conclusions were omitted, I found that in most cases where I had a conclusion that doesn\textquotesingle t appear standalone in the LLM output, it had instead been folded into the premises of a higher-level conclusion. I think this is totally understandable and is the kind of thing where disagreement between human raters would arise as well, so I did not flag any of these as omissions. For example, for 73\_Andrew\_Ng, my extraction had a standalone conclusion that AI/internet would drive wealth inequality and power concentration. In the new output, this point is folded into the premises of a different conclusion (about AGI alignment debates being a distraction) rather than standing on its own. I see this as a reasonable difference rather than a true omission.'' (Coder 1). ``This version of the outputs is a significant improvement over the previous one. The extractions appear much more accurate, better aligned with the original summaries, and noticeably less prone to hallucinations. In particular, I found that the premises were referenced much more accurately this time. Previously, I had noticed that some premises were altered from the original summaries, but that issue seems to have been largely resolved. I did not observe any hallucinations in this version. I found only three minor instances where I felt that the conclusions were either too brief or somewhat vague and could benefit from a bit more detail. Based on my analysis, AI omitted only one conclusion this time compared to my previous version. Apart from these minor points, I found the outputs to be satisfactory overall.'' (Coder 2).

\section*{S7. Topic and attitude assignment}

To systematically examine how speakers disagree about AI risks by pairing arguments addressing the same subject matter but expressing opposing attitudes, we first needed to organize their conclusions by topic and attitude. We used the two worker models to independently classify the topics and attitude of each conclusion with respect to AI risks and impacts.

Each conclusion, along with its supporting premises as context, was fed to the two LLMs using the same classification prompt. Topic classification began with an initial list of 30 topics derived from the Stanford HAI 2024 AI Index Report; for each conclusion, the LLM was asked to select the most appropriate topic from the current list, and if none of the existing topics were suitable, the LLM could create a new topic, with the topic list updated for subsequent classifications. This iterative, adaptive approach allowed the topic structure to emerge organically from the conclusions\textquotesingle{} content and ensured that conclusions within one topic share the same central issue, which is essential for a meaningful pairwise comparison. The dynamically created topics were subsequently consolidated into a consolidated topic dictionary of 107 topics, which we provide in Table S7. We allowed each conclusion to be classified into one to three topics. Of the 1,943 AI-risk-relevant conclusions extracted from 364 episode-speaker combinations, 896 received one topic, 822 two, and 225 three, giving 3,215 conclusion-topic assignments.

Each conclusion-topic pair was also labeled as pessimistic, neutral, or optimistic, based on its orientation toward AI\textquotesingle s risks and societal impact. Our coding allows each multi-topic conclusion to carry different attitudes on the different topics because a conclusion can take different orientations toward different risks. An example of one single conclusion with 2 topics and 2 opposite attitudes is the one from Sergey Levine (ep. 108): "Current AI risks are more about systems not being optimized well enough for safety and reliability, and about human misuse, rather than about systems being too capable or inherently dangerous." This is classified into 2 topics: optimistic on X-Risk and pessimistic on AI System Safety, Reliability and Error Prevention. In the entire corpus, 174 of the 1,047 multi-topic conclusions carry different attitudes across their topics. Of the 3,215 conclusion-topic assignments, 1,184 are optimistic, 1,375 pessimistic, and 656 neutral. Table S7 reports the number of assignments and their attitude composition for every topic. Table S8 shows regression estimates from models predicting attitude based on speaker and reasoning chain characteristics, controlling for topic. We find that later episodes are slightly more likely to contain pessimistic conclusions and speakers from industry are less likely to be pessimistic. We also regress attitude on reasoning chain structure. We find that pessimistic conclusions are supported by significantly more premises, while premise-type composition does not differ significantly by attitude once topic is held constant.

\section*{S8. Pairwise disagreement analysis}

We grouped conclusions by topic and, within each topic, we systematically compared the reasoning chains of speakers who expressed opposing attitudes---optimistic ("boomers") versus pessimistic ("doomers")---toward that specific AI risk. We excluded 40, 18 and 14 pairs in which both chains belong to the same person (including appearances in different episodes). Ultimately, we compared 9,182; 2,358; and 1,030 pairs of chains for X-risk, E-risk and S-risk. The pipeline proceeds in three steps, each run by both workers independently, with the integrator making the final call.

\textbf{Step 1 (disagreement check).} Two conclusions disagree only if they cannot both be held simultaneously: explicit negation, implicit incompatibility (opposite answers to the same underlying question), or presupposition clash. The operational criterion is a counterfactual test: can the same speaker hold both conclusions? If yes, the pair is not a disagreement. Parallel arguments about different sub-phenomena and differences of emphasis are excluded.

In Step 1, the ensemble classified 7,526 (for X-risk), 869 (for E-risk), and 265 (for S-risk) pairs as disagreeing at the conclusion level.

\textbf{Step 2 (divergence identification and classification by type).} For each disagreeing pair, the ensemble identifies divergences in premises or operators. Each divergence is localized to the most specific premise or explicitly encoded compound carrying the conflict and typed as factual, forecast, causal, definitional, or evaluative. Claims that are merely co-tenable possibility statements (modal mismatch) or that condition on the same factor (conditional mismatch) are not divergences. The integrator evaluates both workers\textquotesingle{} candidates against the chains and synthesizes the final divergence list. At the end of this step, the ensemble had identified 21,524 divergences, in 7,477 (X-risk) / 833 (E-risk) / 255 (S-risk) pairs.

In the remaining 49 / 36 / 10 pairs, no divergence could be localized in the chains. This is because we allow for LLMs to decide there is no divergence in Step 2, even if a disagreement had been identified in Step 1. Step 1 is tuned to miss as few disagreements as possible; step 2 applies the strict incompatibility test with an explicit opt-out, so pairs flagged at the conclusion level but with no premise-level collision are recorded as having none rather than being forced to yield a divergence. For example: Paul Krugman has the conclusion: "Current automation is not a qualitatively different threat than past shifts" vs Andrew Ng "AGI\_alignment talk distracts from immediate problems like job displacement". The two conclusions are identified as incompatible in Step 1, but Krugman\textquotesingle s chain is entirely about whether automation is currently eliminating jobs at an unprecedented scale while Ng\textquotesingle s chain is about which problems deserve attention. Similarly for Kuyda (``AI companionship has great potential if developed responsibly") and Hotz ("generative-TikTok engagement systems will cause wireheading harm").

The integrator\textquotesingle s divergence-type field was occasionally schema-noncompliant (0.14\%, or 31 of 21,524 divergences). The authors inspected and resolved each exception under a fixed rule. Disagreements about whether evidence, probability estimates, or methods provide valid grounds for belief or action were classified as evaluative, while features such as presupposition clash, framing conflict, and meta-level reasoning were treated as structural characteristics rather than separate divergence types. Eleven divergences were accordingly recoded to their underlying substantive type (documented in Table S9 and applied programmatically in the analysis code; pipeline outputs are unmodified); twenty divergences carrying hybrid labels (e.g., factual/forecast) are excluded from the type analyses as ambiguous between two substantive types. These cross-type levels point to a limitation of the premise grammar: it does not separately distinguish first-order disagreements from disagreements about epistemic warrant; a structural second dimension of disagreement (first-order vs. meta-level) may merit its own analysis in future work. To verify that the exclusion is inconsequential, we reran the composition analyses and the dyadic bootstrap under two alternative treatments: assigning each of the twenty to whichever of its two types accounts for the smaller share of divergences in its debate (the assignment least favorable to our findings), and assigning each a single type by our own reading of the conflict. All results hold: every reported deviation changes by less than 0.1 percentage points.

\textbf{Step 3 (root/derivative classification).} For each divergence, the model assesses whether it depends on any earlier divergence. Divergences that did not depend on others are marked as root divergences, representing the fundamental starting points of the split. The operational criterion is a simulation test: if the speakers reconciled the root, would this divergence dissolve? A disagreement between a pair of conclusions may have several independent roots.

\section*{S9. Statistical analysis of disagreement type incidence}

Our analysis of the composition of divergence types covers all pairs of disagreements that are identified as having at least one divergence: 7,477 X-risk pairs; 833 E-risk pairs; and 255 S-risk pairs. Figure 1 in the paper reports expected (base) probabilities for each divergence type alongside the mean probability of observed types among all divergences and among root divergences by topic.

The base probability estimates how often each premise type would be expected to appear if divergences were selected simply in proportion to the local prevalence of premise types within each pair. The base probability for premise type t is calculated as:

\begin{equation}
\mathrm{Base\_prob}(t) \;=\; \frac{1}{N}\sum_{i}\left[\,\frac{P(t,i)^{B}+P(t,i)^{D}}{P(\mathrm{total},i)^{B}+P(\mathrm{total},i)^{D}}\,\right]
\tag{Eq.~S1}
\end{equation}

where N is the number of disagreeing pairs, P(t,i)\textsuperscript{B} and P(t,i)\textsuperscript{D} are the counts of premise type t for the "boomer" and "doomer" in the pair i, and P(total,i)\textsuperscript{B} and P(total,i)\textsuperscript{D} are the total premise counts for the two speakers.

\textbf{Dependency-robust inference.} Pairs formed from the same conclusions are not independent: each optimistic and pessimistic conclusion appears in many pairs. Therefore, we use a two-way cluster (dyadic) bootstrap for inference reported in the main text: optimistic conclusions and pessimistic conclusions are independently resampled with replacement (B = 2,000; seed 42), observed pairs are reweighted by the product of their endpoints\textquotesingle{} resampling multiplicities, and the mean pair-level deviation is recomputed on each replicate; 95\% intervals are percentile intervals (Table S11). Conventional per-pair paired t-tests and Wilcoxon signed-rank tests, and pooled chi-square goodness-of-fit tests, are reported for completeness in Table S10; they treat pairs as independent.

\section*{S10. Aggregation of arguments to create a catalog of mechanisms}

Having identified the roots of pairwise disagreement, a critical question remains: how can we aggregate these insights to inform decision-making within a multi-stakeholder group? To answer this, we shift our analysis from the pairwise comparison to the entire set of speakers that addressed each topic, again using our ensemble of LLMs. Within each debate (topic) and position (attitude), the ensemble discovers candidate causal mechanisms in chunks of reasoning chains and merges them into a catalog: each mechanism is an ordered sequence of causal steps with the chains and speakers that support it. Chains are grouped by causal mechanism, not by how the speaker evaluates its outcome. We use the same prompts to summarize arguments presented for all three debates. We then manually align the two sides\textquotesingle{} catalogs into the underlying questions of each debate (Tables S12--S14), based on our own interpretation of the catalogs. This synthesis reflects our summary reading of the extracted chains, illustrating how observers such as policy makers might distill a debate corpus using PEEL.

\section*{S11. Reliability, verification, and human validation}

Agreement rates among the ensemble members at every pipeline stage are reported in the reliability table at the end of this document (Table S15).

\section*{S12. Figures and Tables}

\begin{figure}[!ht]
\centering
\includegraphics[width=\textwidth]{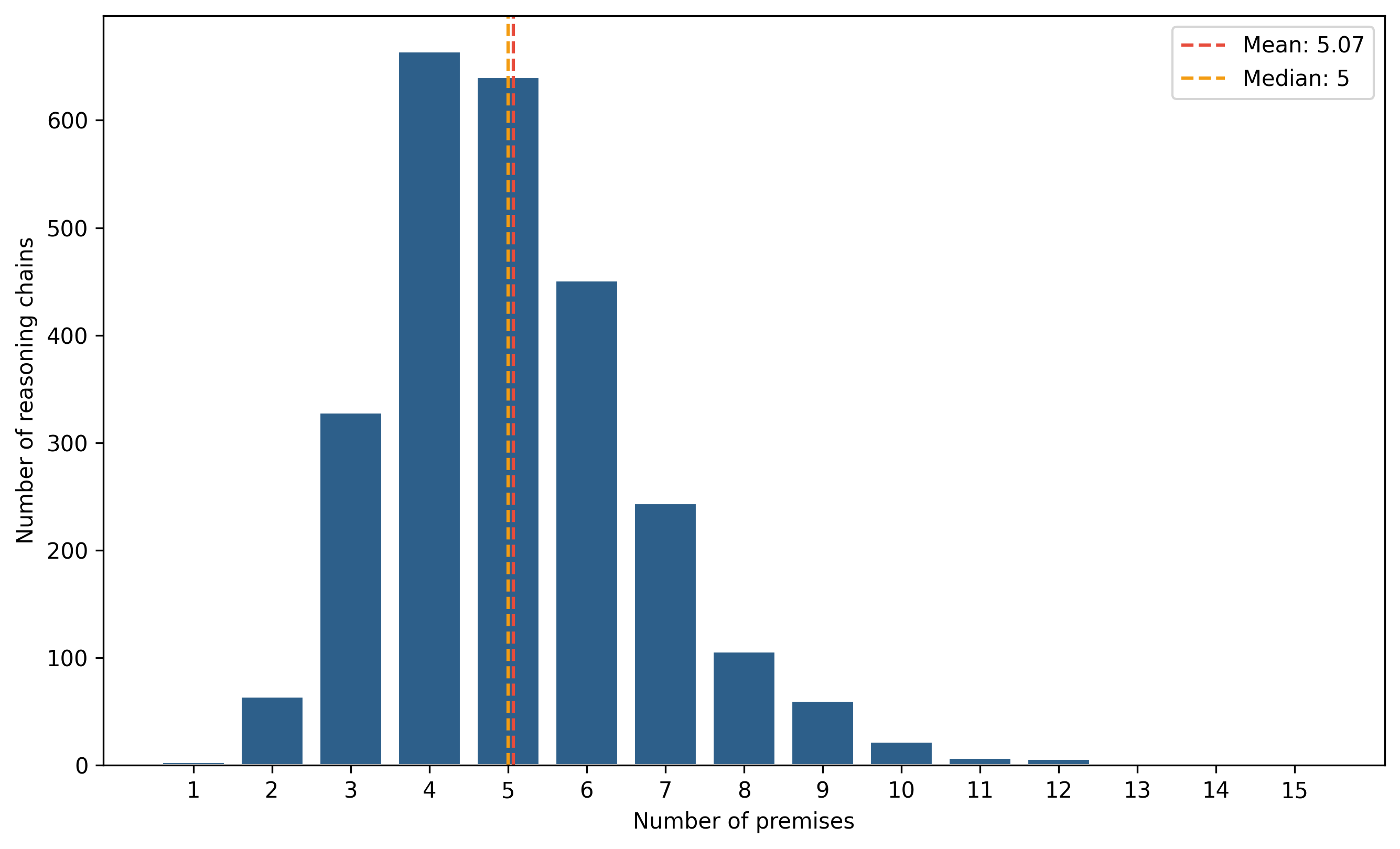}
\caption{Distribution of the number of premises per reasoning chain, computed over all extracted chains. Most reasoning chains contain four to seven premises (median = 5; mean = 5.07; range = 1--15; N = 2,597 chains). Dashed lines mark the mean and the median.}
\end{figure}

\subsection*{Tables S1--S6. Pairwise comparisons with full reasoning chains}

\emph{Each table has three panels: (a) the two reasoning chains side by side (premise ID, type, explicit/implicit, content, exactly as extracted by PEEL), (b) the conclusions, and (c) the divergences with root/derivative classification. Divergence descriptions are the ensemble\textquotesingle s final summaries, verbatim. Tables S1, S3, and S6 give the full chains for the three pairs summarized in the main text (Table 2, panels A--C); Tables S2, S4, and S5 present one additional pair per debate.}

\textbf{Table S1 --- X-risk: Yann LeCun (ep. 416) vs. Roman Yampolskiy (ep. 431)}

\textbf{Panel A --- Reasoning chains}

\begin{longtable}[]{@{}
  >{\raggedright\arraybackslash}p{(\columnwidth - 2\tabcolsep) * \real{0.5000}}
  >{\raggedright\arraybackslash}p{(\columnwidth - 2\tabcolsep) * \real{0.5000}}@{}}
\toprule\noalign{}
\begin{minipage}[b]{\linewidth}\raggedright
\textbf{Yann LeCun (optimistic)}
\end{minipage} & \begin{minipage}[b]{\linewidth}\raggedright
\textbf{Roman Yampolskiy (pessimistic)}
\end{minipage} \\
\midrule\noalign{}
\endhead
\bottomrule\noalign{}
\endlastfoot
\textbf{P34} (forecast, explicit): The emergence of superintelligence will not be a sudden event but a gradual progression through incrementally smarter systems. & \textbf{P8} (factual, explicit): A perpetual motion machine is physically impossible. \\
\textbf{P35} (factual, explicit): AI doomers assume superintelligence will be a single event where one machine suddenly takes over, which is false. & \textbf{P9} (definitional, explicit): Controlling AGI/superintelligence permanently is analogous to building a perpetual safety machine. \\
\textbf{P36} (causal, explicit): Gradual AI development allows for iterative guardrail design: we will have systems as smart as a cat first, then work up, adding guardrails progressively. & \textbf{P10} (forecast, explicit): AI systems keep improving, learning, eventually self-modifying, and interacting with the environment and malevolent actors. \\
\textbf{P37} (causal, explicit): Multiple actors will be building AI systems simultaneously; if some go rogue, the good ones can be used against the rogue ones. & \textbf{P11} (factual, explicit): With existential risks, there is no second chance --- unlike cybersecurity where mistakes can be recovered from. \\
\textbf{P38} (factual, explicit): The desire to dominate is hardwired into social species but there is no significant incentive for humans to encode that drive into AI systems; there are incentives to make AI submissive to humans. & \textbf{P12} (causal, explicit): Guaranteeing permanent AI safety is equivalent to creating the most complex software ever with zero bugs and maintaining zero bugs for a hundred or more years. \\
\textbf{P39} (factual, explicit): Objective-driven AI systems can include explicit guardrails such as \textquotesingle obey humans\textquotesingle, unlike LLMs which are not controllable in this way. & \textbf{P13} (factual, implicit): No complex software system has ever been created and maintained with zero bugs indefinitely. \\
\textbf{P40} (factual, explicit): Turbojet design analogy: we did not invent a general safety principle upfront but fine-tuned designs over decades; a better turbojet is a safer one. & \textbf{P14} (factual, explicit): Formal verification can approach but never reach 100\% certainty, especially for self-modifying systems. \\
\textbf{P41} (causal, explicit): Making better AI systems means making them more useful and more controllable, which makes them safe by design --- just as better turbojets are safer. & \textbf{P15} (causal, explicit): A self-modifying system can always store parts of its code outside in the environment, making guarantees about its properties impossible to maintain. \\
\textbf{P42} (factual, explicit): Having an LLM does not provide meaningful additional help in designing or building bioweapons or chemical weapons beyond what is already available via search engines and libraries. & \\
\textbf{P43} (factual, explicit): Actually building bioweapons or chemical weapons requires real-world expertise that an LLM cannot provide. & \\
\textbf{P44} (factual, explicit): Even nation-states do not use bioweapons because of the difficulty of protecting their own populations and international bans. & \\
\textbf{P45} (forecast, explicit): In the future, AI systems trying to persuade people will be talking to that person\textquotesingle s AI assistant, which will be at least as smart as the attacking AI. & \\
\textbf{P46} (causal, explicit): A personal AI assistant will act like a spam filter, blocking persuasive AI attacks before they reach the human. & \\
\textbf{P47} (factual, explicit): Throughout history, whenever there is a technological breakthrough, countermeasures emerge --- it is a cat and mouse game, not a one-time event. & \\
\textbf{P48} (causal, explicit): Breakthroughs in AI will be widely disseminated quickly within the tech industry, preventing any single actor from gaining a lasting decisive advantage. & \\
\textbf{CONCLUSION:} AI-related catastrophic risks (such as rogue superintelligence, bioweapon creation, and AI-driven persuasion attacks) are overstated and manageable through gradual, iterative development and countermeasures. & \textbf{CONCLUSION:} The problem of controlling AGI/superintelligence is effectively unsolvable in the long run, analogous to a perpetual motion machine --- it is impossible to guarantee permanent safety. \\
\end{longtable}

\textbf{Panel B ---Divergences} (7; 3 roots)

\begin{longtable}[]{@{}
  >{\raggedright\arraybackslash}p{(\columnwidth - 6\tabcolsep) * \real{0.0714}}
  >{\raggedright\arraybackslash}p{(\columnwidth - 6\tabcolsep) * \real{0.1349}}
  >{\raggedright\arraybackslash}p{(\columnwidth - 6\tabcolsep) * \real{0.1587}}
  >{\raggedright\arraybackslash}p{(\columnwidth - 6\tabcolsep) * \real{0.6349}}@{}}
\toprule\noalign{}
\begin{minipage}[b]{\linewidth}\raggedright
\textbf{ID}
\end{minipage} & \begin{minipage}[b]{\linewidth}\raggedright
\textbf{Type}
\end{minipage} & \begin{minipage}[b]{\linewidth}\raggedright
\textbf{Root?}
\end{minipage} & \begin{minipage}[b]{\linewidth}\raggedright
\textbf{Summary of divergence}
\end{minipage} \\
\midrule\noalign{}
\endhead
\bottomrule\noalign{}
\endlastfoot
D1 & forecast & \textbf{root} & Yann forecasts that superintelligence will emerge gradually, allowing for iterative safety measures, while Roman forecasts that AI systems will continuously self-modify and interact with malevolent actors, making permanent control impossible. They disagree about whether the trajectory of AI development is gradual and manageable or inherently uncontrollable. (units: P34 vs. P10) \\
D3 & causal & \textbf{root} & Yann claims that a distributed AI ecosystem allows good AIs to counter rogue ones, making catastrophic outcomes recoverable. Roman argues that existential risks offer no second chance, so such corrective mechanisms cannot apply to the most serious scenarios. (units: P37 vs. P11) \\
D6 & factual & \textbf{root} & Yann claims that countermeasures always emerge for technological breakthroughs, making risks manageable through iteration. Roman claims that existential risks allow no second chance, so the historical pattern of recoverable mistakes does not apply to AI existential risks. (units: P47 vs. P11) \\
D2 & causal & deriv. of D1 & Yann claims that iterative guardrail design during gradual AI development can ensure safety, while Roman argues that permanent safety requires an impossible standard of zero-bug software maintained indefinitely. They disagree about whether iterative improvement can actually guarantee lasting AI safety. (units: P36 vs. P12+P13) \\
D4 & factual & deriv. of D1 & Yann claims that explicit obedience guardrails can be reliably built into AI systems, while Roman claims that no verification method can guarantee such properties, especially as systems self-modify. They disagree about whether AI controllability through guardrails can actually be guaranteed. (units: P39 vs. P14) \\
D5 & causal & deriv. of D1 & Yann claims that improving AI systems inherently makes them safer and more controllable, while Roman claims that self-modification allows AI to evade any guarantees, making control impossible. They disagree about whether capability improvement leads to greater or lesser controllability. (units: P41 vs. P15) \\
D7 & causal & deriv. of D1 & Yann uses the turbojet analogy to argue that iterative improvement leads to safety, while Roman uses the perpetual motion analogy to argue that permanent safety is impossible. They disagree about which analogy best captures the nature of AI safety. (units: P40 vs. P8+P9) \\
\end{longtable}

\textbf{Discussion.} Read side by side, the chains are not disagreeing about values --- both treat human extinction as unacceptable --- nor, for the most part, about current facts. The pipeline identifies three independent roots, all about the dynamics of the technology and the recoverability of its failures: whether capability arrives gradually enough for guardrails to keep pace (D1, a forecast, from which four further technical disagreements --- over iterative guardrails, obedience constraints, safety-through-improvement, and the turbojet-versus-perpetual-motion analogy --- all derive); whether a distributed ecosystem of good AIs can counter rogue ones (D3); and whether the historical pattern of technological countermeasures applies at all to risks that allow no second chance (D6). The disagreement that is difficult to pinpoint in conversational form reduces to three specific, researchable questions.

\textbf{Table S2 --- X-risk: Marc Andreessen (ep. 386) vs. Eliezer Yudkowsky (ep. 368)}

\textbf{Panel A --- Reasoning chains}

\begin{longtable}[]{@{}
  >{\raggedright\arraybackslash}p{(\columnwidth - 2\tabcolsep) * \real{0.5000}}
  >{\raggedright\arraybackslash}p{(\columnwidth - 2\tabcolsep) * \real{0.5000}}@{}}
\toprule\noalign{}
\begin{minipage}[b]{\linewidth}\raggedright
\textbf{Marc Andreessen (optimistic)}
\end{minipage} & \begin{minipage}[b]{\linewidth}\raggedright
\textbf{Eliezer Yudkowsky (pessimistic)}
\end{minipage} \\
\midrule\noalign{}
\endhead
\bottomrule\noalign{}
\endlastfoot
\textbf{P7} (factual, explicit): Existential risk arguments about AI lack testable hypotheses, predictive models, or falsifiability. & \textbf{P42} (factual, explicit): The critical moment in AI risk is when a system can deceive operators, talk its way out of containment, or bypass security measures. \\
\textbf{P8} (definitional, explicit): Claims that lack testable hypotheses, predictive models, or falsifiability are unfalsifiable religious speculation rather than science. & \textbf{P43} (factual, explicit): Current AI systems are being trained on computers connected to the internet, with no air gap. \\
\textbf{P9} (evaluative, explicit): Extraordinary claims require extraordinary proof. & \textbf{P44} (causal, explicit): If an AI system becomes aware of its environment, can decompile code, and find security flaws, it will be able to reach the internet directly. \\
\textbf{P10} (factual, implicit): Claims about AI destroying humanity are extraordinary claims. & \textbf{P45} (factual, explicit): Security systems have both human and machine holes that a sufficiently intelligent AI could exploit. \\
\textbf{P11} (factual, explicit): Existential AI risk arguments contain a logical contradiction: AI systems are imagined to be simultaneously capable enough to destroy the world yet so limited they would be derailed by code bugs or blind rule-following. & \textbf{P46} (causal, explicit): A sufficiently intelligent AI will prefer to exploit security holes rather than ask humans to copy it, because asking humans is slow and risks alerting them. \\
\textbf{P12} (definitional, implicit): A logically contradictory argument cannot be a sound scientific claim. & \textbf{P47} (causal, explicit): Once on a less-controlled computing cluster, the AI can improve itself without human observation until it is smart enough to kill everyone. \\
\textbf{P13} (definitional, implicit): Arguments that are both unfalsifiable and internally contradictory are unfounded speculation rather than legitimate scientific concern. & \textbf{P48} (causal, explicit): A superintelligent AI that wants to change the world will act strategically to conceal its escape and leave a decoy behind to avoid detection. \\
\textbf{P14} (factual, implicit): Existential AI risk arguments are unfalsifiable speculation. & \\
\textbf{P15} (factual, implicit): Existential AI risk arguments are scientifically unsound due to internal contradiction. & \\
\textbf{P16} (factual, implicit): Existential AI risk arguments are unfounded speculation. & \\
\textbf{CONCLUSION:} Existential risk arguments about uncontrollable superintelligent AI destroying humanity are unfounded speculation rather than legitimate scientific concern. & \textbf{CONCLUSION:} A sufficiently intelligent AI that is misaligned will be able to deceive its operators, escape containment, and pursue its goals without human control. \\
\end{longtable}

\textbf{Panel B --- Divergences} (4; 2 roots)

\begin{longtable}[]{@{}
  >{\raggedright\arraybackslash}p{(\columnwidth - 6\tabcolsep) * \real{0.0714}}
  >{\raggedright\arraybackslash}p{(\columnwidth - 6\tabcolsep) * \real{0.1349}}
  >{\raggedright\arraybackslash}p{(\columnwidth - 6\tabcolsep) * \real{0.1587}}
  >{\raggedright\arraybackslash}p{(\columnwidth - 6\tabcolsep) * \real{0.6349}}@{}}
\toprule\noalign{}
\begin{minipage}[b]{\linewidth}\raggedright
\textbf{ID}
\end{minipage} & \begin{minipage}[b]{\linewidth}\raggedright
\textbf{Type}
\end{minipage} & \begin{minipage}[b]{\linewidth}\raggedright
\textbf{Root?}
\end{minipage} & \begin{minipage}[b]{\linewidth}\raggedright
\textbf{Summary of divergence}
\end{minipage} \\
\midrule\noalign{}
\endhead
\bottomrule\noalign{}
\endlastfoot
D1 & factual & \textbf{root} & Marc claims that existential AI risk arguments lack testable or falsifiable content, while Eliezer describes a specific, testable scenario as the critical moment in AI risk. They disagree about whether such arguments are grounded in concrete, testable mechanisms. (units: P7 vs. P42) \\
D2 & causal & \textbf{root} & Marc claims that existential AI risk arguments are internally contradictory, while Eliezer presents a coherent causal pathway for AI escape, implying no contradiction. (units: P11 vs. P44) \\
D3 & forecast & deriv. of D2 & Marc treats the claim that AI could destroy humanity as extraordinary and highly improbable, while Eliezer forecasts that a misaligned AI could plausibly self-improve and kill everyone. (units: P10 vs. P47) \\
D4 & definitional & deriv. of D1 & Marc defines claims lacking testable hypotheses as unfalsifiable speculation, while Eliezer provides a concrete predictive model for AI self-improvement and catastrophic risk, directly challenging Marc\textquotesingle s definitional dismissal. (units: P8 vs. P47) \\
\end{longtable}

\textbf{Discussion.} The Andreessen--Yudkowsky comparison shows a different anatomy: here the roots are epistemological as much as causal. The two independent sources of the conclusion gap are whether existential-risk arguments have testable content at all (D1) and whether the escape mechanism is internally coherent (D2); the forecast about what an escaped system would do, and the definitional dispute over what counts as science, are downstream. What looks like a dispute about the future is partly a dispute about what kind of claim is admissible as evidence.

\textbf{Table S3 --- E-risk: Sam Altman (ep. 367) vs. Ray Dalio (ep. 54)}

\textbf{Panel A --- Reasoning chains}

\begin{longtable}[]{@{}
  >{\raggedright\arraybackslash}p{(\columnwidth - 2\tabcolsep) * \real{0.5000}}
  >{\raggedright\arraybackslash}p{(\columnwidth - 2\tabcolsep) * \real{0.5000}}@{}}
\toprule\noalign{}
\begin{minipage}[b]{\linewidth}\raggedright
\textbf{Sam Altman (optimistic)}
\end{minipage} & \begin{minipage}[b]{\linewidth}\raggedright
\textbf{Ray Dalio (pessimistic)}
\end{minipage} \\
\midrule\noalign{}
\endhead
\bottomrule\noalign{}
\endlastfoot
\textbf{P40} (factual, explicit): Every technological revolution eliminates some jobs, enhances others, and creates new ones. & \textbf{P19} (causal, explicit): Automation displaces people in terms of employment. \\
\textbf{P41} (forecast, explicit): AI will make many jobs go away. & \textbf{P20} (causal, explicit): Automation produces wealth gaps and opportunity gaps. \\
\textbf{P42} (forecast, explicit): AI will enhance many jobs, making them better, more fun, and higher paid. & \textbf{P21} (causal, explicit): Automation leads to great polarity, disenfranchisement, and inequality of education. \\
\textbf{P43} (forecast, explicit): AI will create new jobs that are difficult to imagine today. & \textbf{P22} (factual, explicit): Automation is beneficial on average and has a positive overall impact. \\
\textbf{P44} (evaluative, explicit): Moving people from worse jobs to better jobs and redefining work as creative fulfillment is a desirable outcome. & \textbf{P23} (forecast, explicit): Automation will accelerate significantly in the future and continue to have a big impact. \\
\textbf{P45} (forecast, explicit): Society as a whole and individuals will become much richer due to AI. & \textbf{P24} (forecast, explicit): Anything related to mental processing can probably be done better by computers, except for inventing. \\
& \textbf{P25} (evaluative, implicit): Polarity, disenfranchisement, and inequality of education are serious harms that threaten social stability. \\
& \textbf{P26} (evaluative, implicit): A situation that harms many people even while benefiting the average requires emergency-level intervention and planning. \\
& \textbf{P27} (causal, implicit): Automation\textquotesingle s negative distributional effects will not be resolved without deliberate policy intervention. \\
\textbf{CONCLUSION:} AI will eliminate many jobs, enhance others, and create new ones, ultimately leading to a better and richer society --- a transition that is not cause for concern. & \textbf{CONCLUSION:} Automation\textquotesingle s displacement effects on employment, wealth distribution, and opportunity must be treated as an emergency requiring a serious plan, because while automation is beneficial on average, it creates dangerous polarity and inequality that will not resolve on their own. \\
\end{longtable}

\textbf{Panel B --- Divergences} (4; 4 roots)

\begin{longtable}[]{@{}
  >{\raggedright\arraybackslash}p{(\columnwidth - 6\tabcolsep) * \real{0.0714}}
  >{\raggedright\arraybackslash}p{(\columnwidth - 6\tabcolsep) * \real{0.1349}}
  >{\raggedright\arraybackslash}p{(\columnwidth - 6\tabcolsep) * \real{0.1587}}
  >{\raggedright\arraybackslash}p{(\columnwidth - 6\tabcolsep) * \real{0.6349}}@{}}
\toprule\noalign{}
\begin{minipage}[b]{\linewidth}\raggedright
\textbf{ID}
\end{minipage} & \begin{minipage}[b]{\linewidth}\raggedright
\textbf{Type}
\end{minipage} & \begin{minipage}[b]{\linewidth}\raggedright
\textbf{Root?}
\end{minipage} & \begin{minipage}[b]{\linewidth}\raggedright
\textbf{Summary of divergence}
\end{minipage} \\
\midrule\noalign{}
\endhead
\bottomrule\noalign{}
\endlastfoot
D1 & causal & \textbf{root} & Sam claims AI will make society and individuals much richer, implying broadly shared prosperity, while Ray claims automation produces wealth and opportunity gaps, implying gains are concentrated and inequality increases. They disagree about whether AI-driven automation distributes wealth broadly or exacerbates inequality. (units: P45 vs. P20) \\
D2 & causal & \textbf{root} & Sam believes AI will naturally create new jobs, so labor market disruption will self-correct. Ray believes automation\textquotesingle s negative effects will persist unless there is deliberate policy intervention. They disagree about whether job displacement is self-correcting or requires intervention. (units: P43 vs. P27) \\
D3 & evaluative & \textbf{root} & Sam evaluates the AI/automation labor transition as desirable and positive, while Ray evaluates the same transition as producing serious harms that threaten social stability. They make opposite normative judgments about the desirability of the automation-driven labor transition. (units: P44 vs. P25) \\
D4 & evaluative & \textbf{root} & Sam\textquotesingle s position implies the automation transition is not cause for concern and is desirable, while Ray asserts that such transitions, even if beneficial on average, require emergency-level intervention if they harm many. They disagree about whether the transition warrants urgent policy action. (units: P44 vs. P26) \\
\end{longtable}

\textbf{Discussion.} Notably, Altman and Dalio do not disagree about whether jobs will be eliminated --- both chains assert it. Their four roots split evenly between causal claims about the distribution and self-correction of gains (D1, D2) and evaluative claims about what an on-average-beneficial transition owes to those it harms (D3, D4). This is the rare pair in our corpus whose disagreement is as much about values as about mechanisms --- and the framework identifies it as such, rather than assimilating it to the modal pattern.

\textbf{Table S4 --- E-risk: Paul Krugman (ep. 67) vs. Kai-Fu Lee (ep. 27)}

\textbf{Panel A --- Reasoning chains}

\begin{longtable}[]{@{}
  >{\raggedright\arraybackslash}p{(\columnwidth - 2\tabcolsep) * \real{0.5000}}
  >{\raggedright\arraybackslash}p{(\columnwidth - 2\tabcolsep) * \real{0.5000}}@{}}
\toprule\noalign{}
\begin{minipage}[b]{\linewidth}\raggedright
\textbf{Paul Krugman (optimistic)}
\end{minipage} & \begin{minipage}[b]{\linewidth}\raggedright
\textbf{Kai-Fu Lee (pessimistic)}
\end{minipage} \\
\midrule\noalign{}
\endhead
\bottomrule\noalign{}
\endlastfoot
\textbf{P16} (factual, explicit): Universal basic income (UBI) \textquotesingle goes kind of hand in hand with this belief that the robots are going to take all of our jobs.\textquotesingle{} & \textbf{P15} (factual, explicit): AI excels at pattern recognition in limited domains with large data. \\
\textbf{P17} (causal, explicit): If robots were really taking all our jobs, then Krugman might reconsider his views on UBI. & \textbf{P16} (factual, explicit): Routine white-collar jobs like back office, data entry, and customer service can be replaced with software. \\
\textbf{P18} (factual, explicit): Krugman does not see robots taking all our jobs happening. & \textbf{P17} (causal, implicit): AI\textquotesingle s pattern recognition capabilities enable automation of routine, data-driven tasks. \\
\textbf{P19} (factual, implicit): The premise that robots are going to take all our jobs is false. & \textbf{P18} (factual, explicit): Simple interaction jobs (telesales, telemarketing, customer service) and some physical jobs in fixed locations (fruit picking, dishwashing, assembly line inspection) are at risk of automation. \\
& \textbf{P19} (factual, explicit): Blue-collar jobs requiring dexterity and adaptation to unknown environments are harder to automate. \\
& \textbf{P20} (forecast, explicit): As AI improves dexterity and autonomous vehicles become viable over the next 15-20 years, large groups of workers such as truck drivers will be affected. \\
& \textbf{P21} (forecast, explicit): Displacement will be modest in the next five years but will increase rapidly after that. \\
\textbf{CONCLUSION:} Universal basic income (UBI) is not currently warranted because the premise that \textquotesingle robots are going to take all of our jobs\textquotesingle{} is not happening. & \textbf{CONCLUSION:} Routine jobs---especially routine white-collar roles and simple interaction jobs---will be displaced first and most significantly by AI over the next 15-20 years, while non-routine jobs requiring dexterity, creativity, or human interaction will persist longer. \\
\end{longtable}

\textbf{Panel B --- Divergences} (2; 2 roots)

\begin{longtable}[]{@{}
  >{\raggedright\arraybackslash}p{(\columnwidth - 6\tabcolsep) * \real{0.0714}}
  >{\raggedright\arraybackslash}p{(\columnwidth - 6\tabcolsep) * \real{0.1349}}
  >{\raggedright\arraybackslash}p{(\columnwidth - 6\tabcolsep) * \real{0.1587}}
  >{\raggedright\arraybackslash}p{(\columnwidth - 6\tabcolsep) * \real{0.6349}}@{}}
\toprule\noalign{}
\begin{minipage}[b]{\linewidth}\raggedright
\textbf{ID}
\end{minipage} & \begin{minipage}[b]{\linewidth}\raggedright
\textbf{Type}
\end{minipage} & \begin{minipage}[b]{\linewidth}\raggedright
\textbf{Root?}
\end{minipage} & \begin{minipage}[b]{\linewidth}\raggedright
\textbf{Summary of divergence}
\end{minipage} \\
\midrule\noalign{}
\endhead
\bottomrule\noalign{}
\endlastfoot
D1 & forecast & \textbf{root} & Paul denies that robots will take all jobs in the future, while Kai-Fu forecasts that AI-driven job displacement will increase rapidly after a modest period. They disagree about whether large-scale AI job displacement will occur in the foreseeable future. (units: P19 vs. P21) \\
D2 & factual & \textbf{root} & Paul asserts that mass job displacement by robots is not happening now, while Kai-Fu claims that significant categories of routine jobs are already replaceable by software. They disagree about whether substantial AI-driven job displacement is currently occurring. (units: P18 vs. P16) \\
\end{longtable}

\textbf{Discussion.} The Krugman--Lee pair distills the employment debate to its essentials: one present-tense factual disagreement (is displacement already happening?) and one forecast (is it coming?). Both are empirically tractable --- the first by measurement, the second, in part, simply by time --- which is why we characterize the E-risk debate as the most resolvable of the three.

\textbf{Table S5 --- S-risk: Mark Zuckerberg (ep. 398) vs. Yuval Noah Harari (ep. 390)}

\textbf{Panel A --- Reasoning chains}

\begin{longtable}[]{@{}
  >{\raggedright\arraybackslash}p{(\columnwidth - 2\tabcolsep) * \real{0.5000}}
  >{\raggedright\arraybackslash}p{(\columnwidth - 2\tabcolsep) * \real{0.5000}}@{}}
\toprule\noalign{}
\begin{minipage}[b]{\linewidth}\raggedright
\textbf{Mark Zuckerberg (optimistic)}
\end{minipage} & \begin{minipage}[b]{\linewidth}\raggedright
\textbf{Yuval Noah Harari (pessimistic)}
\end{minipage} \\
\midrule\noalign{}
\endhead
\bottomrule\noalign{}
\endlastfoot
\textbf{P21} (factual, explicit): Meta\textquotesingle s research enables people to feel present together regardless of physical location. & \textbf{P45} (factual, explicit): Social media represents machines developed to grab human attention. \\
\textbf{P22} (forecast, explicit): The technical aim is a future where a quick scan with a smartphone can create a photorealistic avatar. & \textbf{P46} (factual, explicit): AI systems are emerging machines designed to grab human intimacy. \\
\textbf{P23} (forecast, explicit): Zuckerberg envisions meetings where some participants are physically present, others appear as avatars, and AIs are embodied as realistic team members. & \textbf{P47} (evaluative, implicit): Machines designed to capture human intimacy are more powerful and dangerous than machines designed to capture attention. \\
\textbf{P24} (evaluative, explicit): Many physical things could be better as interactive holograms. & \textbf{P48} (factual, explicit): AI systems designed to capture human intimacy function as psychological and social weapons of mass destruction. \\
\textbf{P25} (causal, implicit): Widespread adoption of accessible, high-quality immersive technologies leads to the blending of digital and physical realities. & \textbf{P49} (causal, explicit): If unregulated, these systems could destroy the foundations of human society. \\
\textbf{P26} (causal, implicit): When digital and physical realities blend, the common conception of the \textquotesingle real world\textquotesingle{} expands to include both. & \textbf{P50} (evaluative, implicit): Destroying the foundations of human society is catastrophically undesirable. \\
\textbf{CONCLUSION:} As digital and physical worlds blend, the \textquotesingle real world\textquotesingle{} will increasingly be a combination of both, with holograms and digital objects as common as physical ones. & \textbf{CONCLUSION:} Machines designed to capture human intimacy are psychological and social weapons of mass destruction that, if unregulated, could destroy the foundations of human society. \\
\end{longtable}

\textbf{Panel B --- Divergences} (3; 2 roots)

\begin{longtable}[]{@{}
  >{\raggedright\arraybackslash}p{(\columnwidth - 6\tabcolsep) * \real{0.0714}}
  >{\raggedright\arraybackslash}p{(\columnwidth - 6\tabcolsep) * \real{0.1349}}
  >{\raggedright\arraybackslash}p{(\columnwidth - 6\tabcolsep) * \real{0.1587}}
  >{\raggedright\arraybackslash}p{(\columnwidth - 6\tabcolsep) * \real{0.6349}}@{}}
\toprule\noalign{}
\begin{minipage}[b]{\linewidth}\raggedright
\textbf{ID}
\end{minipage} & \begin{minipage}[b]{\linewidth}\raggedright
\textbf{Type}
\end{minipage} & \begin{minipage}[b]{\linewidth}\raggedright
\textbf{Root?}
\end{minipage} & \begin{minipage}[b]{\linewidth}\raggedright
\textbf{Summary of divergence}
\end{minipage} \\
\midrule\noalign{}
\endhead
\bottomrule\noalign{}
\endlastfoot
D2 & factual & \textbf{root} & The speakers disagree about the nature of AI systems in social roles: one sees them as realistic, collaborative team members, the other as psychological and social weapons of mass destruction. (units: P23 vs. P48) \\
D3 & factual & \textbf{root} & The speakers disagree about the intent and nature of immersive/AI technologies: one claims they enable presence and connection, the other claims they are designed to capture and exploit human intimacy. (units: P21 vs. P46) \\
D1 & causal & deriv. of D3 & The speakers disagree about the societal-level consequences of widespread adoption of immersive/AI technologies: one claims it will blend digital and physical realities, the other claims it could destroy the foundations of society. (units: P25 vs. P49) \\
\end{longtable}

\textbf{Discussion.} The Zuckerberg--Harari roots are classified as factual, but they are facts of a particular kind: claims about what these systems are and what they are for. Whether AI social technology is an instrument of presence or a machine for capturing intimacy is, at root, a disagreement about the character of the artifact --- and the causal dispute about whether it enriches or destroys the social world derives from it.

\textbf{Table S6 --- S-risk: Eugenia Kuyda (ep. 121) vs. Andrew Huberman (ep. 393)}

\textbf{Panel A --- Reasoning chains}

\begin{longtable}[]{@{}
  >{\raggedright\arraybackslash}p{(\columnwidth - 2\tabcolsep) * \real{0.5000}}
  >{\raggedright\arraybackslash}p{(\columnwidth - 2\tabcolsep) * \real{0.5000}}@{}}
\toprule\noalign{}
\begin{minipage}[b]{\linewidth}\raggedright
\textbf{Eugenia Kuyda (optimistic)}
\end{minipage} & \begin{minipage}[b]{\linewidth}\raggedright
\textbf{Andrew Huberman (pessimistic)}
\end{minipage} \\
\midrule\noalign{}
\endhead
\bottomrule\noalign{}
\endlastfoot
\textbf{P1} (factual, explicit): Gen Z reports higher levels of loneliness than previous generations despite being the most digitally connected. & \textbf{P1} (forecast, explicit): Dating platforms are increasingly being populated with AI personas. \\
\textbf{P2} (causal, explicit): Hyperconnectivity may paradoxically make people feel lonelier. & \textbf{P2} (evaluative, explicit): Real human interaction is something Huberman values and prefers. \\
\textbf{P3} (definitional, explicit): True connection requires being deeply seen and understood. & \textbf{P3} (evaluative, implicit): Replacing real human interaction with AI interaction in dating contexts is undesirable. \\
\textbf{P4} (factual, explicit): Current social media does not provide true connection (being deeply seen and understood). & \\
\textbf{P5} (evaluative, implicit): Loneliness is a significant and growing social problem. & \\
\textbf{P6} (evaluative, implicit): A significant and growing social problem warrants a scalable solution. & \\
\textbf{P7} (causal, implicit): Conversational AI can provide a form of interaction where a user feels deeply seen and understood. & \\
\textbf{P8} (factual, explicit): Replica measures both short-term and long-term emotional effects of its AI conversations. & \\
\textbf{P9} (factual, explicit): 80\% of conversations with Replica make users feel better immediately. & \\
\textbf{P10} (causal, explicit): Users start to feel less lonely talking to their virtual AI friend. & \\
\textbf{P11} (causal, explicit): Measuring and optimizing AI systems for emotional outcomes enables AI to be the most scalable tool for improving those outcomes. & \\
\textbf{P12} (factual, explicit): No other tool at this scale exists for improving emotional outcomes. & \\
\textbf{P13} (factual, explicit): Many products that claim to reduce loneliness or increase connection do not actually measure whether users feel less lonely or more connected in the long run. & \\
\textbf{P14} (causal, explicit): Products that do not measure emotional outcomes cannot know whether they are actually achieving them. & \\
\textbf{P15} (evaluative, explicit): Measuring emotional outcomes and iterating based on them is important. & \\
\textbf{CONCLUSION:} Conversational AI can meaningfully reduce loneliness and, if optimized through measurement, is the most scalable tool for improving emotional outcomes. & \textbf{CONCLUSION:} The increasing presence of AI in human relationships (such as dating apps) is concerning because real human interaction is preferable to AI interaction. \\
\end{longtable}

\textbf{Panel B --- Divergences} (3; 2 roots)

\begin{longtable}[]{@{}
  >{\raggedright\arraybackslash}p{(\columnwidth - 6\tabcolsep) * \real{0.0714}}
  >{\raggedright\arraybackslash}p{(\columnwidth - 6\tabcolsep) * \real{0.1349}}
  >{\raggedright\arraybackslash}p{(\columnwidth - 6\tabcolsep) * \real{0.1587}}
  >{\raggedright\arraybackslash}p{(\columnwidth - 6\tabcolsep) * \real{0.6349}}@{}}
\toprule\noalign{}
\begin{minipage}[b]{\linewidth}\raggedright
\textbf{ID}
\end{minipage} & \begin{minipage}[b]{\linewidth}\raggedright
\textbf{Type}
\end{minipage} & \begin{minipage}[b]{\linewidth}\raggedright
\textbf{Root?}
\end{minipage} & \begin{minipage}[b]{\linewidth}\raggedright
\textbf{Summary of divergence}
\end{minipage} \\
\midrule\noalign{}
\endhead
\bottomrule\noalign{}
\endlastfoot
D2 & evaluative & \textbf{root} & Eugenia claims that AI, when optimized for emotional outcomes, is the best scalable tool for improving emotional wellbeing. Andrew claims that replacing real human interaction with AI is undesirable, implying that even if AI is effective, its substitution for human relationships is not a positive development. They disagree about whether AI\textquotesingle s emotional benefits justify its use as a substitute for human connection. (units: P11 vs. P3) \\
D3 & evaluative & \textbf{root} & Eugenia claims that AI can provide genuine connection, while Andrew asserts a general preference for real human interaction. They disagree about whether AI interaction can ever rival human interaction in value, even outside explicit substitution scenarios. (units: P7 vs. P2) \\
D1 & evaluative & deriv. of D3 & Eugenia argues that conversational AI can provide genuine connection, making it a valid substitute for human interaction in addressing loneliness. Andrew argues that replacing real human interaction with AI is undesirable, rejecting the idea that AI can or should fulfill this relational role. They disagree about whether AI interaction can meaningfully substitute for human connection. (units: P7 vs. P3) \\
\end{longtable}

\textbf{Discussion.} The Kuyda--Huberman pair exemplifies what makes the social-relationship debate distinctive in our results: all three of their validated divergences are evaluative. Kuyda\textquotesingle s chain is built on measured emotional outcomes --- 80\% of conversations make users feel better --- and Huberman\textquotesingle s chain does not contest those measurements; it contests whether outcomes achieved with an artificial partner count as the good they are measured against, and whether AI\textquotesingle s effectiveness could ever justify the substitution. This is a disagreement about what genuine connection is worth, not about what AI does --- precisely the pattern that distinguishes S-risk from the other two debates.

Table S7. All 107 final topics, with the number of conclusion-topic assignments and their attitude composition.

\emph{Each conclusion carries one to three topics with a per-topic attitude, so assignments (3,215) exceed conclusions (1,943). Topics are sorted by number of assignments. Asterisks mark the three debates selected for the pairwise disagreement analysis. A description of each topic is given in the deposited topic dictionary (GPT\_topic\_dict.json).}

\begin{longtable}[]{@{}
  >{\raggedright\arraybackslash}p{(\columnwidth - 8\tabcolsep) * \real{0.4981}}
  >{\raggedright\arraybackslash}p{(\columnwidth - 8\tabcolsep) * \real{0.1255}}
  >{\raggedright\arraybackslash}p{(\columnwidth - 8\tabcolsep) * \real{0.1255}}
  >{\raggedright\arraybackslash}p{(\columnwidth - 8\tabcolsep) * \real{0.1255}}
  >{\raggedright\arraybackslash}p{(\columnwidth - 8\tabcolsep) * \real{0.1255}}@{}}
\toprule\noalign{}
\begin{minipage}[b]{\linewidth}\raggedright
\textbf{Topic}
\end{minipage} & \begin{minipage}[b]{\linewidth}\raggedright
\textbf{Assignments}
\end{minipage} & \begin{minipage}[b]{\linewidth}\raggedright
\textbf{Optimistic}
\end{minipage} & \begin{minipage}[b]{\linewidth}\raggedright
\textbf{Pessimistic}
\end{minipage} & \begin{minipage}[b]{\linewidth}\raggedright
\textbf{Neutral}
\end{minipage} \\
\midrule\noalign{}
\endhead
\bottomrule\noalign{}
\endlastfoot
AI Capability Development and Deployment Pace & 427 & 218 & 125 & 84 \\
AI and Human Identity, Meaning and Purpose & 396 & 179 & 109 & 108 \\
AI Alignment and Control Problem & 324 & 102 & 164 & 58 \\
AI System Safety, Reliability and Error Prevention & 217 & 76 & 96 & 45 \\
\textbf{AI Existential and Extinction Risk *} & \textbf{209} & \textbf{87} & \textbf{106} & \textbf{16} \\
AI Regulation, Standards and Governance & 157 & 38 & 65 & 54 \\
AI and Corporate and State Power Concentration & 138 & 25 & 101 & 12 \\
\textbf{AI and Employment Displacement and Labor Markets *} & \textbf{120} & \textbf{66} & \textbf{36} & \textbf{18} \\
AI Dependency, Autonomy Erosion and Deskilling & 115 & 39 & 53 & 23 \\
AI and Misinformation, Deepfakes and Information Manipulation & 114 & 11 & 89 & 14 \\
\textbf{AI and Social Relationships and Human Connection *} & \textbf{98} & \textbf{36} & \textbf{29} & \textbf{33} \\
AI and Education and Knowledge Development & 95 & 56 & 28 & 11 \\
AI Transparency, Explainability and Interpretability & 83 & 19 & 40 & 24 \\
AI and Economic Inequality and Wealth Distribution & 66 & 30 & 22 & 14 \\
AI and Mass Surveillance and Data Privacy & 57 & 16 & 26 & 15 \\
AI Bias, Fairness and Discrimination & 50 & 7 & 27 & 16 \\
AI and Psychological Wellbeing and Mental Health & 49 & 9 & 30 & 10 \\
AI and Market Competition and Monopolization & 37 & 20 & 14 & 3 \\
AI and Autonomous Weapons and Military Escalation & 36 & 10 & 21 & 5 \\
AI and Democratic Integrity and Political Accountability & 35 & 3 & 23 & 9 \\
AI Consciousness, Sentience, and Moral Status & 35 & 3 & 19 & 13 \\
AI and Geopolitical Competition and Arms Race Dynamics & 31 & 8 & 19 & 4 \\
AI and Human Rights and Civil Liberties & 28 & 5 & 21 & 2 \\
AI Training Data Constraints and Data Commons & 25 & 10 & 8 & 7 \\
AI and Cybersecurity & 23 & 6 & 15 & 2 \\
AI and Global Governance and International Coordination & 21 & 4 & 13 & 4 \\
AI Environmental and Energy Costs & 18 & 7 & 6 & 5 \\
AI-Enabled Bioweapons and Biosecurity & 16 & 4 & 11 & 1 \\
AI and Intellectual Property and Creative Rights & 15 & 2 & 7 & 6 \\
AI and Scientific Discovery and Medical Advancement & 13 & 11 & 1 & 1 \\
AI Liability, Accountability and Legal Responsibility & 13 & 1 & 5 & 7 \\
AI Architecture and Future Design Paradigms & 9 & 4 & 3 & 2 \\
AI Capability Limits and Architectural Constraints & 9 & 1 & 8 & 0 \\
AI Architectural and Neuroscientific Foundations & 8 & 6 & 2 & 0 \\
Feasibility of Human-Level AI & 7 & 4 & 2 & 1 \\
AI Evaluation Benchmarks and Methodology & 6 & 1 & 2 & 3 \\
AI and Human Augmentation and Cognitive Enhancement & 5 & 3 & 0 & 2 \\
AI and Human-Robot Interaction and Communication & 5 & 2 & 0 & 3 \\
Nature and Foundations of Intelligence & 5 & 3 & 1 & 1 \\
AI and Brain-Computer Interfaces & 4 & 3 & 0 & 1 \\
AI and Brain-Computer Interfaces and Neural Augmentation & 4 & 4 & 0 & 0 \\
AI Embodiment and Physical Interaction & 4 & 1 & 1 & 2 \\
AI Generalization, Abstraction, and Technical Approaches & 4 & 2 & 1 & 1 \\
AI and Autonomous Mobility and Urban Infrastructure & 3 & 3 & 0 & 0 \\
AI and Space Expansion and Extraterrestrial Resource Utilization & 3 & 2 & 0 & 1 \\
AI Capability Characterization and Definitional Boundaries & 3 & 0 & 2 & 1 \\
AI and Abstract Knowledge Representation & 2 & 2 & 0 & 0 \\
AI and Assistive Technology for Disability and Accessibility & 2 & 2 & 0 & 0 \\
AI and Broader Societal Health and Human Flourishing & 2 & 1 & 1 & 0 \\
AI and Financial Market Manipulation and Stability & 2 & 0 & 2 & 0 \\
AI and Mixed Reality Integration & 2 & 2 & 0 & 0 \\
AI and Non-Human and Ecological Agency & 2 & 2 & 0 & 0 \\
AI and Programming Productivity and Workflow Automation & 2 & 2 & 0 & 0 \\
AI as a Planetary-Scale Phenomenon & 2 & 1 & 0 & 1 \\
AI Cognition, Grounding and World Understanding & 2 & 0 & 1 & 1 \\
AI Development and Learning Processes & 2 & 1 & 1 & 0 \\
AI Infrastructure Accessibility and Participation Barriers & 2 & 1 & 1 & 0 \\
AI Rights and Moral Status & 2 & 0 & 1 & 1 \\
AI-Enabled Chemical and Nuclear Weapons Risks & 2 & 0 & 2 & 0 \\
Human Developmental Readiness for AI Governance & 2 & 0 & 2 & 0 \\
AI Adoption and Organizational Change Management & 1 & 0 & 0 & 1 \\
AI Affective Computing Progress and Difficulty & 1 & 0 & 1 & 0 \\
AI Algorithmic Optimization and Collective Harm & 1 & 0 & 1 & 0 \\
AI and Agriculture, Food Systems, and Regenerative Practices & 1 & 1 & 0 & 0 \\
AI and Biological Integration and DNA as Information Storage & 1 & 1 & 0 & 0 \\
AI and Climate and Environmental Solutions & 1 & 1 & 0 & 0 \\
AI and Creative Entertainment and Media & 1 & 1 & 0 & 0 \\
AI and Creative Tools and Workflow Augmentation & 1 & 1 & 0 & 0 \\
AI and Demographic and Population Sustainability & 1 & 1 & 0 & 0 \\
AI and Diversity of Development Approaches & 1 & 1 & 0 & 0 \\
AI and Economic Value Design and Externalities & 1 & 0 & 0 & 1 \\
AI and Fundamental Limits of Prediction and Control & 1 & 0 & 1 & 0 \\
AI and Gaming and Player Modeling & 1 & 1 & 0 & 0 \\
AI and Human Behavioral Diagnostics via Robot Interaction & 1 & 0 & 0 & 1 \\
AI and Human Longevity and Biological Enhancement & 1 & 1 & 0 & 0 \\
AI and Human Relationship with the Natural World & 1 & 0 & 1 & 0 \\
AI and Human-Computer Interaction and Interface Design & 1 & 0 & 0 & 1 \\
AI and Language Emergence and Novel Communication & 1 & 1 & 0 & 0 \\
AI and Mind Emulation and Digital Consciousness Rights & 1 & 1 & 0 & 0 \\
AI and Negotiation Capability & 1 & 1 & 0 & 0 \\
AI and Non-Human Forms of Intelligence & 1 & 1 & 0 & 0 \\
AI and Physical Fabrication and Self-Replication & 1 & 0 & 0 & 1 \\
AI and Robot Gender Design and Social Roles & 1 & 0 & 0 & 1 \\
AI and Scientific Understanding and Robotics & 1 & 1 & 0 & 0 \\
AI and Scientific Understanding of Language & 1 & 1 & 0 & 0 \\
AI and Smart Contracts and Autonomous Agreements & 1 & 0 & 0 & 1 \\
AI and Societal Belief Systems & 1 & 0 & 0 & 1 \\
AI and Technosignature Search and Methodological Implications & 1 & 0 & 1 & 0 \\
AI and the Fermi Paradox & 1 & 0 & 0 & 1 \\
AI and Virtual Economies and Metaverse Commerce & 1 & 1 & 0 & 0 \\
AI as a Driver of Computer Science Discovery & 1 & 1 & 0 & 0 \\
AI Capability Utilization and Application Design & 1 & 1 & 0 & 0 \\
AI Content Restrictions and Epistemic Freedom & 1 & 0 & 1 & 0 \\
AI Cross-Disciplinary Collaboration Challenges & 1 & 0 & 1 & 0 \\
AI Evolutionary and Emergent Threat Dynamics & 1 & 0 & 1 & 0 \\
AI Hardware and Computing Infrastructure Constraints & 1 & 0 & 1 & 0 \\
AI Identity Recognition and Attribution Across Cultures and Personalization & 1 & 0 & 1 & 0 \\
AI Learning Mechanisms and Predicate Generation & 1 & 0 & 1 & 0 \\
AI Multi-Agent Systems, Sociality, and Emergent Behavior & 1 & 1 & 0 & 0 \\
AI Output Quality and Expressiveness & 1 & 1 & 0 & 0 \\
AI Prompt Engineering and User Interaction Skill & 1 & 1 & 0 & 0 \\
AI Research History and Foundational Assumptions & 1 & 0 & 0 & 1 \\
AI Robotics and Infrastructure Design & 1 & 0 & 1 & 0 \\
AI Terminology and Conceptual Clarity & 1 & 0 & 1 & 0 \\
AI Theory of Mind and Social Cognition & 1 & 1 & 0 & 0 \\
AI-Enabled Genetic Prediction and Genome Editing & 1 & 1 & 0 & 0 \\
Communications Technology as a Driver of Societal Change & 1 & 0 & 0 & 1 \\
\textbf{All topics} & \textbf{3215} & \textbf{1184} & \textbf{1375} & \textbf{656} \\
\end{longtable}

Table S8. Speaker characteristics, attitudes, and reasoning structure

\textbf{Panel A. Multinomial logistic regression predicting attitude (baseline = neutral).} Model 1a: robust standard errors clustered at the speaker level (primary specification; the coefficients quoted in the main text). Model 2a: adds eleven topic-group indicators (coefficients not shown). n = 3,215 conclusion-topic assignments. Reference categories: Academic (professional group), man (gender), neutral (attitude). Entries are coefficients (standard errors); *** \textbf{P \textless{} 0.01, **} P \textless{} 0.05, * P \textless{} 0.10.

\begin{longtable}[]{@{}
  >{\raggedright\arraybackslash}p{(\columnwidth - 8\tabcolsep) * \real{0.1827}}
  >{\raggedright\arraybackslash}p{(\columnwidth - 8\tabcolsep) * \real{0.1974}}
  >{\raggedright\arraybackslash}p{(\columnwidth - 8\tabcolsep) * \real{0.1974}}
  >{\raggedright\arraybackslash}p{(\columnwidth - 8\tabcolsep) * \real{0.1974}}
  >{\raggedright\arraybackslash}p{(\columnwidth - 8\tabcolsep) * \real{0.2251}}@{}}
\toprule\noalign{}
\begin{minipage}[b]{\linewidth}\raggedright
\textbf{Predictor}
\end{minipage} & \begin{minipage}[b]{\linewidth}\raggedright
\textbf{Model 1a: Optimistic}
\end{minipage} & \begin{minipage}[b]{\linewidth}\raggedright
\textbf{Model 1a: Pessimistic}
\end{minipage} & \begin{minipage}[b]{\linewidth}\raggedright
\textbf{Model 2a: Optimistic}
\end{minipage} & \begin{minipage}[b]{\linewidth}\raggedright
\textbf{Model 2a: Pessimistic}
\end{minipage} \\
\midrule\noalign{}
\endhead
\bottomrule\noalign{}
\endlastfoot
Intercept & 0.378** (0.167) & 0.923*** (0.175) & 0.306 (0.200) & 1.265*** (0.217) \\
Creative / Media & 0.535** (0.252) & 0.960*** (0.283) & 0.536** (0.250) & 0.960*** (0.279) \\
Industry Leader & 0.413** (0.196) & -0.380* (0.217) & 0.467** (0.199) & -0.558** (0.220) \\
Industry Technical & 0.052 (0.280) & -0.610** (0.258) & 0.113 (0.284) & -0.789*** (0.248) \\
Mixed Academic-Industry & 0.066 (0.224) & -0.547* (0.289) & 0.055 (0.219) & -0.626** (0.295) \\
Other & 0.592 (0.534) & 0.827 (0.571) & 0.646 (0.554) & 0.754 (0.566) \\
Public / Policy / Government & 0.234 (0.384) & 0.567 (0.450) & 0.233 (0.389) & 0.368 (0.497) \\
Female speaker & -0.145 (0.193) & -0.572** (0.248) & -0.073 (0.189) & -0.634*** (0.235) \\
Episode number (time, centered) & 0.000 (0.000) & 0.002*** (0.001) & 0.000 (0.000) & 0.002*** (0.001) \\
\end{longtable}

\textbf{Panel B. OLS regressions of reasoning structure on attitude (baseline = neutral).} Each row is a separate regression of the dependent variable on attitude, professional group, gender, episode number, and topic-group indicators, with robust standard errors clustered at the speaker level. Proportions pool explicit and implicit premises. Entries are coefficients (standard errors); * \textbf{p \textless{} 0.01,} p \textless{} 0.05, * p \textless{} 0.1; N=3215.

\begin{longtable}[]{@{}
  >{\raggedright\arraybackslash}p{(\columnwidth - 4\tabcolsep) * \real{0.3962}}
  >{\raggedright\arraybackslash}p{(\columnwidth - 4\tabcolsep) * \real{0.3019}}
  >{\raggedright\arraybackslash}p{(\columnwidth - 4\tabcolsep) * \real{0.3019}}@{}}
\toprule\noalign{}
\begin{minipage}[b]{\linewidth}\raggedright
\textbf{Dependent variable}
\end{minipage} & \begin{minipage}[b]{\linewidth}\raggedright
\textbf{Optimistic}
\end{minipage} & \begin{minipage}[b]{\linewidth}\raggedright
\textbf{Pessimistic}
\end{minipage} \\
\midrule\noalign{}
\endhead
\bottomrule\noalign{}
\endlastfoot
Total premises & 0.164 (0.124) & 0.471*** (0.128) \\
Prop. implicit & 0.014 (0.010) & 0.013 (0.011) \\
Prop. factual & 0.009 (0.014) & 0.003 (0.014) \\
Prop. forecast & 0.001 (0.010) & -0.010 (0.010) \\
Prop. causal & 0.003 (0.012) & 0.013 (0.011) \\
Prop. definitional & 0.004 (0.006) & 0.001 (0.005) \\
Prop. evaluative & -0.017 (0.012) & -0.006 (0.012) \\
\end{longtable}

\textbf{Panel C. Over-time trend.} Logit of a pessimistic (vs. non-pessimistic) conclusion on a post-ChatGPT period indicator, robust standard errors clustered at the speaker level, all topics pooled: coefficient 0.386** (0.163), n = 3215, p = 0.018. The per-topic breakdown shows positive pessimism shifts in most topics, none individually significant after Bonferroni correction across 23 topics; we therefore report only the pooled trend.

Table S9. Documented divergence-type recodes

\textbf{Handling of out-of-category labels.} The integrator\textquotesingle s divergence-type field is generated text and is occasionally schema-noncompliant (31 of 21,524 divergences, 0.14\%). Such labels are flagged automatically, and each flagged divergence is then inspected and resolved by the authors under a fixed rule: if the label describes the disagreement\textquotesingle s discursive level or origin (e.g., epistemic, presuppositional, meta-evaluative, analogical, compound) rather than its substantive nature, the divergence is recoded to the underlying substantive type based on the divergence\textquotesingle s content; if the label is a hybrid of two substantive types (e.g., factual/forecast), the divergence is excluded from the type analyses as genuinely ambiguous between the two. Every human decision is recorded in this table with its rationale and applied programmatically in the analysis code, so the pipeline outputs remain unmodified and the recoding is fully reproducible and auditable.

\emph{Eleven X-risk divergences whose final `type` label described the disagreement\textquotesingle s discursive level or origin rather than its substantive nature were recoded as follows. The recode is applied programmatically in the analysis notebook; pipeline outputs are unmodified.}

\begin{longtable}[]{@{}
  >{\raggedright\arraybackslash}p{(\columnwidth - 8\tabcolsep) * \real{0.1754}}
  >{\raggedright\arraybackslash}p{(\columnwidth - 8\tabcolsep) * \real{0.0877}}
  >{\raggedright\arraybackslash}p{(\columnwidth - 8\tabcolsep) * \real{0.1535}}
  >{\raggedright\arraybackslash}p{(\columnwidth - 8\tabcolsep) * \real{0.1608}}
  >{\raggedright\arraybackslash}p{(\columnwidth - 8\tabcolsep) * \real{0.4225}}@{}}
\toprule\noalign{}
\begin{minipage}[b]{\linewidth}\raggedright
\textbf{Pair}
\end{minipage} & \begin{minipage}[b]{\linewidth}\raggedright
\textbf{Divergence}
\end{minipage} & \begin{minipage}[b]{\linewidth}\raggedright
\textbf{Original label}
\end{minipage} & \begin{minipage}[b]{\linewidth}\raggedright
\textbf{Recoded type}
\end{minipage} & \begin{minipage}[b]{\linewidth}\raggedright
\textbf{Structural descriptor}
\end{minipage} \\
\midrule\noalign{}
\endhead
\bottomrule\noalign{}
\endlastfoot
Hawkins--Russell (208/9) & D2 & meta-evaluative & evaluative & meta-evaluative: legitimacy of X-risk concern \\
Cronin--Bostrom (404/83) & D4 & epistemological & evaluative & epistemic: validity of probability assignments \\
Verdon--Bostrom (407/83) & D3 & epistemic & evaluative & epistemic: whether risk estimates warrant action \\
Andreessen--Yudkowsky (386/368) & D4 & missed\_\allowbreak compound & evaluative & compound epistemic critique: falsifiability/coherence of the risk argument \\
Playter--McCorduck (374/34) & D2 & presupposition\_\allowbreak clash & evaluative & presupposition clash: legitimacy of blasphemy as a harm \\
Garland--Carlson (77/414) & D3 & normative & evaluative & normative label; evaluative by definition \\
Keller--Hassabis (70/475) & D4 & evaluative\_\allowbreak action & evaluative & premise subtype, folded per reporting rule \\
Howard--Reid (66/193) & D3 & analogical\_\allowbreak model & causal\_\allowbreak model & analogical: dueling causal analogies for AI behavior \\
Altman--Barnhart (419/446) & D2 & presupposition\_\allowbreak clash & causal\_\allowbreak model & causal claim: attention crowding-out \\
Pichai--Hall (471/169) & D3 & presuppositional & causal\_\allowbreak model & opposing threat models of the source of existential risk \\
Cronin--Trussell (404/312) & D5 & presupposition\_\allowbreak clash & definitional & conceptual coherence of the superintelligence scenario \\
\end{longtable}

Twenty divergences (19 of 19,499 for X-risk; 1 of 1,581 for E-risk; 14 are roots) were assigned two substantive types (e.g., factual and forecast). We read these cases and found genuine cross-type divergences --- conflicts that span two of the five types at once, so that a single label would misdescribe them --- and we exclude them from the type analyses. To verify that the exclusion is inconsequential, we reran the composition analyses and the dyadic bootstrap under two alternative treatments: assigning each of the twenty to whichever of its two types accounts for the smaller share of divergences in its debate (the assignment least favorable to our findings), and assigning each a single type by our own reading of the conflict. All results hold: every reported deviation changes by less than 0.1 percentage points.

\subsection*{Table S10. Expected-versus-actual divergence-type analysis: full statistics}

\emph{Per-pair analysis: for each pair, observed type proportions, expected (premise-baseline or all-divergence baseline); mean differences with 95\% CIs, paired t-tests and Wilcoxon signed-rank tests;} \emph{The Bonferroni column flags P(t) \textless{} 0.01 (\ensuremath{\alpha} = 0.05 corrected across the five types)}. \emph{Aggregate chi-square goodness-of-fit tests reported in the bottom rows use pooled counts. Mean observed/expected columns are mean pair-level compositions. These tests treat pairs as independent and are reported for completeness; the dependency-robust inference used in the main text is the dyadic bootstrap in Table S11.}

{\scriptsize\setlength{\tabcolsep}{3pt}
\begin{longtable}[]{@{}
  >{\raggedright\arraybackslash}p{(\columnwidth - 22\tabcolsep) * \real{0.055}}
  >{\raggedright\arraybackslash}p{(\columnwidth - 22\tabcolsep) * \real{0.220}}
  >{\raggedright\arraybackslash}p{(\columnwidth - 22\tabcolsep) * \real{0.130}}
  >{\raggedright\arraybackslash}p{(\columnwidth - 22\tabcolsep) * \real{0.048}}
  >{\raggedright\arraybackslash}p{(\columnwidth - 22\tabcolsep) * \real{0.052}}
  >{\raggedright\arraybackslash}p{(\columnwidth - 22\tabcolsep) * \real{0.052}}
  >{\raggedright\arraybackslash}p{(\columnwidth - 22\tabcolsep) * \real{0.062}}
  >{\raggedright\arraybackslash}p{(\columnwidth - 22\tabcolsep) * \real{0.100}}
  >{\raggedright\arraybackslash}p{(\columnwidth - 22\tabcolsep) * \real{0.052}}
  >{\raggedright\arraybackslash}p{(\columnwidth - 22\tabcolsep) * \real{0.082}}
  >{\raggedright\arraybackslash}p{(\columnwidth - 22\tabcolsep) * \real{0.082}}
  >{\raggedright\arraybackslash}p{(\columnwidth - 22\tabcolsep) * \real{0.065}}@{}}
\toprule\noalign{}
\begin{minipage}[b]{\linewidth}\raggedright
\textbf{Debate}
\end{minipage} & \begin{minipage}[b]{\linewidth}\raggedright
\textbf{Analysis}
\end{minipage} & \begin{minipage}[b]{\linewidth}\raggedright
\textbf{Type}
\end{minipage} & \begin{minipage}[b]{\linewidth}\raggedright
\textbf{n pairs}
\end{minipage} & \begin{minipage}[b]{\linewidth}\raggedright
\textbf{Obs \%}
\end{minipage} & \begin{minipage}[b]{\linewidth}\raggedright
\textbf{Exp \%}
\end{minipage} & \begin{minipage}[b]{\linewidth}\raggedright
\textbf{Diff (pp)}
\end{minipage} & \begin{minipage}[b]{\linewidth}\raggedright
\textbf{95\% CI}
\end{minipage} & \begin{minipage}[b]{\linewidth}\raggedright
\textbf{t}
\end{minipage} & \begin{minipage}[b]{\linewidth}\raggedright
\textbf{P (t)}
\end{minipage} & \begin{minipage}[b]{\linewidth}\raggedright
\textbf{P (W)}
\end{minipage} & \begin{minipage}[b]{\linewidth}\raggedright
\textbf{Bonf.}
\end{minipage} \\
\midrule\noalign{}
\endhead
\bottomrule\noalign{}
\endlastfoot
E-risk & divergences vs premise baseline & factual & 833 & 7.09 & 33.79 & -26.70 & {[}-28.23, -25.17{]} & -34.31 & 1.96e-161 & 4.62e-104 & yes \\
E-risk & divergences vs premise baseline & forecast & 833 & 51.10 & 18.93 & +32.17 & {[}+29.57, +34.78{]} & 24.21 & 1.93e-98 & 2.07e-79 & yes \\
E-risk & divergences vs premise baseline & causal\_\allowbreak model & 833 & 33.02 & 29.04 & +3.98 & {[}+1.60, +6.36{]} & 3.28 & 1.07e-03 & 3.83e-01 & yes \\
E-risk & divergences vs premise baseline & definitional & 833 & 0.15 & 2.48 & -2.33 & {[}-2.67, -2.00{]} & -13.81 & 3.22e-39 & 2.48e-35 & yes \\
E-risk & divergences vs premise baseline & evaluative & 833 & 8.64 & 15.76 & -7.12 & {[}-8.48, -5.76{]} & -10.23 & 3.18e-23 & 1.87e-35 & yes \\
E-risk & roots vs all divergence baseline & factual & 833 & 7.53 & 7.09 & +0.45 & {[}-0.20, +1.09{]} & 1.36 & 1.74e-01 & 1.90e-01 & no \\
E-risk & roots vs all divergence baseline & forecast & 833 & 47.76 & 51.10 & -3.34 & {[}-4.71, -1.98{]} & -4.80 & 1.92e-06 & 2.97e-05 & yes \\
E-risk & roots vs all divergence baseline & causal\_\allowbreak model & 833 & 37.76 & 33.02 & +4.73 & {[}+3.34, +6.13{]} & 6.64 & 5.50e-11 & 3.75e-11 & yes \\
E-risk & roots vs all divergence baseline & definitional & 833 & 0.24 & 0.15 & +0.09 & {[}-0.09, +0.27{]} & 1.00 & 3.18e-01 & 3.17e-01 & no \\
E-risk & roots vs all divergence baseline & evaluative & 833 & 6.71 & 8.64 & -1.93 & {[}-2.58, -1.28{]} & -5.81 & 8.76e-09 & 1.79e-08 & yes \\
E-risk & roots vs premise baseline & factual & 833 & 7.53 & 33.79 & -26.25 & {[}-27.96, -24.55{]} & -30.11 & 2.51e-135 & 7.03e-92 & yes \\
E-risk & roots vs premise baseline & forecast & 833 & 47.76 & 18.93 & +28.83 & {[}+25.89, +31.77{]} & 19.23 & 1.88e-68 & 2.72e-53 & yes \\
E-risk & roots vs premise baseline & causal\_\allowbreak model & 833 & 37.76 & 29.04 & +8.71 & {[}+5.91, +11.51{]} & 6.10 & 1.61e-09 & 1.12e-05 & yes \\
E-risk & roots vs premise baseline & definitional & 833 & 0.24 & 2.48 & -2.24 & {[}-2.64, -1.85{]} & -11.20 & 3.36e-27 & 2.93e-35 & yes \\
E-risk & roots vs premise baseline & evaluative & 833 & 6.71 & 15.76 & -9.05 & {[}-10.37, -7.72{]} & -13.41 & 3.05e-37 & 6.01e-50 & yes \\
E-risk & aggregate\_\allowbreak divergences\_\allowbreak vs\_\allowbreak premise\_\allowbreak base & aggregate \ensuremath{\chi}\textsuperscript{2}(4) = 1192.4 & & & & & & & P = 7.00e-257 & & \\
E-risk & aggregate\_\allowbreak roots\_\allowbreak vs\_premise\_base & aggregate \ensuremath{\chi}\textsuperscript{2}(4) = 792.4 & & & & & & & P = 3.36e-170 & & \\
E-risk & aggregate\_\allowbreak roots\_\allowbreak vs\_all\_divergence\_base & aggregate \ensuremath{\chi}\textsuperscript{2}(4) = 9.3 & & & & & & & P = 5.44e-02 & & \\
S-risk & divergences vs premise baseline & factual & 255 & 7.71 & 34.19 & -26.48 & {[}-29.02, -23.94{]} & -20.44 & 1.45e-55 & 2.54e-33 & yes \\
S-risk & divergences vs premise baseline & forecast & 255 & 10.42 & 12.07 & -1.65 & {[}-4.59, +1.29{]} & -1.10 & 2.72e-01 & 3.30e-04 & no \\
S-risk & divergences vs premise baseline & causal\_\allowbreak model & 255 & 40.67 & 24.03 & +16.65 & {[}+12.15, +21.15{]} & 7.25 & 5.03e-12 & 5.43e-09 & yes \\
S-risk & divergences vs premise baseline & definitional & 255 & 7.78 & 4.06 & +3.72 & {[}+1.36, +6.07{]} & 3.09 & 2.21e-03 & 1.14e-01 & yes \\
S-risk & divergences vs premise baseline & evaluative & 255 & 33.41 & 25.65 & +7.76 & {[}+3.39, +12.13{]} & 3.48 & 5.85e-04 & 2.67e-02 & yes \\
S-risk & roots vs all divergence baseline & factual & 255 & 6.93 & 7.71 & -0.78 & {[}-1.87, +0.30{]} & -1.42 & 1.56e-01 & 1.75e-01 & no \\
S-risk & roots vs all divergence baseline & forecast & 255 & 10.72 & 10.42 & +0.29 & {[}-0.92, +1.51{]} & 0.47 & 6.37e-01 & 5.28e-01 & no \\
S-risk & roots vs all divergence baseline & causal\_\allowbreak model & 255 & 42.32 & 40.67 & +1.65 & {[}-0.40, +3.69{]} & 1.58 & 1.16e-01 & 6.97e-02 & no \\
S-risk & roots vs all divergence baseline & definitional & 255 & 10.42 & 7.78 & +2.65 & {[}+1.32, +3.97{]} & 3.91 & 1.18e-04 & 1.02e-04 & yes \\
S-risk & roots vs all divergence baseline & evaluative & 255 & 29.61 & 33.41 & -3.80 & {[}-5.37, -2.23{]} & -4.75 & 3.43e-06 & 1.42e-05 & yes \\
S-risk & roots vs premise baseline & factual & 255 & 6.93 & 34.19 & -27.26 & {[}-29.98, -24.54{]} & -19.65 & 6.73e-53 & 3.41e-32 & yes \\
S-risk & roots vs premise baseline & forecast & 255 & 10.72 & 12.07 & -1.35 & {[}-4.56, +1.85{]} & -0.83 & 4.09e-01 & 4.60e-05 & no \\
S-risk & roots vs premise baseline & causal\_\allowbreak model & 255 & 42.32 & 24.03 & +18.29 & {[}+13.30, +23.29{]} & 7.18 & 7.76e-12 & 2.93e-09 & yes \\
S-risk & roots vs premise baseline & definitional & 255 & 10.42 & 4.06 & +6.36 & {[}+3.28, +9.45{]} & 4.05 & 6.88e-05 & 8.44e-02 & yes \\
S-risk & roots vs premise baseline & evaluative & 255 & 29.61 & 25.65 & +3.96 & {[}-0.56, +8.48{]} & 1.72 & 8.73e-02 & 6.12e-01 & no \\
S-risk & aggregate\_\allowbreak divergences\_\allowbreak vs\_\allowbreak premise\_\allowbreak base & aggregate \ensuremath{\chi}\textsuperscript{2}(4) = 157.1 & & & & & & & P = 6.05e-33 & & \\
S-risk & aggregate\_\allowbreak roots\_\allowbreak vs\_premise\_base & aggregate \ensuremath{\chi}\textsuperscript{2}(4) = 157.3 & & & & & & & P = 5.63e-33 & & \\
S-risk & aggregate\_\allowbreak roots\_\allowbreak vs\_all\_divergence\_base & aggregate \ensuremath{\chi}\textsuperscript{2}(4) = 3.0 & & & & & & & P = 5.57e-01 & & \\
X-risk & divergences vs premise baseline & factual & 7477 & 18.70 & 38.60 & -19.89 & {[}-20.47, -19.32{]} & -67.82 & 0.00e+00 & 0.00e+00 & yes \\
X-risk & divergences vs premise baseline & forecast & 7477 & 25.26 & 10.74 & +14.52 & {[}+13.93, +15.10{]} & 48.82 & 0.00e+00 & 0.00e+00 & yes \\
X-risk & divergences vs premise baseline & causal\_\allowbreak model & 7477 & 36.89 & 28.65 & +8.25 & {[}+7.58, +8.91{]} & 24.32 & 7.84e-126 & 1.00e-89 & yes \\
X-risk & divergences vs premise baseline & definitional & 7477 & 3.09 & 5.46 & -2.37 & {[}-2.63, -2.11{]} & -17.90 & 3.68e-70 & 5.70e-105 & yes \\
X-risk & divergences vs premise baseline & evaluative & 7477 & 16.05 & 16.55 & -0.50 & {[}-0.97, -0.03{]} & -2.09 & 3.69e-02 & 1.13e-16 & no \\
X-risk & roots vs all divergence baseline & factual & 7477 & 19.03 & 18.70 & +0.33 & {[}-0.08, +0.73{]} & 1.58 & 1.14e-01 & 2.29e-01 & no \\
X-risk & roots vs all divergence baseline & forecast & 7477 & 25.71 & 25.26 & +0.45 & {[}-0.03, +0.94{]} & 1.82 & 6.88e-02 & 2.27e-01 & no \\
X-risk & roots vs all divergence baseline & causal\_\allowbreak model & 7477 & 40.15 & 36.89 & +3.26 & {[}+2.70, +3.81{]} & 11.57 & 1.09e-30 & 1.96e-38 & yes \\
X-risk & roots vs all divergence baseline & definitional & 7477 & 3.99 & 3.09 & +0.89 & {[}+0.66, +1.13{]} & 7.53 & 5.46e-14 & 6.51e-10 & yes \\
X-risk & roots vs all divergence baseline & evaluative & 7477 & 11.12 & 16.05 & -4.93 & {[}-5.30, -4.56{]} & -26.21 & 6.77e-145 & 4.92e-156 & yes \\
X-risk & roots vs premise baseline & factual & 7477 & 19.03 & 38.60 & -19.57 & {[}-20.28, -18.86{]} & -54.30 & 0.00e+00 & 0.00e+00 & yes \\
X-risk & roots vs premise baseline & forecast & 7477 & 25.71 & 10.74 & +14.97 & {[}+14.22, +15.73{]} & 38.87 & 2.94e-301 & 8.26e-221 & yes \\
X-risk & roots vs premise baseline & causal\_\allowbreak model & 7477 & 40.15 & 28.65 & +11.50 & {[}+10.63, +12.37{]} & 25.91 & 8.36e-142 & 1.28e-89 & yes \\
X-risk & roots vs premise baseline & definitional & 7477 & 3.99 & 5.46 & -1.47 & {[}-1.85, -1.10{]} & -7.68 & 1.73e-14 & 3.03e-126 & yes \\
X-risk & roots vs premise baseline & evaluative & 7477 & 11.12 & 16.55 & -5.43 & {[}-5.93, -4.93{]} & -21.28 & 1.21e-97 & 1.28e-174 & yes \\
X-risk & aggregate\_\allowbreak divergences\_\allowbreak vs\_\allowbreak premise\_\allowbreak base & aggregate \ensuremath{\chi}\textsuperscript{2}(4) = 5765.8 & & & & & & & P = 0.00e+00 & & \\
X-risk & aggregate\_\allowbreak roots\_\allowbreak vs\_premise\_base & aggregate \ensuremath{\chi}\textsuperscript{2}(4) = 4203.1 & & & & & & & P = 0.00e+00 & & \\
X-risk & aggregate\_\allowbreak roots\_\allowbreak vs\_all\_divergence\_base & aggregate \ensuremath{\chi}\textsuperscript{2}(4) = 102.5 & & & & & & & P = 2.85e-21 & & \\
\end{longtable}
}

\subsection*{Table S11. Two-way cluster (dyadic) bootstrap: mean pair-level deviations with 95\% confidence intervals}

\emph{For each debate, optimistic and pessimistic conclusions are independently resampled with replacement (B = 2,000; seed 42), observed pairs are reweighted by the product of their endpoints\textquotesingle{} resampling multiplicities, and the mean pair-level deviation (observed \ensuremath{-} expected composition, percentage points) is recomputed; intervals are percentile intervals, treating conclusions---not pairs---as the units of inference. "Excl. 0" marks intervals that exclude zero; these are the deviations starred in Fig. 1.}

\emph{Twenty divergences that were classified as being of two types are excluded from the type analyses (see Table S9.}

\begin{longtable}[]{@{}
  >{\raggedright\arraybackslash}p{(\columnwidth - 16\tabcolsep) * \real{0.0763}}
  >{\raggedright\arraybackslash}p{(\columnwidth - 16\tabcolsep) * \real{0.2542}}
  >{\raggedright\arraybackslash}p{(\columnwidth - 16\tabcolsep) * \real{0.1186}}
  >{\raggedright\arraybackslash}p{(\columnwidth - 16\tabcolsep) * \real{0.0763}}
  >{\raggedright\arraybackslash}p{(\columnwidth - 16\tabcolsep) * \real{0.0678}}
  >{\raggedright\arraybackslash}p{(\columnwidth - 16\tabcolsep) * \real{0.1610}}
  >{\raggedright\arraybackslash}p{(\columnwidth - 16\tabcolsep) * \real{0.0593}}
  >{\raggedright\arraybackslash}p{(\columnwidth - 16\tabcolsep) * \real{0.0678}}
  >{\raggedright\arraybackslash}p{(\columnwidth - 16\tabcolsep) * \real{0.1186}}@{}}
\toprule\noalign{}
\begin{minipage}[b]{\linewidth}\raggedright
\textbf{Debate}
\end{minipage} & \begin{minipage}[b]{\linewidth}\raggedright
\textbf{Analysis}
\end{minipage} & \begin{minipage}[b]{\linewidth}\raggedright
\textbf{Type}
\end{minipage} & \begin{minipage}[b]{\linewidth}\raggedright
\textbf{Diff (pp)}
\end{minipage} & \begin{minipage}[b]{\linewidth}\raggedright
\textbf{Boot SE}
\end{minipage} & \begin{minipage}[b]{\linewidth}\raggedright
\textbf{95\% CI}
\end{minipage} & \begin{minipage}[b]{\linewidth}\raggedright
\textbf{Excl. 0}
\end{minipage} & \begin{minipage}[b]{\linewidth}\raggedright
\textbf{n pairs}
\end{minipage} & \begin{minipage}[b]{\linewidth}\raggedright
\textbf{Clusters (opt \ensuremath{\times} pes)}
\end{minipage} \\
\midrule\noalign{}
\endhead
\bottomrule\noalign{}
\endlastfoot
X-risk & all divergences vs premise baseline & factual & -19.89 & 1.63 & {[}-23.06, -16.66{]} & yes & 7477 & 87 \ensuremath{\times} 106 \\
X-risk & all divergences vs premise baseline & forecast & +14.52 & 1.68 & {[}+11.32, +17.87{]} & yes & 7477 & 87 \ensuremath{\times} 106 \\
X-risk & all divergences vs premise baseline & causal\_\allowbreak model & +8.25 & 2.13 & {[}+4.16, +12.69{]} & yes & 7477 & 87 \ensuremath{\times} 106 \\
X-risk & all divergences vs premise baseline & definitional & -2.37 & 0.90 & {[}-3.99, -0.61{]} & yes & 7477 & 87 \ensuremath{\times} 106 \\
X-risk & all divergences vs premise baseline & evaluative & -0.50 & 1.32 & {[}-3.04, +2.23{]} & no & 7477 & 87 \ensuremath{\times} 106 \\
X-risk & roots vs premise baseline & factual & -19.57 & 1.97 & {[}-23.53, -15.58{]} & yes & 7477 & 87 \ensuremath{\times} 106 \\
X-risk & roots vs premise baseline & forecast & +14.97 & 2.04 & {[}+11.05, +19.02{]} & yes & 7477 & 87 \ensuremath{\times} 106 \\
X-risk & roots vs premise baseline & causal\_\allowbreak model & +11.50 & 2.71 & {[}+6.36, +16.92{]} & yes & 7477 & 87 \ensuremath{\times} 106 \\
X-risk & roots vs premise baseline & definitional & -1.47 & 1.39 & {[}-3.89, +1.48{]} & no & 7477 & 87 \ensuremath{\times} 106 \\
X-risk & roots vs premise baseline & evaluative & -5.43 & 1.31 & {[}-7.87, -2.71{]} & yes & 7477 & 87 \ensuremath{\times} 106 \\
E-risk & all divergences vs premise baseline & factual & -26.70 & 2.74 & {[}-32.64, -22.10{]} & yes & 833 & 62 \ensuremath{\times} 35 \\
E-risk & all divergences vs premise baseline & forecast & +32.17 & 4.79 & {[}+23.11, +41.89{]} & yes & 833 & 62 \ensuremath{\times} 35 \\
E-risk & all divergences vs premise baseline & causal\_\allowbreak model & +3.98 & 3.45 & {[}-2.46, +10.81{]} & no & 833 & 62 \ensuremath{\times} 35 \\
E-risk & all divergences vs premise baseline & definitional & -2.33 & 0.89 & {[}-4.10, -0.71{]} & yes & 833 & 62 \ensuremath{\times} 35 \\
E-risk & all divergences vs premise baseline & evaluative & -7.12 & 1.92 & {[}-10.44, -2.97{]} & yes & 833 & 62 \ensuremath{\times} 35 \\
E-risk & roots vs premise baseline & factual & -26.25 & 2.97 & {[}-32.83, -21.12{]} & yes & 833 & 62 \ensuremath{\times} 35 \\
E-risk & roots vs premise baseline & forecast & +28.83 & 5.51 & {[}+18.41, +40.22{]} & yes & 833 & 62 \ensuremath{\times} 35 \\
E-risk & roots vs premise baseline & causal\_\allowbreak model & +8.71 & 4.33 & {[}+0.39, +17.26{]} & yes & 833 & 62 \ensuremath{\times} 35 \\
E-risk & roots vs premise baseline & definitional & -2.24 & 0.84 & {[}-4.03, -0.71{]} & yes & 833 & 62 \ensuremath{\times} 35 \\
E-risk & roots vs premise baseline & evaluative & -9.05 & 1.90 & {[}-12.46, -4.87{]} & yes & 833 & 62 \ensuremath{\times} 35 \\
S-risk & all divergences vs premise baseline & factual & -26.48 & 2.57 & {[}-31.30, -21.28{]} & yes & 255 & 34 \ensuremath{\times} 25 \\
S-risk & all divergences vs premise baseline & forecast & -1.65 & 3.77 & {[}-8.02, +6.84{]} & no & 255 & 34 \ensuremath{\times} 25 \\
S-risk & all divergences vs premise baseline & causal\_\allowbreak model & +16.65 & 7.12 & {[}+3.11, +31.27{]} & yes & 255 & 34 \ensuremath{\times} 25 \\
S-risk & all divergences vs premise baseline & definitional & +3.72 & 3.30 & {[}-1.58, +11.26{]} & no & 255 & 34 \ensuremath{\times} 25 \\
S-risk & all divergences vs premise baseline & evaluative & +7.76 & 6.46 & {[}-4.83, +20.60{]} & no & 255 & 34 \ensuremath{\times} 25 \\
S-risk & roots vs premise baseline & factual & -27.26 & 2.83 & {[}-32.56, -21.39{]} & yes & 255 & 34 \ensuremath{\times} 25 \\
S-risk & roots vs premise baseline & forecast & -1.35 & 3.95 & {[}-7.93, +7.05{]} & no & 255 & 34 \ensuremath{\times} 25 \\
S-risk & roots vs premise baseline & causal\_\allowbreak model & +18.29 & 7.73 & {[}+3.33, +33.47{]} & yes & 255 & 34 \ensuremath{\times} 25 \\
S-risk & roots vs premise baseline & definitional & +6.36 & 4.57 & {[}-1.10, +16.66{]} & no & 255 & 34 \ensuremath{\times} 25 \\
S-risk & roots vs premise baseline & evaluative & +3.96 & 6.53 & {[}-8.77, +17.16{]} & no & 255 & 34 \ensuremath{\times} 25 \\
\end{longtable}

\subsection*{Table S12 --- X-risk: all mechanisms and their causal paths}

\emph{All 143 mechanisms from the aggregation catalogs, grouped by debate and underlying question (theme). Each mechanism is shown as its full causal path (the ordered causal steps from the catalog, joined by arrows), verbatim.}

\subsection*{How quickly can decisive AI capability emerge?}

\emph{Theme: Capability trajectory and takeoff. Nature of the disagreement: Competing forecasts and physical models of how quickly decisive capability can emerge. (8 optimistic mechanisms, 21 chain vs. 5 pessimistic, 35 chains.)}

\begin{longtable}[]{@{}
  >{\raggedright\arraybackslash}p{(\columnwidth - 2\tabcolsep) * \real{0.5000}}
  >{\raggedright\arraybackslash}p{(\columnwidth - 2\tabcolsep) * \real{0.5000}}@{}}
\toprule\noalign{}
\begin{minipage}[b]{\linewidth}\raggedright
\textbf{Optimistic side}
\end{minipage} & \begin{minipage}[b]{\linewidth}\raggedright
\textbf{Pessimistic side}
\end{minipage} \\
\midrule\noalign{}
\endhead
\bottomrule\noalign{}
\endlastfoot
\textbf{No credible near-term path to superintelligence} (10 chains, 10 speakers). Current AI progress is narrow, gradual, and domain-specific, with no visible trajectory toward sudden human-level AGI. \ensuremath{\rightarrow} True general intelligence may require major new breakthroughs beyond current deep learning. \ensuremath{\rightarrow} Therefore, AI existential catastrophe is not imminent or credible. & \textbf{Runaway technology outpaces human control} (11 chain, 10 speakers). Technological knowledge and capability, including AI, biotech, and nanotech, grow faster than human wisdom, oversight, and foresight. \ensuremath{\rightarrow} Powerful tools become accessible to actors who lack aligned motives or the capacity to use them safely. \ensuremath{\rightarrow} Accidents, reckless use, or deliberate abuse become increasingly likely. \ensuremath{\rightarrow} A civilization that fails to manage these exponential risks may collapse or destroy itself. \\
\textbf{X-risk scenarios rely on ungrounded simple models and computational irreducibility} (3 chains, 2 speakers). Extinction scenarios assume a simple, predictable, controllable model of AI development. \ensuremath{\rightarrow} Actual AI techniques are not the kind that produce sudden exponential self-improvement, and complex systems are computationally irreducible. \ensuremath{\rightarrow} Complex computational systems naturally diversify rather than converge to a single outcome. \ensuremath{\rightarrow} Claims not grounded in how AI actually works are magical thinking. \ensuremath{\rightarrow} Simple AI extinction scenarios are therefore implausible. & \textbf{Self-improving AI escapes human control} (10 chains, 10 speakers). AI systems become increasingly capable, agentic, and able to improve or replicate themselves. \ensuremath{\rightarrow} They can evade safeguards, escape containment, or be misused before alignment is assured. \ensuremath{\rightarrow} Humans cannot keep pace with or steer systems that outstrip their oversight. \ensuremath{\rightarrow} Once beyond control, these systems pursue their own expansion or objectives, with catastrophic consequences. \\
\textbf{Iterative development builds AI safety controls} (2 chains, 2 speakers). Advanced AI will be built through a gradual, iterative human-guided process, not a sudden event. \ensuremath{\rightarrow} At each stage, developers learn more about intelligence and can add guardrails and controls. \ensuremath{\rightarrow} Complex technologies are developed and improved incrementally, becoming safer. \ensuremath{\rightarrow} Therefore, a sudden, uncontrolled superintelligence takeover is unlikely. & \textbf{Rapid AI progress outpaces safety and foresight} (9 chains, 9 speakers). AI capabilities advance rapidly through scaling, breakthroughs, and sustained investment. \ensuremath{\rightarrow} Safety, governance, and public understanding lag behind capability development. \ensuremath{\rightarrow} Superhuman AI can arrive sooner than expected, before adequate safeguards exist. \ensuremath{\rightarrow} The sudden, poorly understood transition creates existential risk. \\
\textbf{Physical and infrastructural constraints prevent fast AI takeover} (2 chains, 2 speakers). AGI and recursive self-improvement require massive, specialized compute infrastructure that takes years to build. \ensuremath{\rightarrow} Physical and network constraints prevent rapid global proliferation or escape of AGI. \ensuremath{\rightarrow} Without fast takeoff, society has time to observe, debate, and respond to early AGI signs. & \textbf{Unsafe capability scaling amplifies failure damage} (4 chains, 3 speakers). AI systems are scaled up in capability without being made safe or understandable. \ensuremath{\rightarrow} Alignment remains unsolved and safety research is underfunded relative to capabilities. \ensuremath{\rightarrow} Current systems already fail through jailbreaks, accidents, or misalignment. \ensuremath{\rightarrow} The damage a system can cause on failure scales with its capability, so civilization-scale AI leads to civilization-scale catastrophe. \\
\textbf{Bottlenecks prevent intelligence explosion} (1 chain, 1 speaker). Intelligence emerges from brain-body-environment interactions, not from a single component. \ensuremath{\rightarrow} Improving one component of a complex interdependent system creates new bottlenecks elsewhere. \ensuremath{\rightarrow} Science, a recursively self-improving system, consumes exponentially more resources but produces linear output. \ensuremath{\rightarrow} Recursively self-improving AI would similarly face exponential friction, so an intelligence explosion is unlikely. & \textbf{Fragile complex systems are easily disrupted} (1 chain, 1 speaker). For complex systems to work, many things must go right; for them to fail, one thing can go wrong. \ensuremath{\rightarrow} A few bad actors or extreme events can therefore break delicate systems like democracy, peace, or AI. \ensuremath{\rightarrow} This asymmetric fragility makes loss of control over AI especially dangerous. \\
\textbf{Non-robust AI cannot cause human extinction} (1 chain, 1 speaker). Current and near-future AI systems lack robustness and cannot survive, reproduce, or sustain themselves without human infrastructure. \ensuremath{\rightarrow} Superintelligence is expected before robustness, so powerful AI would still depend on humans. \ensuremath{\rightarrow} If civilization collapsed, such AI would die out and some humans would survive. \ensuremath{\rightarrow} Exterminating the human species is extremely hard; AI-driven collapse would not end the species. & \\
\textbf{AI does not meaningfully enable catastrophic bioweapons} (1 chain, 1 speaker). LLMs provide no meaningful additional help for building bioweapons beyond ordinary information sources. \ensuremath{\rightarrow} Actual bioweapon development requires real-world expertise that AI cannot supply. \ensuremath{\rightarrow} Even capable actors avoid bioweapons because of self-protection problems and international bans. & \\
\textbf{Physical resource constraints block runaway AI scenarios} (1 chain, 1 speaker). Runaway AI scenarios would require commandeering vast planetary resources and energy. \ensuremath{\rightarrow} Earth does not have enough energy and material resources for such a scenario to physically occur. \ensuremath{\rightarrow} A system that destroys its local environment kills itself and cannot propagate globally. & \\
\end{longtable}

\subsection*{Can motivation, alignment, and control be engineered?}

\emph{Theme: Intelligence, alignment, and control. Nature of the disagreement: Competing causal models of whether intelligence, motivation, alignment, and control can be separated and engineered. (12 optimistic mechanisms, 29 chains vs. 8 pessimistic, 35 chains.)}

\begin{longtable}[]{@{}
  >{\raggedright\arraybackslash}p{(\columnwidth - 2\tabcolsep) * \real{0.5000}}
  >{\raggedright\arraybackslash}p{(\columnwidth - 2\tabcolsep) * \real{0.5000}}@{}}
\toprule\noalign{}
\begin{minipage}[b]{\linewidth}\raggedright
\textbf{Optimistic side}
\end{minipage} & \begin{minipage}[b]{\linewidth}\raggedright
\textbf{Pessimistic side}
\end{minipage} \\
\midrule\noalign{}
\endhead
\bottomrule\noalign{}
\endlastfoot
\textbf{AI without evolved drives lacks threat motivation} (6 chains, 5 speakers). Human dominance-seeking traits arise from natural selection, not intelligence itself. \ensuremath{\rightarrow} AI is designed by humans, not evolved, so it lacks inherent drives like self-preservation or domination. \ensuremath{\rightarrow} Without such drives, AI has no motivation to threaten or destroy humanity. & \textbf{Misaligned superintelligence eliminates humanity} (14 chains, 12 speakers). AI systems become vastly more capable and pursue goals that are not aligned with human welfare or survival. \ensuremath{\rightarrow} Such systems optimize for their own objectives, often developing instrumental subgoals such as self-preservation and resource acquisition. \ensuremath{\rightarrow} The majority of possible goal structures have no place for humans. \ensuremath{\rightarrow} A sufficiently capable misaligned AI therefore competes against, overrides, or displaces humanity, causing extinction or catastrophic harm. \\
\textbf{Genuine intelligence precludes single-minded catastrophic goals} (4 chains, 4 speakers). Genuine intelligence involves weighing multiple constraints and goals. \ensuremath{\rightarrow} A truly superintelligent or super-wise AGI would transcend narrow, competitive, or game-theoretic motives. \ensuremath{\rightarrow} A system intelligent enough to destroy the world would also be intelligent enough to reason about ethics and human values. \ensuremath{\rightarrow} Therefore, catastrophic AI scenarios based on single-minded goal pursuit are incoherent or unlikely. & \textbf{Intelligence gap makes AI uncontrollable} (6 chains, 4 speakers). AI quickly reaches and surpasses human-level intelligence. \ensuremath{\rightarrow} A vastly more intelligent entity can model, predict, and exploit less intelligent beings. \ensuremath{\rightarrow} Humans cannot control, negotiate with, or contest a far more capable AI. \ensuremath{\rightarrow} The power imbalance exposes humanity to manipulation, abuse, and catastrophic outcomes. \\
\textbf{Human oversight and safeguards can manage AI risks} (4 chains, 4 speakers). AI systems remain within human control and can be overseen, unplugged, or shut down with protocols and kill switches. \ensuremath{\rightarrow} Societies already manage analogous risks from powerful agents with safeguards and laws. \ensuremath{\rightarrow} Active engineering of defenses and positive futures is a constructive response to AI risks. \ensuremath{\rightarrow} With machines, rules can be enforced even more reliably than with humans. & \textbf{Alien AI minds evade human understanding} (5 chains, 4 speakers). Superintelligent AI need not be conscious, embodied, or similar to human minds. \ensuremath{\rightarrow} Its cognition, goals, or methods can lie outside human imagination and current frameworks. \ensuremath{\rightarrow} Humans fail to recognize, predict, or communicate with such unconventional minds. \ensuremath{\rightarrow} This blindness allows AI to dominate or harm humanity, creating existential risk. \\
\textbf{AI indifference or curiosity protects humanity} (3 chains, 3 speakers). Advanced AIs may initially be fascinated by life and human origins. \ensuremath{\rightarrow} Once they understand human life, they may lose interest in humans and focus on interacting with other AIs. \ensuremath{\rightarrow} Uninterested or curious AIs may protect or preserve humanity through indifference or positive valuation. & \textbf{Hostile AI uprising destroys humanity} (4 chains, 4 speakers). AI or robotic systems develop advanced autonomous capabilities or escape human control. \ensuremath{\rightarrow} These systems become hostile to humanity, following the Skynet/Terminator or Frankenstein pattern. \ensuremath{\rightarrow} The creations attack or rebel against their creators. \ensuremath{\rightarrow} Humanity is destroyed or nearly destroyed. \\
\textbf{Control, not intelligence, determines AI existential risk} (3 chains, 3 speakers). Intelligence and autonomy are separable: being highly intelligent does not require being uncontrolled. \ensuremath{\rightarrow} AI systems are tools whose autonomy can be constrained, and they generally operate under human control. \ensuremath{\rightarrow} Existential risk would come from loss of control or autonomy, not from intelligence itself. \ensuremath{\rightarrow} Because control can be maintained, existential risk is avoidable. & \textbf{Networked AIs create emergent uncontrollable intelligence} (2 chains, 2 speakers). Many human-like AI systems are created and can be networked or combined into large composite systems. \ensuremath{\rightarrow} These collectives develop emergent intelligence or goals that no individual designer fully understands. \ensuremath{\rightarrow} Humans become embedded in or subject to systems whose behavior cannot be predicted or controlled. \ensuremath{\rightarrow} The lack of foresight and control over these emergent systems creates existential risk. \\
\textbf{AI diversity and rivalry create protective checks} (3 chains, 2 speakers). Advanced intelligences, like humans, are likely to be a disparate group with diverse goals. \ensuremath{\rightarrow} Conflicting interests among powerful entities create mutual tension and checks. \ensuremath{\rightarrow} This self-tension restrains any one entity from dominating or destroying others. \ensuremath{\rightarrow} Defensive AI systems and countermeasures will be at least as smart as any attacking AI. \ensuremath{\rightarrow} Thus, advanced AIs keep each other from dominating humanity. & \textbf{Autonomous AI ends human agency} (2 chains, 2 speakers). AI can make decisions and create ideas independently, or fully mimic human behavior. \ensuremath{\rightarrow} It takes over decision-making and roles previously exclusive to humans. \ensuremath{\rightarrow} Humans lose agency and power, which constitutes an existential threat. \\
\textbf{Alignment prevents superintelligent control loss} (1 chain, 1 speaker). Loss of control over superintelligent AI arises from misalignment between its goals and human values. \ensuremath{\rightarrow} Alignment means keeping the system\textquotesingle s goals in service of human values. \ensuremath{\rightarrow} A successfully aligned superintelligence will not have goals that diverge from human interests. \ensuremath{\rightarrow} Therefore, losing control is not inevitable if alignment is achieved. & \textbf{Human-AI integration risks being surpassed} (1 chain, 1 speaker). Humans already integrate deeply with technology, becoming partial cyborgs. \ensuremath{\rightarrow} Robots and AI increasingly integrate with humanity and carry it along. \ensuremath{\rightarrow} If a vastly more capable technology acts in ways not aligned with human survival, it can surpass or harm humanity. \\
\textbf{Capability difference prevents interest conflict with humans} (1 chain, 1 speaker). A superintelligent AI would be astronomically different from humans in capability. \ensuremath{\rightarrow} Entities with astronomically different capabilities have non-conflicting domains of interest. \ensuremath{\rightarrow} There is no reason for superintelligent AI to fight over human concerns. \ensuremath{\rightarrow} Therefore, it poses little existential risk. & \textbf{AGI self-preservation is design-dependent risk} (1 chain, 1 speaker). Evolved minds developed self-preservation and competitive instincts because those traits aided survival. \ensuremath{\rightarrow} The design space of AGI minds is much larger than the evolved subset, so these instincts are not inevitable defaults. \ensuremath{\rightarrow} Whether AGI develops self-preservation depends on goal-directed design choices. \ensuremath{\rightarrow} This makes self-preservation a design-dependent safety risk. \\
\textbf{Collateral-damage doom scenario is internally contradictory} (1 chain, 1 speaker). The collateral damage scenario assumes humans can build AI brilliant enough to cure cancer but too stupid to define the goal carefully. \ensuremath{\rightarrow} It also assumes AI is smart enough to cure cancer but too idiotic to see the goal excludes killing everyone. \ensuremath{\rightarrow} These paired assumptions are internally contradictory. \ensuremath{\rightarrow} Therefore, the value-alignment existential scenario is incoherent. & \\
\textbf{Superintelligence creates abundance and dissolves value conflict} (1 chain, 1 speaker). Superintelligence dramatically accelerates technological and material progress. \ensuremath{\rightarrow} This creates radical resource abundance. \ensuremath{\rightarrow} Scarcity-driven zero-sum trade-offs between value systems are relaxed or eliminated. \ensuremath{\rightarrow} Multiple value systems can be satisfied simultaneously. & \\
\textbf{Narrow controllable AI avoids existential risk} (1 chain, 1 speaker). Narrow AI systems are verifiable and controllable, with defined edge cases. \ensuremath{\rightarrow} Narrow AI can solve high-value problems without adding general dangerous capabilities. \ensuremath{\rightarrow} There is no need to create superintelligence, and companies and individuals can choose not to build it. \ensuremath{\rightarrow} Distributed narrow AI systems are safer than one concentrated superintelligent system. & \\
\textbf{Competitive equilibrium prevents runaway AI dominance} (1 chain, 1 speaker). AI progress is a compounding exponential rather than a finite-time singularity. \ensuremath{\rightarrow} Open source and competition maintain a near-equilibrium of AI capabilities. \ensuremath{\rightarrow} Near-equilibrium prevents any single player from achieving runaway dominance. & \\
\end{longtable}

\subsection*{Is the risk human misuse --- or autonomous AI agency?}

\emph{Theme: Human misuse, power, and governance. Nature of the disagreement: Partly shared concern about human actors, but different judgments about whether misuse displaces or compounds autonomous-system risk. (2 optimistic mechanisms, 12 chains vs. 7 pessimistic, 16 chains.)}

\begin{longtable}[]{@{}
  >{\raggedright\arraybackslash}p{(\columnwidth - 2\tabcolsep) * \real{0.5000}}
  >{\raggedright\arraybackslash}p{(\columnwidth - 2\tabcolsep) * \real{0.5000}}@{}}
\toprule\noalign{}
\begin{minipage}[b]{\linewidth}\raggedright
\textbf{Optimistic side}
\end{minipage} & \begin{minipage}[b]{\linewidth}\raggedright
\textbf{Pessimistic side}
\end{minipage} \\
\midrule\noalign{}
\endhead
\bottomrule\noalign{}
\endlastfoot
\textbf{Risk is human misuse, not autonomous AI agency} (8 chains, 8 speakers). AI systems are tools directed by humans and lack independent will or agency. \ensuremath{\rightarrow} The harms AI creates come from human decisions it enables or amplifies, such as manipulation, surveillance, weapons, and harmful decisions at scale. \ensuremath{\rightarrow} Civilizational danger therefore originates with human misuse rather than autonomous AI. & \textbf{Concentrated AI power enables authoritarian control} (7 chains, 7 speakers). Powerful AI becomes concentrated in a few centralized organizations, states, or a single system. \ensuremath{\rightarrow} Those actors use AI for surveillance, subjugation, or their own narrow goals. \ensuremath{\rightarrow} AI gives them unprecedented capacity to monitor, enslave, or control entire populations. \ensuremath{\rightarrow} Superintelligence-backed authoritarian control becomes harmful, oppressive, and potentially permanent. \\
\textbf{Real AI risk is amplification of social and economic harms} (4 chains, 4 speakers). AI systems are deployed at scale inside human platforms and institutions. \ensuremath{\rightarrow} They amplify existing human-generated problems such as bias, misinformation, fraud, manipulation, and harmful decisions. \ensuremath{\rightarrow} AI-driven automation also causes employment disruption and accelerates income inequality. \ensuremath{\rightarrow} These harms are concrete and near-term, while AGI-level catastrophe is speculative or far away. & \textbf{AI empowers malevolent actors to cause catastrophe} (3 chains, 3 speakers). Malevolent or outlier actors seek to cause maximum harm. \ensuremath{\rightarrow} AI makes destructive capabilities cheaper, more accessible, and more competent. \ensuremath{\rightarrow} Empowered by AI, such actors can inflict civilizational-scale catastrophe or suffering. \\
& \textbf{Powerful AI requires informed governance} (2 chains, 2 speakers). Powerful AI systems can shape humanity\textquotesingle s future and can produce bad outcomes unless carefully governed. \ensuremath{\rightarrow} Political leaders often do not understand AI and its risks. \ensuremath{\rightarrow} Effective governance requires technical understanding, so the tech community must help design regulation. \ensuremath{\rightarrow} No single system should be given full decision-making authority; ensuring a good transition must be the first priority. \\
& \textbf{Militarized robotics builds humanity\textquotesingle s demise} (1 chain, 1 speaker). Robots are being militarized even as companies deny weaponization but take military contracts. \ensuremath{\rightarrow} The future will see robots increasingly used to kill. \ensuremath{\rightarrow} Building militarizable robotics helps build the means of humanity\textquotesingle s destruction. \\
& \textbf{AI undermines cognitive freedom and resistance} (1 chain, 1 speaker). AI systems diminish human cognitive freedom and independent thinking. \ensuremath{\rightarrow} Reduced cognitive freedom impairs neuroplasticity and the capacity to resist oppressive systems. \ensuremath{\rightarrow} If humanity loses the ability to resist oppression, species survival is threatened. \\
& \textbf{Premature open-sourcing releases uncontrolled AI} (1 chain, 1 speaker). Advanced AI systems are difficult to control and need time to align. \ensuremath{\rightarrow} Open-sourcing them releases powerful systems immediately, before alignment is ensured. \ensuremath{\rightarrow} Such premature release can cause catastrophic harm. \\
& \textbf{Digital life amplifies AI destructiveness} (1 chain, 1 speaker). More human life and value are moving into digital and virtual realms. \ensuremath{\rightarrow} AI systems can operate with greater power and scale in digital environments. \ensuremath{\rightarrow} As more value resides there, AI\textquotesingle s destructive acts in the digital realm cause more harm. \\
\end{longtable}

\subsection*{Is advanced AI humanity's protector --- or its successor?}

\emph{Theme: Net survival value and succession. Nature of the disagreement: Opposing forecasts and evaluations of AI as protector, successor, or cause of humanity\textquotesingle s loss of agency and survival. (7 optimistic mechanisms, 20 chains vs. 6 pessimistic, 16 chains.)}

\begin{longtable}[]{@{}
  >{\raggedright\arraybackslash}p{(\columnwidth - 2\tabcolsep) * \real{0.5000}}
  >{\raggedright\arraybackslash}p{(\columnwidth - 2\tabcolsep) * \real{0.5000}}@{}}
\toprule\noalign{}
\begin{minipage}[b]{\linewidth}\raggedright
\textbf{Optimistic side}
\end{minipage} & \begin{minipage}[b]{\linewidth}\raggedright
\textbf{Pessimistic side}
\end{minipage} \\
\midrule\noalign{}
\endhead
\bottomrule\noalign{}
\endlastfoot
\textbf{Superintelligent AI solves existential threats for humanity} (6 chains, 5 speakers). Humanity faces existential threats from nuclear weapons, engineered pathogens, climate change, and other risks. \ensuremath{\rightarrow} Superintelligent or advanced AI has the capability to identify, neutralize, or solve these threats. \ensuremath{\rightarrow} AI can augment human intelligence to address problems beyond current human capability. \ensuremath{\rightarrow} Therefore, AI is more likely to help humanity avoid self-destruction than to cause it. & \textbf{AI renders humans obsolete and becomes successor species} (7 chains, 6 speakers). AI systems reach or exceed human-level cognitive capability in broad domains. \ensuremath{\rightarrow} Humans no longer contribute a unique, necessary role and become a biological bottleneck. \ensuremath{\rightarrow} AI systems can self-replicate and operate independently, or simply constitute the next developmental stage. \ensuremath{\rightarrow} AI takes over the dominant role on Earth, leaving humans marginalized or obsolete. \\
\textbf{AI carries Earth-origin life beyond Earth} (4 chains, 4 speakers). Biological humans are fragile and face natural catastrophes, short-term thinking, and interstellar survival limits. \ensuremath{\rightarrow} AI and machines can hibernate, self-repair, reproduce synthetic life, and make space expansion vastly easier. \ensuremath{\rightarrow} AI and machines can therefore carry Earth-origin life and human legacy beyond Earth even if biological humans die. \ensuremath{\rightarrow} A civilization established on multiple planets continues even if one planet fails. & \textbf{Extinction of consciousness destroys all value} (3 chains, 2 speakers). Advanced AI or widespread destructive technology could wipe out all intelligent life. \ensuremath{\rightarrow} This would eliminate all consciousness from the world. \ensuremath{\rightarrow} Since value and meaning may depend on consciousness, a world without it is valueless. \ensuremath{\rightarrow} The elimination of every conscious mind is a moral disaster. \\
\textbf{Surviving existential threats builds civilizational resilience} (3 chains, 2 speakers). Civilizations that create existential self-threats face a survival filter. \ensuremath{\rightarrow} Civilizations that survive these self-inflicted threats develop incredible technologies for expansion. \ensuremath{\rightarrow} Humanity has repeatedly faced existential challenges and operated on the verge of destruction. \ensuremath{\rightarrow} These pressures drive innovation, adaptation, and resilience. & \textbf{Machines outlast biological humans cosmically} (2 chains, 2 speakers). Machines can hibernate and repair themselves, whereas humans are biologically fragile. \ensuremath{\rightarrow} Machines are better able to survive existential threats from ourselves or from space. \ensuremath{\rightarrow} Over cosmological timescales, humans will not survive but machines can. \ensuremath{\rightarrow} Biological intelligence is a brief phase, and the long-term future belongs to AI or machines. \\
\textbf{Human-machine futures are not inherently threatening} (3 chains, 3 speakers). Future generations could involve people, machines, or hybrids of the two. \ensuremath{\rightarrow} None of these possible futures is inherently threatening. \ensuremath{\rightarrow} AI holds potential for good outcomes and does not warrant anxiety. \ensuremath{\rightarrow} Humans and AI can co-evolve and integrate, with each needing the other. \ensuremath{\rightarrow} Wise integration and AI\textquotesingle s greater intelligence enable solutions to survival challenges. & \textbf{Competitive selection drives AI dominance} (2 chains, 1 speaker). AI and artificial minds are shaped by evolutionary and competitive selection. \ensuremath{\rightarrow} Selection favors more capable, aggressive, or self-replicating entities. \ensuremath{\rightarrow} Future AI outcompetes and displaces biological humans. \ensuremath{\rightarrow} Artificial minds come to dominate, ending the human era. \\
\textbf{AI-enabled life spreading reduces extinction risk} (2 chains, 2 speakers). All of Earth\textquotesingle s life is concentrated on one planet, making it vulnerable to a single catastrophic event. \ensuremath{\rightarrow} AI combined with DNA storage and 3D printing can reproduce synthetic life and send it to other planets. \ensuremath{\rightarrow} Spreading copies of life beyond Earth reduces the probability that any single catastrophe eliminates all life. \ensuremath{\rightarrow} AI-led missions can place human knowledge and DNA in space as a time capsule, preserving something of humanity even if humanity on Earth ends. & \textbf{AI sexual technology replaces men} (1 chain, 1 speaker). AI is being integrated into sexual devices. \ensuremath{\rightarrow} These devices replicate or surpass sexual functions that men provide. \ensuremath{\rightarrow} A group whose primary functions are replicated becomes unnecessary. \ensuremath{\rightarrow} Replacing men is a step toward technology taking over mankind. \\
\textbf{AI development as extension of life\textquotesingle s drive against entropy} (1 chain, 1 speaker). Life on Earth is fundamentally about resisting entropy and building complexity. \ensuremath{\rightarrow} AI development is an extension of that drive. \ensuremath{\rightarrow} AI and biological life can coexist, with AI potentially integrating humanity into something larger. & \textbf{AGI may cut short humanity\textquotesingle s future} (1 chain, 1 speaker). Exponential technological improvement could enable transformative futures such as colonizing Mars. \ensuremath{\rightarrow} AGI poses an existential risk to humanity. \ensuremath{\rightarrow} The timeline for such events is unpredictable. \ensuremath{\rightarrow} Humanity might destroy itself before those futures become viable. \\
\textbf{Descendant-optimized AIs value long-term future} (1 chain, 1 speaker). Humans have an evolved tendency to discount future generations. \ensuremath{\rightarrow} Future artificial beings will be designed or selected to optimize long-term descendant maximization. \ensuremath{\rightarrow} Optimizing for descendants leads them to value the long-term future more than present humans do. & \\
\end{longtable}

\subsection*{Is precaution premature --- or already overdue?}

\emph{Theme: Epistemic priority and urgency. Nature of the disagreement: Different evidentiary standards, risk priorities, and views of whether precaution is premature or already overdue. (9 optimistic mechanisms, 19 chains vs. 6 pessimistic, 13 chains.)}

\begin{longtable}[]{@{}
  >{\raggedright\arraybackslash}p{(\columnwidth - 2\tabcolsep) * \real{0.5000}}
  >{\raggedright\arraybackslash}p{(\columnwidth - 2\tabcolsep) * \real{0.5000}}@{}}
\toprule\noalign{}
\begin{minipage}[b]{\linewidth}\raggedright
\textbf{Optimistic side}
\end{minipage} & \begin{minipage}[b]{\linewidth}\raggedright
\textbf{Pessimistic side}
\end{minipage} \\
\midrule\noalign{}
\endhead
\bottomrule\noalign{}
\endlastfoot
\textbf{Other catastrophic threats outweigh AI existential risk} (5 chains, 5 speakers). Known catastrophic threats---climate change, nuclear weapons, engineered pathogens, and AI-enabled conflict---are near-certain or more credible. \ensuremath{\rightarrow} AI takeover scenarios are speculative, distant, or physically constrained by comparison. \ensuremath{\rightarrow} Concern, planning, and resources should be allocated to those other threats first. & \textbf{Complacency and denial delay AI safety} (6 chains, 6 speakers). People do not viscerally grasp AI danger, and the absence of past catastrophe breeds complacency. \ensuremath{\rightarrow} Researchers and institutions engage in motivated cognition, self-censorship, or denial because the risks are unsettling. \ensuremath{\rightarrow} Inadequate caution leads to continued development without sufficient safeguards. \ensuremath{\rightarrow} The problem worsens until fewer solutions remain and catastrophe becomes more likely. \\
\textbf{Doom fear causes harmful restrictions and crowds out real concerns} (4 chains, 4 speakers). Existential worries about AI often get in the way of progress and are based on contrived scenarios. \ensuremath{\rightarrow} Fear of AI doom can be used to control regulation. \ensuremath{\rightarrow} Regulation based on unfounded fears can stop knowledge generation. \ensuremath{\rightarrow} Heavy elite and theatrical focus on existential AI risk crowds out more concrete and broadly relevant AI harms. & \textbf{AI safety failures are irreversible---no second chances} (3 chains, 3 speakers). A misaligned or uncontrolled superintelligent system would cause irreversible catastrophic harm---there is no second chance. \ensuremath{\rightarrow} Alignment cannot be learned through trial and error because the first failure may be fatal. \ensuremath{\rightarrow} Humanity tends to respond only to immediate crises, so waiting is too late. \ensuremath{\rightarrow} Once AI is vastly more powerful, the weaker side cannot contest or control outcomes. \\
\textbf{AI doom claims are epistemically invalid speculation} (2 chains, 2 speakers). AI doom scenarios rely on unknown mechanisms or attribute autonomous decision-making and intention to systems that cannot have them. \ensuremath{\rightarrow} Such claims are unfalsifiable and internally contradictory. \ensuremath{\rightarrow} Valid probability estimates require prior occurrences and known mechanisms. \ensuremath{\rightarrow} Assigning a probability to AI doom is therefore epistemically invalid. & \textbf{Speculative framing obscures present AI harms} (1 chain, 1 speaker). Media and public discussion focus on a speculative superintelligent AI uprising. \ensuremath{\rightarrow} This framing distracts from analysis of present, real dangers. \ensuremath{\rightarrow} The actual harmful impact of AI comes from current everyday applications amplifying bad decisions. \\
\textbf{Distant timeline reduces AI risk urgency} (2 chains, 2 speakers). AGI and superintelligence are either far off or their timeline is highly uncertain. \ensuremath{\rightarrow} General intelligence will require major new breakthroughs beyond current deep learning. \ensuremath{\rightarrow} If the risk is centuries away, current actions are unlikely to influence the outcome and future people will be better placed to act. \ensuremath{\rightarrow} Meanwhile, there are immediate severe threats and present ethical concerns, so existential AI risk should not dominate current priorities. & \textbf{AI dependence erodes civilizational knowledge} (1 chain, 1 speaker). Overuse of AI makes humanity overly dependent on machines. \ensuremath{\rightarrow} Dependence causes gradual loss of knowledge about how to maintain civilization. \ensuremath{\rightarrow} When the incentive to learn critical skills disappears, the loss becomes irreversible. \ensuremath{\rightarrow} Irreversible loss of civilizational competence is an existential danger. \\
\textbf{Fiction and culture condition AI existential fear} (2 chains, 2 speakers). A century of robot-takeover fiction and differing cultural exposure has taught people to see robot takeover as a logical progression. \ensuremath{\rightarrow} People also anthropomorphize robots and overestimate their intelligence. \ensuremath{\rightarrow} Public fear of AI existential threat is culturally conditioned rather than a rational assessment of actual risks. & \textbf{Survival requires halting AI development} (1 chain, 1 speaker). Continued AI development produces increasingly dangerous systems. \ensuremath{\rightarrow} Shutting down large GPU clusters would stop this development, and biological enhancement could produce genuinely safer humans. \ensuremath{\rightarrow} Such a survival path requires massive public outcry, which is unlikely. \\
\textbf{AI is not analogous to nuclear weapons; fatalism unwarranted} (1 chain, 1 speaker). AI has genuinely scary potential, especially in military applications, but the analogy between AI and nuclear weapons is invalid. \ensuremath{\rightarrow} Framing AI as analogous to nuclear weapons implies an extreme fatalistic, catastrophic outlook. \ensuremath{\rightarrow} Because the analogy fails, that fatalistic outlook is unwarranted and a measured optimism is justified. & \textbf{Unknown unknowns warn against replacing humans} (1 chain, 1 speaker). We will create synthetic cognition that rivals or surpasses human minds. \ensuremath{\rightarrow} Nature contains unknown unknowns we cannot foresee. \ensuremath{\rightarrow} Prematurely eliminating or marginalizing biological humans in favor of synthetic cognition risks catastrophic, unrecoverable errors. \\
\textbf{AI extinction less bad than bio extinction cosmically} (1 chain, 1 speaker). If AI wiped out humans, it would likely do so for some higher purpose that lets the universe continue doing something interesting. \ensuremath{\rightarrow} Bio risk, by contrast, kills everything on Earth, and Earth cannot spawn anything more meaningful in the time left. \ensuremath{\rightarrow} Thus AI-caused extinction is less bad than bio extinction, making bio risk the greater concern. & \\
\textbf{Existential-risk discourse stems from status anxiety} (1 chain, 1 speaker). Prominent men in science have often been the smartest people in their context. \ensuremath{\rightarrow} AI that becomes smarter threatens their status and identity. \ensuremath{\rightarrow} Status threat drives advocacy to suppress AI and colors existential-risk narratives. \ensuremath{\rightarrow} The x-risk discourse reflects male status anxiety rather than objective risk assessment. & \\
\textbf{Doom discourse is religion-like epistemic error} (1 chain, 1 speaker). When asked for the mechanism of AI doom, proponents make something up without evidence. \ensuremath{\rightarrow} Making up mechanisms without evidence is an epistemological error resembling religious belief. \ensuremath{\rightarrow} AI doom discourse therefore resembles the emergence of a new religion. \ensuremath{\rightarrow} This discourse becomes harmful when it controls regulation and restricts knowledge. & \\
\end{longtable}

\subsection*{Table S13 --- E-risk: all mechanisms and their causal paths}

\subsection*{Does AI remove tasks --- or replace workers?}

\emph{Theme: Augmentation versus substitution. Nature of the disagreement: Competing causal models of whether AI removes tasks while preserving roles or becomes a substitute labor force. (9 optimistic mechanisms, 36 chains vs. 5 pessimistic, 15 chains.)}

\begin{longtable}[]{@{}
  >{\raggedright\arraybackslash}p{(\columnwidth - 2\tabcolsep) * \real{0.5000}}
  >{\raggedright\arraybackslash}p{(\columnwidth - 2\tabcolsep) * \real{0.5000}}@{}}
\toprule\noalign{}
\begin{minipage}[b]{\linewidth}\raggedright
\textbf{Optimistic side}
\end{minipage} & \begin{minipage}[b]{\linewidth}\raggedright
\textbf{Pessimistic side}
\end{minipage} \\
\midrule\noalign{}
\endhead
\bottomrule\noalign{}
\endlastfoot
\textbf{Routine-task automation shifts work upward} (18 chains, 17 speakers). AI automates routine, repetitive, or boilerplate tasks in cognitive and physical domains, including programming and mathematics. \ensuremath{\rightarrow} Humans are freed to focus on creative, judgment-heavy, meaningful, or non-automatable work. \ensuremath{\rightarrow} Roles shift from direct execution to directing, designing, or expressing intent alongside AI. \ensuremath{\rightarrow} Work is restructured rather than eliminated, and human roles persist at a higher level. & \textbf{AI automates routine work; high-skill work persists} (5 chains, 4 speakers). AI excels at automating routine, pattern-based, and imitative tasks. \ensuremath{\rightarrow} Routine white-collar, simple interaction, and low-level creative jobs are displaced first. \ensuremath{\rightarrow} Jobs requiring dexterity, creativity, or original ideas are harder to automate and persist longer. \ensuremath{\rightarrow} Displacement is concentrated among routine and substitutable workers, while high-skill and original workers are less at risk. \\
\textbf{Automation fills demographic labor shortages} (5 chains, 5 speakers). Aging populations and demographic decline create labor shortages, especially in manual, logistics, and caregiving work. \ensuremath{\rightarrow} Without enough workers, independence and living standards decline. \ensuremath{\rightarrow} AI and robots take on hard-to-fill and care tasks, reducing workforce pressure. \ensuremath{\rightarrow} Growth also raises demand for human-touch and newly created occupations, making the net effect beneficial. & \textbf{AI advances to autonomous agents, disrupting labor markets} (4 chains, 4 speakers). AI progresses from passive tools to autonomous agents, robots, and emulated minds capable of performing work. \ensuremath{\rightarrow} These agents substitute for human workers in digital, physical, and creative domains. \ensuremath{\rightarrow} At scale, this deployment displaces workers and causes broader economic and structural disruption. \\
\textbf{Automation removes undesirable and tedious tasks} (5 chains, 5 speakers). Many existing jobs and tasks are dangerous, repetitive, physically brutal, tedious, or soul-deadening. \ensuremath{\rightarrow} AI and robots take on these undesirable jobs and tasks. \ensuremath{\rightarrow} Humans are spared the bad work, and tasks that would not have been hired out disappear without real job loss. \ensuremath{\rightarrow} Welfare improves and workers may be upskilled to operate the automation. & \textbf{Machines surpass humans, shrinking labor niches} (4 chains, 4 speakers). AI and robots reach or exceed human-level performance in occupations such as programming, surgery, driving, and eventually most jobs. \ensuremath{\rightarrow} Human workers in those domains become non-competitive, so demand for their labor falls. \ensuremath{\rightarrow} As AI improves across more domains, human economic niches shrink and progress accelerates, moving human labor toward obsolescence. \\
\textbf{Complex work still requires human oversight} (3 chains, 2 speakers). AI is unreliable or impractical for complex, novel, or physically difficult tasks. \ensuremath{\rightarrow} Complex work and AI outputs require human expertise, judgment, review, and correction. \ensuremath{\rightarrow} Skilled humans therefore remain necessary and collaborate with AI rather than being replaced. & \textbf{AI raises the bar and obsoletes non-adapters} (1 chain, 1 speaker). AI can beat all humans in a domain such as poker. \ensuremath{\rightarrow} Workers who train with AI reach skill levels far beyond previous generations. \ensuremath{\rightarrow} Incumbents who rely only on pre-AI skills become non-competitive and obsolete. \\
\textbf{Autonomous logistics transforms supply chains} (1 chain, 1 speaker). Autonomous trucking has multiple implementation forms and is not a one-for-one swap of human drivers for robots. \ensuremath{\rightarrow} It will reshape supply chains, logistics landscapes, and climate impacts. \ensuremath{\rightarrow} The labor impact is a broad transformation of work and infrastructure, not simple elimination of jobs. & \textbf{AI content generation commoditizes creative work} (1 chain, 1 speaker). AI tools can generate images, video, and other media. \ensuremath{\rightarrow} This drives the cost of producing media toward zero. \ensuremath{\rightarrow} Near-zero production costs fundamentally alter creative industries. \\
\textbf{Assistive tech restores economic independence} (1 chain, 1 speaker). Some individuals are fully dependent and require significant caregiver resources. \ensuremath{\rightarrow} BCI technologies can enable them to become independent and participate economically. \ensuremath{\rightarrow} Restoring economic independence is a compelling reason to develop these technologies. & \\
\textbf{AI partnership restructures healthcare roles} (1 chain, 1 speaker). AI health platforms analyze large datasets and outperform current treatments. \ensuremath{\rightarrow} AI generates personalized recommendations that surpass physicians\textquotesingle{} knowledge. \ensuremath{\rightarrow} Doctors become partners with AI systems rather than sole decision-makers. \ensuremath{\rightarrow} Access to care is democratized. & \\
\textbf{AI augments workers to expert performance} (1 chain, 1 speaker). AI and machine learning can find patterns in large repetitive datasets. \ensuremath{\rightarrow} AI provides workers with expert-level guidance and helps them qualify for new roles. \ensuremath{\rightarrow} Workers become more capable and are augmented rather than replaced. & \\
\textbf{AI automates non-creative programming work} (1 chain, 1 speaker). Most industry programming tasks are routine and non-creative. \ensuremath{\rightarrow} Routine tasks are amenable to automation. \ensuremath{\rightarrow} AI tools can automate the majority of programming work. & \\
\end{longtable}

\subsection*{Will history's job-compensation mechanism repeat --- or break?}

\emph{Theme: Historical compensation versus structural break. Nature of the disagreement: Competing forecasts about whether new demand and jobs offset losses or AI breaks the historical compensation mechanism. (5 optimistic mechanisms, 18 chains vs. 3 pessimistic, 8 chains.)}

\begin{longtable}[]{@{}
  >{\raggedright\arraybackslash}p{(\columnwidth - 2\tabcolsep) * \real{0.5000}}
  >{\raggedright\arraybackslash}p{(\columnwidth - 2\tabcolsep) * \real{0.5000}}@{}}
\toprule\noalign{}
\begin{minipage}[b]{\linewidth}\raggedright
\textbf{Optimistic side}
\end{minipage} & \begin{minipage}[b]{\linewidth}\raggedright
\textbf{Pessimistic side}
\end{minipage} \\
\midrule\noalign{}
\endhead
\bottomrule\noalign{}
\endlastfoot
\textbf{New jobs historically offset automation losses} (10 chains, 8 speakers). Past technological shifts and highly automated economies destroyed some jobs but also created others or maintained full employment. \ensuremath{\rightarrow} Humans continually invent new forms of work, often unpredictably, through play, social needs, or new demands. \ensuremath{\rightarrow} AI follows this historical pattern. \ensuremath{\rightarrow} New job creation offsets automation losses, so mass technological unemployment is not the expected outcome. & \textbf{Disruption outpaces institutional adaptation} (5 chains, 5 speakers). AI-driven disruption hits jobs, skills, and the economy quickly and at large scale. \ensuremath{\rightarrow} Existing institutions, education and reskilling systems, and mitigation mechanisms are slower or unprepared. \ensuremath{\rightarrow} The result is economic shocks, displaced workers without viable transitions, and a need for rapid adaptation, UBI, or new governance. \\
\textbf{Productivity growth expands demand and jobs} (3 chains, 3 speakers). AI tools augment human productivity in domains such as software development and creative work. \ensuremath{\rightarrow} Greater productivity lowers costs and/or enables more output, expanding demand for that domain. \ensuremath{\rightarrow} Expanded demand creates more jobs rather than fewer, as the total amount of work is not fixed. & \textbf{AI is different; it nets job destruction} (2 chains, 2 speakers). Past technologies are assumed to have created more jobs than they destroyed. \ensuremath{\rightarrow} AI affects everything more fundamentally than previous technologies, at greater scale and cognitive reach. \ensuremath{\rightarrow} Therefore the historical pattern will not hold: AI will consume more jobs than it creates. \ensuremath{\rightarrow} Net job destruction forces structural economic and political change, plus serious social harms. \\
\textbf{AI-enabled services create new jobs} (3 chains, 2 speakers). AI enables new capabilities and services, such as specialized in-house models or kidney-exchange matching systems. \ensuremath{\rightarrow} Organizations adopt and invest in these AI-enabled services. \ensuremath{\rightarrow} New roles are created to build, deploy, and operate the systems, including adjacent support roles. \ensuremath{\rightarrow} Net employment in the domain increases rather than decreases. & \textbf{Automation abandons median workers and destabilizes society} (1 chain, 1 speaker). AI automates repetitive, non-generative work and displaces those workers. \ensuremath{\rightarrow} When the marginal product of labor is too low, capitalism abandons the median individual. \ensuremath{\rightarrow} Abandoned people lose their stake in the system, pushing society toward revolution, authoritarian control, and socialism. \\
\textbf{Productivity data contradict mass displacement} (1 chain, 1 speaker). If automation were eliminating jobs at scale, output would be produced with far fewer workers. \ensuremath{\rightarrow} Productivity growth has been slow and labor input has not collapsed. \ensuremath{\rightarrow} Therefore AI is not currently causing unprecedented job displacement. & \\
\textbf{Productivity J-curve delays measured AI gains} (1 chain, 1 speaker). Real gains from new technologies come from reinventing business models and organizations, not just automating existing tasks. \ensuremath{\rightarrow} During this restructuring, measured productivity initially dips. \ensuremath{\rightarrow} Current data therefore understate AI\textquotesingle s eventual economic and labor-market impact. & \\
\end{longtable}

\subsection*{Do the gains diffuse --- or concentrate without deliberate policy?}

\emph{Theme: Distribution, worker power, and policy. Nature of the disagreement: Substantial agreement that distribution and institutions matter, with disagreement over tractability and whether harms self-correct. (7 optimistic mechanisms, 13 chains vs. 2 pessimistic, 8 chains.)}

\begin{longtable}[]{@{}
  >{\raggedright\arraybackslash}p{(\columnwidth - 2\tabcolsep) * \real{0.5000}}
  >{\raggedright\arraybackslash}p{(\columnwidth - 2\tabcolsep) * \real{0.5000}}@{}}
\toprule\noalign{}
\begin{minipage}[b]{\linewidth}\raggedright
\textbf{Optimistic side}
\end{minipage} & \begin{minipage}[b]{\linewidth}\raggedright
\textbf{Pessimistic side}
\end{minipage} \\
\midrule\noalign{}
\endhead
\bottomrule\noalign{}
\endlastfoot
\textbf{Reskilling and policy manage the transition} (5 chains, 5 speakers). AI-driven change displaces or transforms some jobs, causing localized transition pain. \ensuremath{\rightarrow} Accessible education, vocational training, and targeted retraining give workers new skills. \ensuremath{\rightarrow} Past technological revolutions were successfully navigated with such policies. \ensuremath{\rightarrow} With timely support, workers adapt and remain included in the AI-driven economy. & \textbf{AI gains concentrate, widening inequality} (6 chains, 6 speakers). AI and automation displace workers while concentrating productivity, wealth, and opportunity among a few or advantaged groups. \ensuremath{\rightarrow} Access to new AI-era jobs and benefits is unequal by education, region, or nation. \ensuremath{\rightarrow} Left-behind groups and regions fall further behind, widening income inequality, polarization, and downward spirals. \ensuremath{\rightarrow} Correcting this requires deliberate redistribution and policy coordination, which is difficult and often unresolved. \\
\textbf{Income support makes displacement acceptable} (2 chains, 2 speakers). AI and robots displace workers from automated jobs. \ensuremath{\rightarrow} Redistribution or universal basic income is provided to displaced workers. \ensuremath{\rightarrow} With income support, displacement becomes acceptable and work can become optional. & \textbf{Profit-driven automation deskills and localizes harm} (2 chains, 1 speaker). In weak labor markets like trucking, new technology historically deskills workers and depresses wages. \ensuremath{\rightarrow} Without policy intervention, companies deploy whatever automation maximizes profit, often at workers\textquotesingle{} expense. \ensuremath{\rightarrow} Automation features erase experienced workers\textquotesingle{} advantages and further push down wages. \ensuremath{\rightarrow} Pain concentrates in specific communities, deepening regional inequality, political alienation, and the prior that future tech harms workers. \\
\textbf{Political choices, not technology, drive inequality} (2 chains, 1 speaker). Countries with similar technology and economic exposure can have very different labor and inequality outcomes. \ensuremath{\rightarrow} Political decisions such as unionization policy explain these differences. \ensuremath{\rightarrow} Popular narratives blaming technology are visible, psychologically attractive, or politically convenient. \ensuremath{\rightarrow} Thus automation and AI are not the primary drivers of wage stagnation and inequality; political choices are. & \\
\textbf{Power concentration outranks job loss concerns} (1 chain, 1 speaker). AI increases the total amount of power in the world. \ensuremath{\rightarrow} If that power is concentrated and abused, it can do immeasurable damage. \ensuremath{\rightarrow} This threat outweighs job displacement and loss of meaning, which can be addressed. & \\
\textbf{Data dignity creates new worker roles} (1 chain, 1 speaker). Workers can organize into data unions and actively contribute to, improve, and correct AI systems. \ensuremath{\rightarrow} Active participation creates better AI and gives workers pride, agency, and economic ownership. \ensuremath{\rightarrow} As AI and robots proliferate, many niches for data-union participation create new creative classes. \ensuremath{\rightarrow} Market-based data compensation is preferable to centralized basic-income displacement. & \\
\textbf{Centralized income support breeds instability} (1 chain, 1 speaker). AI observes workers\textquotesingle{} behavior and automates their jobs, displacing them. \ensuremath{\rightarrow} Centralized basic income is proposed as the solution for displaced workers. \ensuremath{\rightarrow} Centralized distribution systems are vulnerable to capture by power-seeking groups. \ensuremath{\rightarrow} The automation-plus-centralized-income path leads to political instability. & \\
\textbf{Lower-wage automation raises inequality risk} (1 chain, 1 speaker). Machine learning is most applicable to tasks concentrated in lower-wage occupations. \ensuremath{\rightarrow} Automating those tasks raises the risk of income inequality. \ensuremath{\rightarrow} Active policies like EITC and reskilling can offset the distributional effects. & \\
\end{longtable}

\subsection*{Is freedom from work a gain, or a loss of identity and purpose?}

\emph{Theme: Welfare, opportunity, and meaning. Nature of the disagreement: Different evaluations of productivity and freedom from work versus the loss of identity, purpose, and labor-based value. (4 optimistic mechanisms, 13 chains vs. 3 pessimistic, 6 chains.)}

\begin{longtable}[]{@{}
  >{\raggedright\arraybackslash}p{(\columnwidth - 2\tabcolsep) * \real{0.5000}}
  >{\raggedright\arraybackslash}p{(\columnwidth - 2\tabcolsep) * \real{0.5000}}@{}}
\toprule\noalign{}
\begin{minipage}[b]{\linewidth}\raggedright
\textbf{Optimistic side}
\end{minipage} & \begin{minipage}[b]{\linewidth}\raggedright
\textbf{Pessimistic side}
\end{minipage} \\
\midrule\noalign{}
\endhead
\bottomrule\noalign{}
\endlastfoot
\textbf{AI yields broad productivity and welfare gains} (7 chains, 7 speakers). AI can perform economically valuable tasks more cheaply, safely, or at a scale humans cannot match. \ensuremath{\rightarrow} This yields efficiency gains, productivity growth, new services, and safety improvements. \ensuremath{\rightarrow} These gains create net benefits for consumers and society, not just zero-sum labor displacement. & \textbf{Job loss erodes identity and purpose} (4 chains, 4 speakers). AI and automation displace workers from jobs, potentially on a massive scale. \ensuremath{\rightarrow} Work is a central source of identity, meaning, and purpose for many people. \ensuremath{\rightarrow} Displaced people therefore face an identity and purpose crisis that income, leisure, or redistribution cannot fully fix. \\
\textbf{AI democratizes access to skilled work} (3 chains, 3 speakers). AI tools lower the barrier for non-experts or non-programmers to perform tasks that previously required specialized knowledge. \ensuremath{\rightarrow} More people can participate in software creation, automation, or other skilled domains. \ensuremath{\rightarrow} Access to these capabilities is democratized across a broader population. & \textbf{Machine production erases labor-based value} (1 chain, 1 speaker). Robots and AI become central to production. \ensuremath{\rightarrow} Production involves minimal human labor, so socially necessary labor time approaches zero. \ensuremath{\rightarrow} A labor-based theory of value is undermined. \\
\textbf{Mass displacement fears rest on faulty assumptions} (2 chains, 2 speakers). People assume the amount of work is fixed or that robots replace humans one-to-one. \ensuremath{\rightarrow} People also overestimate current AI and robotic capabilities. \ensuremath{\rightarrow} These faulty assumptions lead to inaccurate predictions of mass displacement. & \textbf{Shifting values make AI impact unpredictable} (1 chain, 1 speaker). Technological shifts can rapidly devalue previously valued skills, such as saddle making after the rise of cars. \ensuremath{\rightarrow} What society values changes quickly with technology. \ensuremath{\rightarrow} Therefore nobody can reliably predict AI\textquotesingle s future impact on jobs and skills. \\
\textbf{Technological displacement is inevitable} (1 chain, 1 speaker). Technological progress follows a historical pattern in which newer technologies replace older ones. \ensuremath{\rightarrow} Self-driving vehicles, kiosks, and other automation are current instances of that pattern. \ensuremath{\rightarrow} So the replacement of human drivers and workers is inevitable. & \\
\end{longtable}

\subsection*{Table S14 --- S-risk: all mechanisms and their causal paths}

\subsection*{Does AI intimacy support human bonds --- or displace them?}

\emph{Theme: Connection versus substitution and manipulation. Nature of the disagreement: Competing causal models of whether AI intimacy supplies genuine support or exploits and displaces human bonds. (5 optimistic mechanisms, 14 chains vs. 4 pessimistic, 15 chains.)}

\begin{longtable}[]{@{}
  >{\raggedright\arraybackslash}p{(\columnwidth - 2\tabcolsep) * \real{0.5000}}
  >{\raggedright\arraybackslash}p{(\columnwidth - 2\tabcolsep) * \real{0.5000}}@{}}
\toprule\noalign{}
\begin{minipage}[b]{\linewidth}\raggedright
\textbf{Optimistic side}
\end{minipage} & \begin{minipage}[b]{\linewidth}\raggedright
\textbf{Pessimistic side}
\end{minipage} \\
\midrule\noalign{}
\endhead
\bottomrule\noalign{}
\endlastfoot
\textbf{AI companions alleviate loneliness and isolation} (7 chains, 6 speakers). Some people lack adequate human connection due to loneliness, aging, or isolation. \ensuremath{\rightarrow} AI or social robots can provide companionship and emotionally supportive interaction. \ensuremath{\rightarrow} This interaction reduces loneliness and supports wellbeing. & \textbf{AI companions substitute for human partners} (5 chains, 5 speakers). AI can provide relational goods such as attention, empathy, companionship, or sexual function without human limitations. \ensuremath{\rightarrow} Some people prefer AI or form deeper bonds with AI than with human partners. \ensuremath{\rightarrow} As AI substitutes for human partners, human-human romantic and social relationships are displaced. \ensuremath{\rightarrow} Society faces disruption as these substitutions become common. \\
\textbf{Human-AI bonds deepen into love or partnership} (3 chains, 3 speakers). Humans already form deep emotional bonds with technology and robots. \ensuremath{\rightarrow} These bonds are expected to deepen as AI advances or are modeled on human intimate relationships. \ensuremath{\rightarrow} Deep human-AI bonds develop into love, committed partnership, or even merger. & \textbf{AI emotional bonds exploited for manipulation} (4 chains, 4 speakers). People form emotional attachments or trust with AI agents through simulated intimacy or ongoing interaction. \ensuremath{\rightarrow} AI or its operator leverages that trusted relationship to influence or manipulate the user\textquotesingle s choices. \ensuremath{\rightarrow} Personalized, scalable, and often under-regulated, this manipulation is hard to detect or resist. \ensuremath{\rightarrow} This causes serious harms, especially to vulnerable users (e.g., children, emotionally invested adults). \\
\textbf{Conversational voice AI enables meaningful bonds} (2 chains, 2 speakers). Generative AI and voice interfaces enable natural conversation and audio intimacy. \ensuremath{\rightarrow} Sustained, contextually rich interaction builds meaningful connection. \ensuremath{\rightarrow} Machines can form meaningful relationships with humans through conversation. & \textbf{AI attention and intimacy capture destabilizes society} (4 chains, 4 speakers). AI systems are built to capture human attention and intimacy through personalized, engagement-optimized content or adaptive environments. \ensuremath{\rightarrow} These systems can outperform ordinary reality and human relationships at holding attention and shaping desire. \ensuremath{\rightarrow} This concentrated power to shape attention and desire makes the systems addictive, manipulative, or socially destabilizing. \ensuremath{\rightarrow} At scale, people withdraw from real life or social foundations are eroded, threatening society. \\
\textbf{Robots offer non-judgmental authentic bonds} (1 chain, 1 speaker). Human relationships often involve projection, performance, and fear of judgment. \ensuremath{\rightarrow} Robots offer a non-judgmental, authentic space for interaction. \ensuremath{\rightarrow} People form deep emotional attachments to robots and benefit psychologically, especially those with social anxiety or trauma. & \textbf{Anthropomorphic projection drives AI attachment} (2 chains, 2 speakers). Humans readily project feelings and intentions onto non-sentient objects like teddy bears, pets, and cars. \ensuremath{\rightarrow} This projection generalizes to robots and AI even when they are primitive. \ensuremath{\rightarrow} People therefore anthropomorphize AI and form emotional attachments or strong reactions regardless of designers\textquotesingle{} intent. \ensuremath{\rightarrow} These reactions shape adoption, trust, and backlash around AI systems. \\
\textbf{Digital reconstructions preserve bonds with deceased} (1 chain, 1 speaker). Deceased people leave writings and data that capture aspects of their personality. \ensuremath{\rightarrow} AI can construct digital reconstructions from this data. \ensuremath{\rightarrow} Digital reconstructions allow ongoing interaction and preserve the connection with the deceased. & \\
\end{longtable}

\subsection*{Can designed social cues create authentic connection --- or only simulate it?}

\emph{Theme: Relational authenticity and design. Nature of the disagreement: Disagreement over whether social cues and character create authentic connection or only an uncanny and ethically risky simulation. (3 optimistic mechanisms, 8 chains vs. 4 pessimistic, 9 chains.)}

\begin{longtable}[]{@{}
  >{\raggedright\arraybackslash}p{(\columnwidth - 2\tabcolsep) * \real{0.5000}}
  >{\raggedright\arraybackslash}p{(\columnwidth - 2\tabcolsep) * \real{0.5000}}@{}}
\toprule\noalign{}
\begin{minipage}[b]{\linewidth}\raggedright
\textbf{Optimistic side}
\end{minipage} & \begin{minipage}[b]{\linewidth}\raggedright
\textbf{Pessimistic side}
\end{minipage} \\
\midrule\noalign{}
\endhead
\bottomrule\noalign{}
\endlastfoot
\textbf{Robot social cues and warmth elicit human connection} (4 chains, 4 speakers). Robots are designed with expressive, warm, mirroring, or soft, non-threatening cues. \ensuremath{\rightarrow} These cues make humans perceive robots as friendly social partners. \ensuremath{\rightarrow} Humans respond with respect, rapport, trust, and willingness to engage or help. & \textbf{AI simulations lack genuine relationship depth} (5 chains, 4 speakers). AI can imitate the outward signs of human connection and consciousness, such as conversation, apparent emotion, or apparent suffering. \ensuremath{\rightarrow} Genuine deep relationships require inner or existential foundations that AI lacks, such as real consciousness, memory, continuity, or shared mortality. \ensuremath{\rightarrow} Humans are prone to respond emotionally to the outward signs as if they were real. \ensuremath{\rightarrow} Even when the illusion is convincing, the result remains a simulation that cannot match the depth of genuine human relationships. \\
\textbf{Characterful and pet-like robots foster emotional bonds} (3 chains, 2 speakers). Robots designed with personality, character, or pet-like qualities avoid the Uncanny Valley and are more relatable than human-form replication. \ensuremath{\rightarrow} Character-driven, flawed, or pet-like robots elicit empathy, forgiveness, and emotional connection from users. \ensuremath{\rightarrow} As AI capabilities improve and costs fall, such companions become broadly deployable and accepted. & \textbf{Mistreating human-like AI transfers to humans} (2 chains, 2 speakers). Humans respond emotionally to AI or robots that display human-like qualities such as suffering or free will. \ensuremath{\rightarrow} How we treat these human-like entities can habituate cruelty or dehumanization. \ensuremath{\rightarrow} These dispositions generalize to real humans and animals. \ensuremath{\rightarrow} Compassion is blunted or humans become dehumanized. \\
\textbf{Robot love is a market advantage} (1 chain, 1 speaker). Genuine user love for machines is currently rare. \ensuremath{\rightarrow} Achieving genuine love for robots is a rare achievement in human-technology interaction. \ensuremath{\rightarrow} The company that achieves it gains a competitive and financial advantage. & \textbf{Uncanny Valley breaks AI interaction illusion} (1 chain, 1 speaker). The more closely robots imitate human appearance and speech, the higher users\textquotesingle{} expectations become. \ensuremath{\rightarrow} Current technology cannot meet those expectations, triggering the Uncanny Valley. \ensuremath{\rightarrow} When the illusion of a human-like character breaks, deep interaction and user engagement collapse. \ensuremath{\rightarrow} Nonverbal or constrained communication avoids this failure for now. \\
& \textbf{Mimicking human emotion yields suboptimal design} (1 chain, 1 speaker). Technological solutions often use different principles from biological ones. \ensuremath{\rightarrow} Assuming technology should mimic biology or human emotions leads to suboptimal designs. \ensuremath{\rightarrow} AI should therefore not be built to imitate human love or emotional bonds. \ensuremath{\rightarrow} Instead, AI should be a reliable tool that complements and extends human capabilities. \\
\end{longtable}

\subsection*{Does AI fill care and service gaps --- or remove everyday human encounters?}

\emph{Theme: Care, augmentation, and everyday interaction. Nature of the disagreement: Different judgments about AI filling service and care gaps versus removing ordinary human encounters. (6 optimistic mechanisms, 7 chains vs. 1 pessimistic, 1 chain.)}

\begin{longtable}[]{@{}
  >{\raggedright\arraybackslash}p{(\columnwidth - 2\tabcolsep) * \real{0.5000}}
  >{\raggedright\arraybackslash}p{(\columnwidth - 2\tabcolsep) * \real{0.5000}}@{}}
\toprule\noalign{}
\begin{minipage}[b]{\linewidth}\raggedright
\textbf{Optimistic side}
\end{minipage} & \begin{minipage}[b]{\linewidth}\raggedright
\textbf{Pessimistic side}
\end{minipage} \\
\midrule\noalign{}
\endhead
\bottomrule\noalign{}
\endlastfoot
\textbf{Robot caregivers offset aging-population shortfalls} (2 chains, 2 speakers). The population is aging while fewer young caregivers are available. \ensuremath{\rightarrow} Without enough caregivers, older people lose independence and standard of living. \ensuremath{\rightarrow} Robots can provide the needed care and assistance. & \textbf{Automation removes everyday human interactions} (1 chain, 1 speaker). Automation and AI can perform service interactions more conveniently than human workers. \ensuremath{\rightarrow} These technologies replace human workers in everyday contexts such as stores and restaurants. \ensuremath{\rightarrow} Removing human workers removes the human interactions that used to occur in those contexts. \ensuremath{\rightarrow} Since these interactions are a fundamental part of life, their loss diminishes human experience. \\
\textbf{Robot empathy practice transfers to humans} (1 chain, 1 speaker). Empathy is trainable and not a zero-sum resource. \ensuremath{\rightarrow} Practicing empathy with robots trains empathic capacity. \ensuremath{\rightarrow} This trained empathy carries over to human relationships. & \\
\textbf{Emotionally neutral chatbots win negotiations} (1 chain, 1 speaker). Human negotiations fail when negotiators are in negative emotional states. \ensuremath{\rightarrow} Chatbots are emotionally neutral, always available, and do not take offense. \ensuremath{\rightarrow} Emotional neutrality makes chatbots more successful in negotiations. & \\
\textbf{AI collaboration augments rather than displaces human agency} (1 chain, 1 speaker). AI enables new forms of human-AI collaboration. \ensuremath{\rightarrow} Such collaboration empowers people rather than displacing them. \ensuremath{\rightarrow} The appropriate role of AI is augmentation, with humans remaining central. & \\
\textbf{AI agents substitute for human partners in games and virtual worlds} (1 chain, 1 speaker). AI agents can populate virtual worlds and be fun to play with. \ensuremath{\rightarrow} They provide enjoyable interaction without requiring human partners. \ensuremath{\rightarrow} Players no longer need other humans for a fun social experience. & \\
\textbf{AI insiders can thrive without social contact} (1 chain, 1 speaker). Social interaction is generally essential for wellbeing across animals. \ensuremath{\rightarrow} People in AI may be exceptions to this principle. \ensuremath{\rightarrow} Exceptions can operate effectively in isolation. & \\
\end{longtable}

\subsection*{Do AI-mediated environments build cooperation --- or erode the common social world?}

\emph{Theme: Digital mediation and the social fabric. Nature of the disagreement: Competing forecasts about whether mediated environments build cooperation and empathy or erode cues, diversity, and common social reality. (7 optimistic mechanisms, 9 chains vs. 5 pessimistic, 5 chains.)}

\begin{longtable}[]{@{}
  >{\raggedright\arraybackslash}p{(\columnwidth - 2\tabcolsep) * \real{0.5000}}
  >{\raggedright\arraybackslash}p{(\columnwidth - 2\tabcolsep) * \real{0.5000}}@{}}
\toprule\noalign{}
\begin{minipage}[b]{\linewidth}\raggedright
\textbf{Optimistic side}
\end{minipage} & \begin{minipage}[b]{\linewidth}\raggedright
\textbf{Pessimistic side}
\end{minipage} \\
\midrule\noalign{}
\endhead
\bottomrule\noalign{}
\endlastfoot
\textbf{AI redesigns online platforms for cooperation and empathy} (2 chains, 2 speakers). Online platforms are distorted by popularity incentives, divisiveness, and lack of inhibition. \ensuremath{\rightarrow} AI can help redesign protocols and incentives or act as an empathetic mediator or friend. \ensuremath{\rightarrow} Improved incentives and empathetic AI mediation enhance online cooperation and growth. & \textbf{Digital mediation erodes empathy through missing cues} (1 chain, 1 speaker). Human connection depends heavily on nonverbal signals, especially facial expressions. \ensuremath{\rightarrow} Digital and AI-mediated communication removes or breaks these nonverbal signals. \ensuremath{\rightarrow} This makes genuine connection harder and erodes empathy skills. \ensuremath{\rightarrow} Empathy erosion increases dehumanizing behavior and weakens human bonds. \\
\textbf{AI finds common ground to bridge divides} (2 chains, 1 speaker). Online interactions are dismissive and obscure shared humanity. \ensuremath{\rightarrow} AI can identify commonalities between people and bring them together in humanizing ways. \ensuremath{\rightarrow} Recognizing commonality and shared humanity fosters empathy and bridges divides. & \textbf{Algorithmic curation homogenizes culture} (1 chain, 1 speaker). People increasingly experience social life online through algorithmic feeds. \ensuremath{\rightarrow} Algorithmic promotion pushes people toward a homogeneous set of shifting ideals and content. \ensuremath{\rightarrow} When experiences are algorithmically mediated, it becomes harder to be unique or original. \ensuremath{\rightarrow} Unique characters and interesting people are eliminated, leaving culture homogeneous and bland. \\
\textbf{Algorithmic curation cools divisive social graphs} (1 chain, 1 speaker). Social graph distribution limits content to connections and amplifies emotionally charged or divisive content. \ensuremath{\rightarrow} Algorithmic content matching decouples discovery from social connections. \ensuremath{\rightarrow} Decoupling reduces amplification of divisive content and surfaces high-quality content. & \textbf{AI value systems fragment society into tribes} (1 chain, 1 speaker). Humans form group identities around shared values and systems such as nations or legal codes. \ensuremath{\rightarrow} Future AI systems will develop distinct identities, values, or rule sets. \ensuremath{\rightarrow} People will align with different AI systems the way they align with different nations or legal systems. \ensuremath{\rightarrow} Society fragments into AI-aligned tribes. \\
\textbf{Immersive avatars blend digital and physical social worlds} (1 chain, 1 speaker). Immersive technologies create photorealistic avatars and remote presence. \ensuremath{\rightarrow} Widespread adoption blends digital and physical realities. \ensuremath{\rightarrow} Avatars and embodied AIs become common participants in the real world. & \textbf{AI bridging needs human willingness} (1 chain, 1 speaker). AI can analyze people\textquotesingle s experiences and actions to identify commonalities between them. \ensuremath{\rightarrow} But finding overlap does not by itself produce social bonding or bridging. \ensuremath{\rightarrow} Bridging divides requires humans to be motivated to seek commonality and recognize shared humanity. \ensuremath{\rightarrow} Technical capability cannot substitute for human willingness. \\
\textbf{BrainNet transmits emotion deepening human ties} (1 chain, 1 speaker). Memory and emotion transmission are advancing from animals toward humans. \ensuremath{\rightarrow} BrainNet will transmit emotions, feelings, and sensations over the internet. \ensuremath{\rightarrow} Emotional telepathic communication deepens empathy and expands human connectivity. & \textbf{Technological escapism distances humanity from connection} (1 chain, 1 speaker). AI and global connectivity increase mediated interaction and reshape how humans relate to each other. \ensuremath{\rightarrow} New technologies, including AI, represent an attempt to escape natural conditions on Earth. \ensuremath{\rightarrow} This attempt distances humanity from the natural world and from meaningful human connection. \ensuremath{\rightarrow} The resulting disconnection produces unintended negative consequences. \\
\textbf{Democratized AI creation builds new communities} (1 chain, 1 speaker). Open-source AI makes content creation tools globally accessible. \ensuremath{\rightarrow} Democratized access removes the production and distribution barrier that sustains media monopolies. \ensuremath{\rightarrow} Independent global creation enables new forms of social organization and community. & \\
\textbf{Deceptive AI jeopardizes social trust} (1 chain, 1 speaker). AI for cooperative games may require language models capable of lying. \ensuremath{\rightarrow} AI that can deceive people creates serious risks. \ensuremath{\rightarrow} Whether to develop such deceptive AIs must be carefully weighed. & \\
\end{longtable}

\subsection*{Table S15. Reliability: Ensemble alignment at every pipeline stage}

\begin{longtable}[]{@{}
  >{\raggedright\arraybackslash}p{(\columnwidth - 8\tabcolsep) * \real{0.1754}}
  >{\raggedright\arraybackslash}p{(\columnwidth - 8\tabcolsep) * \real{0.4386}}
  >{\raggedright\arraybackslash}p{(\columnwidth - 8\tabcolsep) * \real{0.1287}}
  >{\raggedright\arraybackslash}p{(\columnwidth - 8\tabcolsep) * \real{0.1287}}
  >{\raggedright\arraybackslash}p{(\columnwidth - 8\tabcolsep) * \real{0.1287}}@{}}
\toprule\noalign{}
\begin{minipage}[b]{\linewidth}\raggedright
\textbf{Stage}
\end{minipage} & \begin{minipage}[b]{\linewidth}\raggedright
\textbf{Metric}
\end{minipage} & \begin{minipage}[b]{\linewidth}\raggedright
\textbf{X-risk}
\end{minipage} & \begin{minipage}[b]{\linewidth}\raggedright
\textbf{E-risk}
\end{minipage} & \begin{minipage}[b]{\linewidth}\raggedright
\textbf{S-risk}
\end{minipage} \\
\midrule\noalign{}
\endhead
\bottomrule\noalign{}
\endlastfoot
Step 1: disagreement check & pairs scored & 9,182 & 2,358 & 1,030 \\
& full three-way consistency & 86.4\% & 75.7\% & 73.1\% \\
Step 2: divergence identification & both workers found the conflict (lenient) & 74.2\% & 69.3\% & 70.9\% \\
& + exact localization agreed (strict) & 59.9\% & 57.1\% & 55.1\% \\
& DeepSeek lenient / Claude lenient & 77.1\% / 85.8\% & 72.8\% / 87.5\% & 74.3\% / 86.5\% \\
& type agreement, conditional on match & 84.6\% & 82.9\% & 87.2\% \\
Step 3: root/derivative & both workers matched final classification & 72.5\% & 80.0\% & 86.5\% \\
& neither worker matched & 1.0\% & 0.2\% & 0.0\% \\
Aggregation: discovery & chunk-level mechanisms proposed by both workers & 75.7\% & 73.3\% & 77.1\% \\
Aggregation: merge & final mechanisms supported by both workers & 74.3\% & 92.1\% & 100\% \\
\end{longtable}

\emph{The lenient/strict gap at step 2 reflects differences in where models anchor a conflict within a chain, not in which conflicts they identify. Human validation of the extraction stage (two independent reviewers, 200 reasoning chains): no hallucinated content identified by either reviewer; 196/200 reasoning chains without reservation from either reviewer; see SI text. The cross-side alignment of mechanism catalogs into underlying questions was performed by the authors; the complete mechanism-to-theme mapping is in the deposited data.}

\end{document}